\documentclass[traditabstract,longauth]{aa}

\def\setsymbol#1#2{\expandafter\def\csname #1\endcsname{#2}}
\def\getsymbol#1{\csname #1\endcsname}

\def\Planck{\textit{Planck}}

\def\all2013resultspapers{\nocite{planck2013-p01, planck2013-p02, planck2013-p02a, planck2013-p02d, planck2013-p02b, planck2013-p03, planck2013-p03c, planck2013-p03f, planck2013-p03d, planck2013-p03e, planck2013-p01a, planck2013-p06, planck2013-p03a, planck2013-pip88, planck2013-p08, planck2013-p11, planck2013-p12, planck2013-p13, planck2013-p14, planck2013-p15, planck2013-p05b, planck2013-p17, planck2013-p09, planck2013-p09a, planck2013-p20, planck2013-p19, planck2013-pipaberration, planck2013-p05, planck2013-p05a, planck2013-pip56, planck2013-p06b}}

\newbox\tablebox    \newdimen\tablewidth
\def\leaderfil{\leaders\hbox to 5pt{\hss.\hss}\hfil}
\def\endPlancktable{\tablewidth=\columnwidth 
    $$\hss\copy\tablebox\hss$$
    \vskip-\lastskip\vskip -2pt}
\def\endPlancktablewide{\tablewidth=\textwidth 
    $$\hss\copy\tablebox\hss$$
    \vskip-\lastskip\vskip -2pt}
\def\tablenote#1 #2\par{\begingroup \parindent=0.8em
    \abovedisplayshortskip=0pt\belowdisplayshortskip=0pt
    \noindent
    $$\hss\vbox{\hsize\tablewidth \hangindent=\parindent \hangafter=1 \noindent
    \hbox to \parindent{$^#1$\hss}\strut#2\strut\par}\hss$$
    \endgroup}
\def\doubleline{\vskip 3pt\hrule \vskip 1.5pt \hrule \vskip 5pt}

\def\L2{\ifmmode L_2\else $L_2$\fi}

\def\DeltaT{\ifmmode \Delta T\else $\Delta T$\fi}
\def\deltat{\ifmmode \Delta t\else $\Delta t$\fi}
\def\fknee{\ifmmode f_{\rm knee}\else $f_{\rm knee}$\fi}
\def\Fmax{\ifmmode F_{\rm max}\else $F_{\rm max}$\fi}
\def\solar{\ifmmode{\rm M}_{\mathord\odot}\else${\rm M}_{\mathord\odot}$\fi}
\def\Msolar{\ifmmode{\rm M}_{\mathord\odot}\else${\rm M}_{\mathord\odot}$\fi}
\def\Lsolar{\ifmmode{\rm L}_{\mathord\odot}\else${\rm L}_{\mathord\odot}$\fi}

\def\inv{\ifmmode^{-1}\else$^{-1}$\fi}
\def\mo{\ifmmode^{-1}\else$^{-1}$\fi}
\def\sup#1{\ifmmode ^{\rm #1}\else $^{\rm #1}$\fi}
\def\expo#1{\ifmmode \times 10^{#1}\else $\times 10^{#1}$\fi}
\def\,{\thinspace}
\def\lsim{\mathrel{\raise .4ex\hbox{\rlap{$<$}\lower 1.2ex\hbox{$\sim$}}}}
\def\gsim{\mathrel{\raise .4ex\hbox{\rlap{$>$}\lower 1.2ex\hbox{$\sim$}}}}

\def\simprop{\mathrel{\raise .4ex\hbox{\rlap{$\propto$}\lower 1.2ex\hbox{$\sim$}}}}
\def\deg{\ifmmode^\circ\else$^\circ$\fi}
\def\pdeg{\ifmmode $\setbox0=\hbox{$^{\circ}$}\rlap{\hskip.11\wd0 .}$^{\circ}
          \else \setbox0=\hbox{$^{\circ}$}\rlap{\hskip.11\wd0 .}$^{\circ}$\fi}
\def\arcs{\ifmmode {^{\scriptstyle\prime\prime}}
          \else $^{\scriptstyle\prime\prime}$\fi}
\def\arcm{\ifmmode {^{\scriptstyle\prime}}
          \else $^{\scriptstyle\prime}$\fi}
\newdimen\sa  \newdimen\sb
\def\parcs{\sa=.07em \sb=.03em
     \ifmmode \hbox{\rlap{.}}^{\scriptstyle\prime\kern -\sb\prime}\hbox{\kern -\sa}
     \else \rlap{.}$^{\scriptstyle\prime\kern -\sb\prime}$\kern -\sa\fi}
\def\parcm{\sa=.08em \sb=.03em
     \ifmmode \hbox{\rlap{.}\kern\sa}^{\scriptstyle\prime}\hbox{\kern-\sb}
     \else \rlap{.}\kern\sa$^{\scriptstyle\prime}$\kern-\sb\fi}
\def\ra[#1 #2 #3.#4]{#1\sup{h}#2\sup{m}#3\sup{s}\llap.#4}
\def\dec[#1 #2 #3.#4]{#1\deg#2\arcm#3\arcs\llap.#4}
\def\deco[#1 #2 #3]{#1\deg#2\arcm#3\arcs}
\def\rra[#1 #2]{#1\sup{h}#2\sup{m}}

\def\dots{\relax\ifmmode \ldots\else $\ldots$\fi}
\def\WHzsr{\ifmmode $W\,Hz\mo\,sr\mo$\else W\,Hz\mo\,sr\mo\fi}
\def\mHz{\ifmmode $\,mHz$\else \,mHz\fi}
\def\GHz{\ifmmode $\,GHz$\else \,GHz\fi}
\def\mKs{\ifmmode $\,mK\,s$^{1/2}\else \,mK\,s$^{1/2}$\fi}
\def\muKs{\ifmmode \,\mu$K\,s$^{1/2}\else \,$\mu$K\,s$^{1/2}$\fi}
\def\muKRJs{\ifmmode \,\mu$K$_{\rm RJ}$\,s$^{1/2}\else \,$\mu$K$_{\rm RJ}$\,s$^{1/2}$\fi}
\def\muKHz{\ifmmode \,\mu$K\,Hz$^{-1/2}\else \,$\mu$K\,Hz$^{-1/2}$\fi}
\def\MJysr{\ifmmode \,$MJy\,sr\mo$\else \,MJy\,sr\mo\fi}
\def\MJysrmK{\ifmmode \,$MJy\,sr\mo$\,mK$_{\rm CMB}\mo\else \,MJy\,sr\mo\,mK$_{\rm CMB}\mo$\fi}
\def\microns{\ifmmode \,\mu$m$\else \,$\mu$m\fi}

\def\muK{\ifmmode \,\mu$K$\else \,$\mu$\hbox{K}\fi}
\def\microK{\ifmmode \,\mu$K$\else \,$\mu$\hbox{K}\fi}
\def\muW{\ifmmode \,\mu$W$\else \,$\mu$\hbox{W}\fi}
\def\kms{\ifmmode $\,km\,s$^{-1}\else \,km\,s$^{-1}$\fi}
\def\kmsMpc{\ifmmode $\,\kms\,Mpc\mo$\else \,\kms\,Mpc\mo\fi}
\setsymbol{LFI:center:frequency:70GHz:units}{70.3\,GHz}
\setsymbol{LFI:center:frequency:44GHz:units}{44.1\,GHz}
\setsymbol{LFI:center:frequency:30GHz:units}{28.5\,GHz}

\setsymbol{LFI:center:frequency:70GHz}{70.3}
\setsymbol{LFI:center:frequency:44GHz}{44.1}
\setsymbol{LFI:center:frequency:30GHz}{28.5}

\setsymbol{LFI:center:frequency:LFI18:Rad:M:units}{71.7\GHz}
\setsymbol{LFI:center:frequency:LFI19:Rad:M:units}{67.5\GHz}
\setsymbol{LFI:center:frequency:LFI20:Rad:M:units}{69.2\GHz}
\setsymbol{LFI:center:frequency:LFI21:Rad:M:units}{70.4\GHz}
\setsymbol{LFI:center:frequency:LFI22:Rad:M:units}{71.5\GHz}
\setsymbol{LFI:center:frequency:LFI23:Rad:M:units}{70.8\GHz}
\setsymbol{LFI:center:frequency:LFI24:Rad:M:units}{44.4\GHz}
\setsymbol{LFI:center:frequency:LFI25:Rad:M:units}{44.0\GHz}
\setsymbol{LFI:center:frequency:LFI26:Rad:M:units}{43.9\GHz}
\setsymbol{LFI:center:frequency:LFI27:Rad:M:units}{28.3\GHz}
\setsymbol{LFI:center:frequency:LFI28:Rad:M:units}{28.8\GHz}
\setsymbol{LFI:center:frequency:LFI18:Rad:S:units}{70.1\GHz}
\setsymbol{LFI:center:frequency:LFI19:Rad:S:units}{69.6\GHz}
\setsymbol{LFI:center:frequency:LFI20:Rad:S:units}{69.5\GHz}
\setsymbol{LFI:center:frequency:LFI21:Rad:S:units}{69.5\GHz}
\setsymbol{LFI:center:frequency:LFI22:Rad:S:units}{72.8\GHz}
\setsymbol{LFI:center:frequency:LFI23:Rad:S:units}{71.3\GHz}
\setsymbol{LFI:center:frequency:LFI24:Rad:S:units}{44.1\GHz}
\setsymbol{LFI:center:frequency:LFI25:Rad:S:units}{44.1\GHz}
\setsymbol{LFI:center:frequency:LFI26:Rad:S:units}{44.1\GHz}
\setsymbol{LFI:center:frequency:LFI27:Rad:S:units}{28.5\GHz}
\setsymbol{LFI:center:frequency:LFI28:Rad:S:units}{28.2\GHz}

\setsymbol{LFI:center:frequency:LFI18:Rad:M}{71.7}
\setsymbol{LFI:center:frequency:LFI19:Rad:M}{67.5}
\setsymbol{LFI:center:frequency:LFI20:Rad:M}{69.2}
\setsymbol{LFI:center:frequency:LFI21:Rad:M}{70.4}
\setsymbol{LFI:center:frequency:LFI22:Rad:M}{71.5}
\setsymbol{LFI:center:frequency:LFI23:Rad:M}{70.8}
\setsymbol{LFI:center:frequency:LFI24:Rad:M}{44.4}
\setsymbol{LFI:center:frequency:LFI25:Rad:M}{44.0}
\setsymbol{LFI:center:frequency:LFI26:Rad:M}{43.9}
\setsymbol{LFI:center:frequency:LFI27:Rad:M}{28.3}
\setsymbol{LFI:center:frequency:LFI28:Rad:M}{28.8}
\setsymbol{LFI:center:frequency:LFI18:Rad:S}{70.1}
\setsymbol{LFI:center:frequency:LFI19:Rad:S}{69.6}
\setsymbol{LFI:center:frequency:LFI20:Rad:S}{69.5}
\setsymbol{LFI:center:frequency:LFI21:Rad:S}{69.5}
\setsymbol{LFI:center:frequency:LFI22:Rad:S}{72.8}
\setsymbol{LFI:center:frequency:LFI23:Rad:S}{71.3}
\setsymbol{LFI:center:frequency:LFI24:Rad:S}{44.1}
\setsymbol{LFI:center:frequency:LFI25:Rad:S}{44.1}
\setsymbol{LFI:center:frequency:LFI26:Rad:S}{44.1}
\setsymbol{LFI:center:frequency:LFI27:Rad:S}{28.5}
\setsymbol{LFI:center:frequency:LFI28:Rad:S}{28.2}

\setsymbol{LFI:white:noise:sensitivity:70GHz:units}{134.7\muKs}
\setsymbol{LFI:white:noise:sensitivity:44GHz:units}{164.7\muKs}
\setsymbol{LFI:white:noise:sensitivity:30GHz:units}{143.4\muKs}

\setsymbol{LFI:white:noise:sensitivity:70GHz}{134.7}
\setsymbol{LFI:white:noise:sensitivity:44GHz}{164.7}
\setsymbol{LFI:white:noise:sensitivity:30GHz}{143.4}

\setsymbol{LFI:white:noise:sensitivity:LFI18:Rad:M:units}{512.0\muKs}
\setsymbol{LFI:white:noise:sensitivity:LFI19:Rad:M:units}{581.4\muKs}
\setsymbol{LFI:white:noise:sensitivity:LFI20:Rad:M:units}{590.8\muKs}
\setsymbol{LFI:white:noise:sensitivity:LFI21:Rad:M:units}{455.2\muKs}
\setsymbol{LFI:white:noise:sensitivity:LFI22:Rad:M:units}{492.0\muKs}
\setsymbol{LFI:white:noise:sensitivity:LFI23:Rad:M:units}{507.7\muKs}
\setsymbol{LFI:white:noise:sensitivity:LFI24:Rad:M:units}{462.2\muKs}
\setsymbol{LFI:white:noise:sensitivity:LFI25:Rad:M:units}{413.6\muKs}
\setsymbol{LFI:white:noise:sensitivity:LFI26:Rad:M:units}{478.6\muKs}
\setsymbol{LFI:white:noise:sensitivity:LFI27:Rad:M:units}{277.7\muKs}
\setsymbol{LFI:white:noise:sensitivity:LFI28:Rad:M:units}{312.3\muKs}
\setsymbol{LFI:white:noise:sensitivity:LFI18:Rad:S:units}{465.7\muKs}
\setsymbol{LFI:white:noise:sensitivity:LFI19:Rad:S:units}{555.6\muKs}
\setsymbol{LFI:white:noise:sensitivity:LFI20:Rad:S:units}{623.2\muKs}
\setsymbol{LFI:white:noise:sensitivity:LFI21:Rad:S:units}{564.1\muKs}
\setsymbol{LFI:white:noise:sensitivity:LFI22:Rad:S:units}{534.4\muKs}
\setsymbol{LFI:white:noise:sensitivity:LFI23:Rad:S:units}{542.4\muKs}
\setsymbol{LFI:white:noise:sensitivity:LFI24:Rad:S:units}{399.2\muKs}
\setsymbol{LFI:white:noise:sensitivity:LFI25:Rad:S:units}{392.6\muKs}
\setsymbol{LFI:white:noise:sensitivity:LFI26:Rad:S:units}{418.6\muKs}
\setsymbol{LFI:white:noise:sensitivity:LFI27:Rad:S:units}{302.9\muKs}
\setsymbol{LFI:white:noise:sensitivity:LFI28:Rad:S:units}{285.3\muKs}

\setsymbol{LFI:white:noise:sensitivity:LFI18:Rad:M}{512.0}
\setsymbol{LFI:white:noise:sensitivity:LFI19:Rad:M}{581.4}
\setsymbol{LFI:white:noise:sensitivity:LFI20:Rad:M}{590.8}
\setsymbol{LFI:white:noise:sensitivity:LFI21:Rad:M}{455.2}
\setsymbol{LFI:white:noise:sensitivity:LFI22:Rad:M}{492.0}
\setsymbol{LFI:white:noise:sensitivity:LFI23:Rad:M}{507.7}
\setsymbol{LFI:white:noise:sensitivity:LFI24:Rad:M}{462.2}
\setsymbol{LFI:white:noise:sensitivity:LFI25:Rad:M}{413.6}
\setsymbol{LFI:white:noise:sensitivity:LFI26:Rad:M}{478.6}
\setsymbol{LFI:white:noise:sensitivity:LFI27:Rad:M}{277.7}
\setsymbol{LFI:white:noise:sensitivity:LFI28:Rad:M}{312.3}
\setsymbol{LFI:white:noise:sensitivity:LFI18:Rad:S}{465.7}
\setsymbol{LFI:white:noise:sensitivity:LFI19:Rad:S}{555.6}
\setsymbol{LFI:white:noise:sensitivity:LFI20:Rad:S}{623.2}
\setsymbol{LFI:white:noise:sensitivity:LFI21:Rad:S}{564.1}
\setsymbol{LFI:white:noise:sensitivity:LFI22:Rad:S}{534.4}
\setsymbol{LFI:white:noise:sensitivity:LFI23:Rad:S}{542.4}
\setsymbol{LFI:white:noise:sensitivity:LFI24:Rad:S}{399.2}
\setsymbol{LFI:white:noise:sensitivity:LFI25:Rad:S}{392.6}
\setsymbol{LFI:white:noise:sensitivity:LFI26:Rad:S}{418.6}
\setsymbol{LFI:white:noise:sensitivity:LFI27:Rad:S}{302.9}
\setsymbol{LFI:white:noise:sensitivity:LFI28:Rad:S}{285.3}

\setsymbol{LFI:knee:frequency:70GHz:units}{29.5\mHz}
\setsymbol{LFI:knee:frequency:44GHz:units}{56.2\mHz}
\setsymbol{LFI:knee:frequency:30GHz:units}{113.7\mHz}

\setsymbol{LFI:knee:frequency:70GHz}{29.5}
\setsymbol{LFI:knee:frequency:44GHz}{56.2}
\setsymbol{LFI:knee:frequency:30GHz}{113.7}

\setsymbol{LFI:knee:frequency:LFI18:Rad:M:units}{16.3\mHz}
\setsymbol{LFI:knee:frequency:LFI19:Rad:M:units}{15.1\mHz}
\setsymbol{LFI:knee:frequency:LFI20:Rad:M:units}{18.7\mHz}
\setsymbol{LFI:knee:frequency:LFI21:Rad:M:units}{37.2\mHz}
\setsymbol{LFI:knee:frequency:LFI22:Rad:M:units}{12.7\mHz}
\setsymbol{LFI:knee:frequency:LFI23:Rad:M:units}{34.6\mHz}
\setsymbol{LFI:knee:frequency:LFI24:Rad:M:units}{46.2\mHz}
\setsymbol{LFI:knee:frequency:LFI25:Rad:M:units}{24.9\mHz}
\setsymbol{LFI:knee:frequency:LFI26:Rad:M:units}{67.6\mHz}
\setsymbol{LFI:knee:frequency:LFI27:Rad:M:units}{187.4\mHz}
\setsymbol{LFI:knee:frequency:LFI28:Rad:M:units}{122.2\mHz}
\setsymbol{LFI:knee:frequency:LFI18:Rad:S:units}{17.7\mHz}
\setsymbol{LFI:knee:frequency:LFI19:Rad:S:units}{22.0\mHz}
\setsymbol{LFI:knee:frequency:LFI20:Rad:S:units}{8.7\mHz}
\setsymbol{LFI:knee:frequency:LFI21:Rad:S:units}{25.9\mHz}
\setsymbol{LFI:knee:frequency:LFI22:Rad:S:units}{15.8\mHz}
\setsymbol{LFI:knee:frequency:LFI23:Rad:S:units}{129.8\mHz}
\setsymbol{LFI:knee:frequency:LFI24:Rad:S:units}{100.9\mHz}
\setsymbol{LFI:knee:frequency:LFI25:Rad:S:units}{38.9\mHz}
\setsymbol{LFI:knee:frequency:LFI26:Rad:S:units}{58.9\mHz}
\setsymbol{LFI:knee:frequency:LFI27:Rad:S:units}{104.4\mHz}
\setsymbol{LFI:knee:frequency:LFI28:Rad:S:units}{40.7\mHz}

\setsymbol{LFI:knee:frequency:LFI18:Rad:M}{16.3}
\setsymbol{LFI:knee:frequency:LFI19:Rad:M}{15.1}
\setsymbol{LFI:knee:frequency:LFI20:Rad:M}{18.7}
\setsymbol{LFI:knee:frequency:LFI21:Rad:M}{37.2}
\setsymbol{LFI:knee:frequency:LFI22:Rad:M}{12.7}
\setsymbol{LFI:knee:frequency:LFI23:Rad:M}{34.6}
\setsymbol{LFI:knee:frequency:LFI24:Rad:M}{46.2}
\setsymbol{LFI:knee:frequency:LFI25:Rad:M}{24.9}
\setsymbol{LFI:knee:frequency:LFI26:Rad:M}{67.6}
\setsymbol{LFI:knee:frequency:LFI27:Rad:M}{187.4}
\setsymbol{LFI:knee:frequency:LFI28:Rad:M}{122.2}
\setsymbol{LFI:knee:frequency:LFI18:Rad:S}{17.7}
\setsymbol{LFI:knee:frequency:LFI19:Rad:S}{22.0}
\setsymbol{LFI:knee:frequency:LFI20:Rad:S}{8.7}
\setsymbol{LFI:knee:frequency:LFI21:Rad:S}{25.9}
\setsymbol{LFI:knee:frequency:LFI22:Rad:S}{15.8}
\setsymbol{LFI:knee:frequency:LFI23:Rad:S}{129.8}
\setsymbol{LFI:knee:frequency:LFI24:Rad:S}{100.9}
\setsymbol{LFI:knee:frequency:LFI25:Rad:S}{38.9}
\setsymbol{LFI:knee:frequency:LFI26:Rad:S}{58.9}
\setsymbol{LFI:knee:frequency:LFI27:Rad:S}{104.4}
\setsymbol{LFI:knee:frequency:LFI28:Rad:S}{40.7}

\setsymbol{LFI:slope:70GHz:units}{$-1.03$\mHz}
\setsymbol{LFI:slope:44GHz:units}{$-0.89$\mHz}
\setsymbol{LFI:slope:30GHz:units}{$-0.87$\mHz}

\setsymbol{LFI:slope:70GHz}{$-1.03$}
\setsymbol{LFI:slope:44GHz}{$-0.89$}
\setsymbol{LFI:slope:30GHz}{$-0.87$}

\setsymbol{LFI:slope:LFI18:Rad:M:units}{$-1.04$\mHz}
\setsymbol{LFI:slope:LFI19:Rad:M:units}{$-1.09$\mHz}
\setsymbol{LFI:slope:LFI20:Rad:M:units}{$-0.69$\mHz}
\setsymbol{LFI:slope:LFI21:Rad:M:units}{$-1.56$\mHz}
\setsymbol{LFI:slope:LFI22:Rad:M:units}{$-1.01$\mHz}
\setsymbol{LFI:slope:LFI23:Rad:M:units}{$-0.96$\mHz}
\setsymbol{LFI:slope:LFI24:Rad:M:units}{$-0.83$\mHz}
\setsymbol{LFI:slope:LFI25:Rad:M:units}{$-0.91$\mHz}
\setsymbol{LFI:slope:LFI26:Rad:M:units}{$-0.95$\mHz}
\setsymbol{LFI:slope:LFI27:Rad:M:units}{$-0.87$\mHz}
\setsymbol{LFI:slope:LFI28:Rad:M:units}{$-0.88$\mHz}
\setsymbol{LFI:slope:LFI18:Rad:S:units}{$-1.15$\mHz}
\setsymbol{LFI:slope:LFI19:Rad:S:units}{$-1.00$\mHz}
\setsymbol{LFI:slope:LFI20:Rad:S:units}{$-0.95$\mHz}
\setsymbol{LFI:slope:LFI21:Rad:S:units}{$-0.92$\mHz}
\setsymbol{LFI:slope:LFI22:Rad:S:units}{$-1.01$\mHz}
\setsymbol{LFI:slope:LFI23:Rad:S:units}{$-0.95$\mHz}
\setsymbol{LFI:slope:LFI24:Rad:S:units}{$-0.73$\mHz}
\setsymbol{LFI:slope:LFI25:Rad:S:units}{$-1.16$\mHz}
\setsymbol{LFI:slope:LFI26:Rad:S:units}{$-0.79$\mHz}
\setsymbol{LFI:slope:LFI27:Rad:S:units}{$-0.82$\mHz}
\setsymbol{LFI:slope:LFI28:Rad:S:units}{$-0.91$\mHz}

\setsymbol{LFI:slope:LFI18:Rad:M}{$-1.04$}
\setsymbol{LFI:slope:LFI19:Rad:M}{$-1.09$}
\setsymbol{LFI:slope:LFI20:Rad:M}{$-0.69$}
\setsymbol{LFI:slope:LFI21:Rad:M}{$-1.56$}
\setsymbol{LFI:slope:LFI22:Rad:M}{$-1.01$}
\setsymbol{LFI:slope:LFI23:Rad:M}{$-0.96$}
\setsymbol{LFI:slope:LFI24:Rad:M}{$-0.83$}
\setsymbol{LFI:slope:LFI25:Rad:M}{$-0.91$}
\setsymbol{LFI:slope:LFI26:Rad:M}{$-0.95$}
\setsymbol{LFI:slope:LFI27:Rad:M}{$-0.87$}
\setsymbol{LFI:slope:LFI28:Rad:M}{$-0.88$}
\setsymbol{LFI:slope:LFI18:Rad:S}{$-1.15$}
\setsymbol{LFI:slope:LFI19:Rad:S}{$-1.00$}
\setsymbol{LFI:slope:LFI20:Rad:S}{$-0.95$}
\setsymbol{LFI:slope:LFI21:Rad:S}{$-0.92$}
\setsymbol{LFI:slope:LFI22:Rad:S}{$-1.01$}
\setsymbol{LFI:slope:LFI23:Rad:S}{$-0.95$}
\setsymbol{LFI:slope:LFI24:Rad:S}{$-0.73$}
\setsymbol{LFI:slope:LFI25:Rad:S}{$-1.16$}
\setsymbol{LFI:slope:LFI26:Rad:S}{$-0.79$}
\setsymbol{LFI:slope:LFI27:Rad:S}{$-0.82$}
\setsymbol{LFI:slope:LFI28:Rad:S}{$-0.91$}

\setsymbol{LFI:FWHM:70GHz:units}{13\parcm01}
\setsymbol{LFI:FWHM:44GHz:units}{27\parcm92}
\setsymbol{LFI:FWHM:30GHz:units}{32\parcm65}

\setsymbol{LFI:FWHM:70GHz}{13.01}
\setsymbol{LFI:FWHM:44GHz}{27.92}
\setsymbol{LFI:FWHM:30GHz}{32.65}

\setsymbol{LFI:FWHM:LFI18:units}{13\parcm39}
\setsymbol{LFI:FWHM:LFI19:units}{13\parcm01}
\setsymbol{LFI:FWHM:LFI20:units}{12\parcm75}
\setsymbol{LFI:FWHM:LFI21:units}{12\parcm74}
\setsymbol{LFI:FWHM:LFI22:units}{12\parcm87}
\setsymbol{LFI:FWHM:LFI23:units}{13\parcm27}
\setsymbol{LFI:FWHM:LFI24:units}{22\parcm98}
\setsymbol{LFI:FWHM:LFI25:units}{30\parcm46}
\setsymbol{LFI:FWHM:LFI26:units}{30\parcm31}
\setsymbol{LFI:FWHM:LFI27:units}{32\parcm65}
\setsymbol{LFI:FWHM:LFI28:units}{32\parcm66}

\setsymbol{LFI:FWHM:LFI18}{13.39}
\setsymbol{LFI:FWHM:LFI19}{13.01}
\setsymbol{LFI:FWHM:LFI20}{12.75}
\setsymbol{LFI:FWHM:LFI21}{12.74}
\setsymbol{LFI:FWHM:LFI22}{12.87}
\setsymbol{LFI:FWHM:LFI23}{13.27}
\setsymbol{LFI:FWHM:LFI24}{22.98}
\setsymbol{LFI:FWHM:LFI25}{30.46}
\setsymbol{LFI:FWHM:LFI26}{30.31}
\setsymbol{LFI:FWHM:LFI27}{32.65}
\setsymbol{LFI:FWHM:LFI28}{32.66}

\setsymbol{LFI:FWHM:uncertainty:LFI18:units}{0.170\arcm}
\setsymbol{LFI:FWHM:uncertainty:LFI19:units}{0.174\arcm}
\setsymbol{LFI:FWHM:uncertainty:LFI20:units}{0.170\arcm}
\setsymbol{LFI:FWHM:uncertainty:LFI21:units}{0.156\arcm}
\setsymbol{LFI:FWHM:uncertainty:LFI22:units}{0.164\arcm}
\setsymbol{LFI:FWHM:uncertainty:LFI23:units}{0.171\arcm}
\setsymbol{LFI:FWHM:uncertainty:LFI24:units}{0.652\arcm}
\setsymbol{LFI:FWHM:uncertainty:LFI25:units}{1.075\arcm}
\setsymbol{LFI:FWHM:uncertainty:LFI26:units}{1.131\arcm}
\setsymbol{LFI:FWHM:uncertainty:LFI27:units}{1.266\arcm}
\setsymbol{LFI:FWHM:uncertainty:LFI28:units}{1.287\arcm}

\setsymbol{LFI:FWHM:uncertainty:LFI18}{0.170}
\setsymbol{LFI:FWHM:uncertainty:LFI19}{0.174}
\setsymbol{LFI:FWHM:uncertainty:LFI20}{0.170}
\setsymbol{LFI:FWHM:uncertainty:LFI21}{0.156}
\setsymbol{LFI:FWHM:uncertainty:LFI22}{0.164}
\setsymbol{LFI:FWHM:uncertainty:LFI23}{0.171}
\setsymbol{LFI:FWHM:uncertainty:LFI24}{0.652}
\setsymbol{LFI:FWHM:uncertainty:LFI25}{1.075}
\setsymbol{LFI:FWHM:uncertainty:LFI26}{1.131}
\setsymbol{LFI:FWHM:uncertainty:LFI27}{1.266}
\setsymbol{LFI:FWHM:uncertainty:LFI28}{1.287}

\setsymbol{HFI:center:frequency:100GHz:units}{100\,GHz}
\setsymbol{HFI:center:frequency:143GHz:units}{143\,GHz}
\setsymbol{HFI:center:frequency:217GHz:units}{217\,GHz}
\setsymbol{HFI:center:frequency:353GHz:units}{353\,GHz}
\setsymbol{HFI:center:frequency:545GHz:units}{545\,GHz}
\setsymbol{HFI:center:frequency:857GHz:units}{857\,GHz}

\setsymbol{HFI:center:frequency:100GHz}{100}
\setsymbol{HFI:center:frequency:143GHz}{143}
\setsymbol{HFI:center:frequency:217GHz}{217}
\setsymbol{HFI:center:frequency:353GHz}{353}
\setsymbol{HFI:center:frequency:545GHz}{545}
\setsymbol{HFI:center:frequency:857GHz}{857}

\setsymbol{HFI:Ndetectors:100GHz}{8}
\setsymbol{HFI:Ndetectors:143GHz}{11}
\setsymbol{HFI:Ndetectors:217GHz}{12}
\setsymbol{HFI:Ndetectors:353GHz}{12}
\setsymbol{HFI:Ndetectors:545GHz}{3}
\setsymbol{HFI:Ndetectors:857GHz}{4}

\setsymbol{HFI:FWHM:Maps:100GHz:units}{9\parcm88}
\setsymbol{HFI:FWHM:Maps:143GHz:units}{7\parcm18}
\setsymbol{HFI:FWHM:Maps:217GHz:units}{4\parcm87}
\setsymbol{HFI:FWHM:Maps:353GHz:units}{4\parcm65}
\setsymbol{HFI:FWHM:Maps:545GHz:units}{4\parcm72}
\setsymbol{HFI:FWHM:Maps:857GHz:units}{4\parcm39}
\setsymbol{HFI:FWHM:Maps:100GHz}{9.88}
\setsymbol{HFI:FWHM:Maps:143GHz}{7.18}
\setsymbol{HFI:FWHM:Maps:217GHz}{4.87}
\setsymbol{HFI:FWHM:Maps:353GHz}{4.65}
\setsymbol{HFI:FWHM:Maps:545GHz}{4.72}
\setsymbol{HFI:FWHM:Maps:857GHz}{4.39}

\setsymbol{HFI:beam:ellipticity:Maps:100GHz}{1.15}
\setsymbol{HFI:beam:ellipticity:Maps:143GHz}{1.01}
\setsymbol{HFI:beam:ellipticity:Maps:217GHz}{1.06}
\setsymbol{HFI:beam:ellipticity:Maps:353GHz}{1.05}
\setsymbol{HFI:beam:ellipticity:Maps:545GHz}{1.14}
\setsymbol{HFI:beam:ellipticity:Maps:857GHz}{1.19}

\setsymbol{HFI:FWHM:Mars:100GHz:units}{9\parcm37}
\setsymbol{HFI:FWHM:Mars:143GHz:units}{7\parcm04}
\setsymbol{HFI:FWHM:Mars:217GHz:units}{4\parcm68}
\setsymbol{HFI:FWHM:Mars:353GHz:units}{4\parcm43}
\setsymbol{HFI:FWHM:Mars:545GHz:units}{3\parcm80}
\setsymbol{HFI:FWHM:Mars:857GHz:units}{3\parcm67}

\setsymbol{HFI:FWHM:Mars:100GHz}{9.37}
\setsymbol{HFI:FWHM:Mars:143GHz}{7.04}
\setsymbol{HFI:FWHM:Mars:217GHz}{4.68}
\setsymbol{HFI:FWHM:Mars:353GHz}{4.43}
\setsymbol{HFI:FWHM:Mars:545GHz}{3.80}
\setsymbol{HFI:FWHM:Mars:857GHz}{3.67}

\setsymbol{HFI:beam:ellipticity:Mars:100GHz}{1.18}
\setsymbol{HFI:beam:ellipticity:Mars:143GHz}{1.03}
\setsymbol{HFI:beam:ellipticity:Mars:217GHz}{1.14}
\setsymbol{HFI:beam:ellipticity:Mars:353GHz}{1.09}
\setsymbol{HFI:beam:ellipticity:Mars:545GHz}{1.25}
\setsymbol{HFI:beam:ellipticity:Mars:857GHz}{1.03}

\setsymbol{HFI:CMB:relative:calibration:100GHz}{$\lsim 1\%$}
\setsymbol{HFI:CMB:relative:calibration:143GHz}{$\lsim 1\%$}
\setsymbol{HFI:CMB:relative:calibration:217GHz}{$\lsim 1\%$}
\setsymbol{HFI:CMB:relative:calibration:353GHz}{$\lsim 1\%$}
\setsymbol{HFI:CMB:relative:calibration:545GHz}{}
\setsymbol{HFI:CMB:relative:calibration:857GHz}{}

\setsymbol{HFI:CMB:absolute:calibration:100GHz}{$\lsim 2\%$}
\setsymbol{HFI:CMB:absolute:calibration:143GHz}{$\lsim 2\%$}
\setsymbol{HFI:CMB:absolute:calibration:217GHz}{$\lsim 2\%$}
\setsymbol{HFI:CMB:absolute:calibration:353GHz}{$\lsim 2\%$}
\setsymbol{HFI:CMB:absolute:calibration:545GHz}{}
\setsymbol{HFI:CMB:absolute:calibration:857GHz}{}

\setsymbol{HFI:FIRAS:gain:calibration:accuracy:statistical:100GHz}{}
\setsymbol{HFI:FIRAS:gain:calibration:accuracy:statistical:143GHz}{}
\setsymbol{HFI:FIRAS:gain:calibration:accuracy:statistical:217GHz}{}
\setsymbol{HFI:FIRAS:gain:calibration:accuracy:statistical:353GHz}{2.5\%}
\setsymbol{HFI:FIRAS:gain:calibration:accuracy:statistical:545GHz}{1\%}
\setsymbol{HFI:FIRAS:gain:calibration:accuracy:statistical:857GHz}{0.5\%}

\setsymbol{HFI:FIRAS:gain:calibration:accuracy:systematic:100GHz}{}
\setsymbol{HFI:FIRAS:gain:calibration:accuracy:systematic:143GHz}{}
\setsymbol{HFI:FIRAS:gain:calibration:accuracy:systematic:217GHz}{}
\setsymbol{HFI:FIRAS:gain:calibration:accuracy:systematic:353GHz}{}
\setsymbol{HFI:FIRAS:gain:calibration:accuracy:systematic:545GHz}{7\%}
\setsymbol{HFI:FIRAS:gain:calibration:accuracy:systematic:857GHz}{7\%}

\setsymbol{HFI:FIRAS:zero:point:accuracy:100GHz:units}{0.8\MJysr}
\setsymbol{HFI:FIRAS:zero:point:accuracy:143GHz:units}{}
\setsymbol{HFI:FIRAS:zero:point:accuracy:217GHz:units}{}
\setsymbol{HFI:FIRAS:zero:point:accuracy:353GHz:units}{1.4\MJysr}
\setsymbol{HFI:FIRAS:zero:point:accuracy:545GHz:units}{2.2\MJysr}
\setsymbol{HFI:FIRAS:zero:point:accuracy:857GHz:units}{1.7\MJysr}

\setsymbol{HFI:FIRAS:zero:point:accuracy:100GHz}{0.8}
\setsymbol{HFI:FIRAS:zero:point:accuracy:143GHz}{}
\setsymbol{HFI:FIRAS:zero:point:accuracy:217GHz}{}
\setsymbol{HFI:FIRAS:zero:point:accuracy:353GHz}{1.4}
\setsymbol{HFI:FIRAS:zero:point:accuracy:545GHz}{2.2}
\setsymbol{HFI:FIRAS:zero:point:accuracy:857GHz}{1.7}

\setsymbol{HFI:unit:conversion:100GHz:units}{0.2415\MJysrmK}
\setsymbol{HFI:unit:conversion:143GHz:units}{0.3694\MJysrmK}
\setsymbol{HFI:unit:conversion:217GHz:units}{0.4811\MJysrmK}
\setsymbol{HFI:unit:conversion:353GHz:units}{0.2883\MJysrmK}
\setsymbol{HFI:unit:conversion:545GHz:units}{0.05826\MJysrmK}
\setsymbol{HFI:unit:conversion:857GHz:units}{0.002238\MJysrmK}

\setsymbol{HFI:unit:conversion:100GHz}{0.2415}
\setsymbol{HFI:unit:conversion:143GHz}{0.3694}
\setsymbol{HFI:unit:conversion:217GHz}{0.4811}
\setsymbol{HFI:unit:conversion:353GHz}{0.2883}
\setsymbol{HFI:unit:conversion:545GHz}{0.05826}
\setsymbol{HFI:unit:conversion:857GHz}{0.002238}

\setsymbol{HFI:colour:correction:alpha=-2:V1.01:100GHz}{0.9893}
\setsymbol{HFI:colour:correction:alpha=-2:V1.01:143GHz}{0.9759}
\setsymbol{HFI:colour:correction:alpha=-2:V1.01:217GHz}{1.0007}
\setsymbol{HFI:colour:correction:alpha=-2:V1.01:353GHz}{1.0028}
\setsymbol{HFI:colour:correction:alpha=-2:V1.01:545GHz}{1.0019}
\setsymbol{HFI:colour:correction:alpha=-2:V1.01:857GHz}{0.9889}

\setsymbol{HFI:colour:correction:alpha=0:V1.01:100GHz}{1.0008}
\setsymbol{HFI:colour:correction:alpha=0:V1.01:143GHz}{1.0148}
\setsymbol{HFI:colour:correction:alpha=0:V1.01:217GHz}{0.9909}
\setsymbol{HFI:colour:correction:alpha=0:V1.01:353GHz}{0.9888}
\setsymbol{HFI:colour:correction:alpha=0:V1.01:545GHz}{0.9878}
\setsymbol{HFI:colour:correction:alpha=0:V1.01:857GHz}{1.0014}

\providecommand{\sorthelp}[1]{}

\usepackage{epsfig}
\usepackage{graphicx}
\usepackage[dvips]{color}
\usepackage{natbib}
\usepackage[breaklinks,colorlinks, citecolor=blue]{hyperref}
\usepackage{breakurl}
\usepackage{fixltx2e}
\usepackage{amssymb,,amsmath}
\usepackage{longtable}
\usepackage{multirow}
\usepackage{txfonts}
\bibpunct{(}{)}{;}{a}{}{,}

\def\lsim{\mathrel{\raise .4ex\hbox{\rlap{$<$}\lower 1.2ex\hbox{$\sim$}}}}
\def\gsim{\mathrel{\raise .4ex\hbox{\rlap{$>$}\lower 1.2ex\hbox{$\sim$}}}}
\def\arcm{\ifmmode {^{\scriptscriptstyle\prime}}
          \else $^{\scriptscriptstyle\prime}$\fi}
\def\muKRJ{\ifmmode \,\mu$K$_{\rm RJ}$\else \,$\mu$\hbox{K}$_{\rm RJ}$\fi}    
\def\muKCMB{\ifmmode \,\mu$K$_{\rm CMB}$\else \,$\mu$\hbox{K}$_{\rm CMB}$\fi} 
\def\leaderfil{\leaders\hbox to 5pt{\hss.\hss}\hfil}
\def\all2103resultspapers{\nocite{planck2013-p01, planck2013-p02, planck2013-p02a, planck2013-p02d, planck2013-p02b, planck2013-p03, planck2013-p03c, planck2013-p03f, planck2013-p03d, planck2013-p03e, planck2013-p01a, planck2013-p06, planck2013-p03a, planck2013-pip88, planck2013-p08, planck2013-p11, planck2013-p12, planck2013-p13, planck2013-p14, planck2013-p15, planck2013-p05b, planck2013-p17, planck2013-p09, planck2013-p09a, planck2013-p20, planck2013-p19, planck2013-pipaberration, planck2013-p05, planck2013-p05a, planck2013-pip56, planck2013-p06b}}

\DeclareGraphicsExtensions{.pdf, .eps}

\begin{document}

\title{\Planck{} 2013 results. III. LFI systematic uncertainties}

\titlerunning{LFI systematic uncertainties}
\authorrunning{Planck Collaboration}

\author{\small
Planck Collaboration:
N.~Aghanim\inst{60}
\and
C.~Armitage-Caplan\inst{91}
\and
M.~Arnaud\inst{74}
\and
M.~Ashdown\inst{71, 6}
\and
F.~Atrio-Barandela\inst{17}
\and
J.~Aumont\inst{60}
\and
C.~Baccigalupi\inst{85}
\and
A.~J.~Banday\inst{94, 8}
\and
R.~B.~Barreiro\inst{67}
\and
E.~Battaner\inst{95}
% ...
}
\institute{\small
APC, AstroParticule et Cosmologie, Universit\'{e} Paris Diderot, CNRS/IN2P3, CEA/lrfu, Observatoire de Paris, Sorbonne Paris Cit\'{e}, 10, rue Alice Domon et L\'{e}onie Duquet, 75205 Paris Cedex 13, France\\
\and
Aalto University Mets\"{a}hovi Radio Observatory, Mets\"{a}hovintie 114, FIN-02540 Kylm\"{a}l\"{a}, Finland\\
\and
African Institute for Mathematical Sciences, 6-8 Melrose Road, Muizenberg, Cape Town, South Africa\\
\and
Agenzia Spaziale Italiana Science Data Center, Via del Politecnico snc, 00133, Roma, Italy\\
\and
Agenzia Spaziale Italiana, Viale Liegi 26, Roma, Italy\\
\and
Astrophysics Group, Cavendish Laboratory, University of Cambridge, J J Thomson Avenue, Cambridge CB3 0HE, U.K.\\
\and
CITA, University of Toronto, 60 St. George St., Toronto, ON M5S 3H8, Canada\\
\and
CNRS, IRAP, 9 Av. colonel Roche, BP 44346, F-31028 Toulouse cedex 4, France\\
\and
California Institute of Technology, Pasadena, California, U.S.A.\\
\and
Centre for Theoretical Cosmology, DAMTP, University of Cambridge, Wilberforce Road, Cambridge CB3 0WA, U.K.\\
\and
Centro de Estudios de F\'{i}sica del Cosmos de Arag\'{o}n (CEFCA), Plaza San Juan, 1, planta 2, E-44001, Teruel, Spain\\
\and
Computational Cosmology Center, Lawrence Berkeley National Laboratory, Berkeley, California, U.S.A.\\
\and
Consejo Superior de Investigaciones Cient\'{\i}ficas (CSIC), Madrid, Spain\\
\and
DSM/Irfu/SPP, CEA-Saclay, F-91191 Gif-sur-Yvette Cedex, France\\
\and
DTU Space, National Space Institute, Technical University of Denmark, Elektrovej 327, DK-2800 Kgs. Lyngby, Denmark\\
\and
D\'{e}partement de Physique Th\'{e}orique, Universit\'{e} de Gen\`{e}ve, 24, Quai E. Ansermet,1211 Gen\`{e}ve 4, Switzerland\\
\and
Departamento de F\'{\i}sica Fundamental, Facultad de Ciencias, Universidad de Salamanca, 37008 Salamanca, Spain\\
\and
Departamento de F\'{\i}sica, Universidad de Oviedo, Avda. Calvo Sotelo s/n, Oviedo, Spain\\
\and
Departamento de Matem\'{a}ticas, Estad\'{\i}stica y Computaci\'{o}n, Universidad de Cantabria, Avda. de los Castros s/n, Santander, Spain\\
\and
Department of Astronomy and Astrophysics, University of Toronto, 50 Saint George Street, Toronto, Ontario, Canada\\
\and
Department of Astrophysics/IMAPP, Radboud University Nijmegen, P.O. Box 9010, 6500 GL Nijmegen, The Netherlands\\
\and
Department of Electrical Engineering and Computer Sciences, University of California, Berkeley, California, U.S.A.\\
\and
Department of Physics \& Astronomy, University of British Columbia, 6224 Agricultural Road, Vancouver, British Columbia, Canada\\
\and
Department of Physics and Astronomy, Dana and David Dornsife College of Letter, Arts and Sciences, University of Southern California, Los Angeles, CA 90089, U.S.A.\\
\and
Department of Physics and Astronomy, University College London, London WC1E 6BT, U.K.\\
\and
Department of Physics, Florida State University, Keen Physics Building, 77 Chieftan Way, Tallahassee, Florida, U.S.A.\\
\and
Department of Physics, Gustaf H\"{a}llstr\"{o}min katu 2a, University of Helsinki, Helsinki, Finland\\
\and
Department of Physics, Princeton University, Princeton, New Jersey, U.S.A.\\
\and
Department of Physics, University of California, One Shields Avenue, Davis, California, U.S.A.\\
\and
Department of Physics, University of California, Santa Barbara, California, U.S.A.\\
\and
Department of Physics, University of Illinois at Urbana-Champaign, 1110 West Green Street, Urbana, Illinois, U.S.A.\\
\and
Dipartimento di Fisica e Astronomia G. Galilei, Universit\`{a} degli Studi di Padova, via Marzolo 8, 35131 Padova, Italy\\
\and
Dipartimento di Fisica e Scienze della Terra, Universit\`{a} di Ferrara, Via Saragat 1, 44122 Ferrara, Italy\\
\and
Dipartimento di Fisica, Universit\`{a} La Sapienza, P. le A. Moro 2, Roma, Italy\\
\and
Dipartimento di Fisica, Universit\`{a} degli Studi di Milano, Via Celoria, 16, Milano, Italy\\
\and
Dipartimento di Fisica, Universit\`{a} degli Studi di Trieste, via A. Valerio 2, Trieste, Italy\\
\and
Dipartimento di Fisica, Universit\`{a} di Roma Tor Vergata, Via della Ricerca Scientifica, 1, Roma, Italy\\
\and
Discovery Center, Niels Bohr Institute, Blegdamsvej 17, Copenhagen, Denmark\\
\and
Dpto. Astrof\'{i}sica, Universidad de La Laguna (ULL), E-38206 La Laguna, Tenerife, Spain\\
\and
European Space Agency, ESAC, Planck Science Office, Camino bajo del Castillo, s/n, Urbanizaci\'{o}n Villafranca del Castillo, Villanueva de la Ca\~{n}ada, Madrid, Spain\\
\and
European Space Agency, ESTEC, Keplerlaan 1, 2201 AZ Noordwijk, The Netherlands\\
\and
Haverford College Astronomy Department, 370 Lancaster Avenue, Haverford, Pennsylvania, U.S.A.\\
\and
Helsinki Institute of Physics, Gustaf H\"{a}llstr\"{o}min katu 2, University of Helsinki, Helsinki, Finland\\
\and
INAF - Osservatorio Astrofisico di Catania, Via S. Sofia 78, Catania, Italy\\
\and
INAF - Osservatorio Astronomico di Padova, Vicolo dell'Osservatorio 5, Padova, Italy\\
\and
INAF - Osservatorio Astronomico di Roma, via di Frascati 33, Monte Porzio Catone, Italy\\
\and
INAF - Osservatorio Astronomico di Trieste, Via G.B. Tiepolo 11, Trieste, Italy\\
\and
INAF Istituto di Radioastronomia, Via P. Gobetti 101, 40129 Bologna, Italy\\
\and
INAF/IASF Bologna, Via Gobetti 101, Bologna, Italy\\
\and
INAF/IASF Milano, Via E. Bassini 15, Milano, Italy\\
\and
INFN, Sezione di Bologna, Via Irnerio 46, I-40126, Bologna, Italy\\
\and
INFN, Sezione di Roma 1, Universit\`{a} di Roma Sapienza, Piazzale Aldo Moro 2, 00185, Roma, Italy\\
\and
IPAG: Institut de Plan\'{e}tologie et d'Astrophysique de Grenoble, Universit\'{e} Joseph Fourier, Grenoble 1 / CNRS-INSU, UMR 5274, Grenoble, F-38041, France\\
\and
ISDC Data Centre for Astrophysics, University of Geneva, ch. d'Ecogia 16, Versoix, Switzerland\\
\and
IUCAA, Post Bag 4, Ganeshkhind, Pune University Campus, Pune 411 007, India\\
\and
Imperial College London, Astrophysics group, Blackett Laboratory, Prince Consort Road, London, SW7 2AZ, U.K.\\
\and
Infrared Processing and Analysis Center, California Institute of Technology, Pasadena, CA 91125, U.S.A.\\
\and
Institut N\'{e}el, CNRS, Universit\'{e} Joseph Fourier Grenoble I, 25 rue des Martyrs, Grenoble, France\\
\and
Institut Universitaire de France, 103, bd Saint-Michel, 75005, Paris, France\\
\and
Institut d'Astrophysique Spatiale, CNRS (UMR8617) Universit\'{e} Paris-Sud 11, B\^{a}timent 121, Orsay, France\\
\and
Institut d'Astrophysique de Paris, CNRS (UMR7095), 98 bis Boulevard Arago, F-75014, Paris, France\\
\and
Institute for Space Sciences, Bucharest-Magurale, Romania\\
\and
Institute of Astronomy and Astrophysics, Academia Sinica, Taipei, Taiwan\\
\and
Institute of Astronomy, University of Cambridge, Madingley Road, Cambridge CB3 0HA, U.K.\\
\and
Institute of Theoretical Astrophysics, University of Oslo, Blindern, Oslo, Norway\\
\and
Instituto de Astrof\'{\i}sica de Canarias, C/V\'{\i}a L\'{a}ctea s/n, La Laguna, Tenerife, Spain\\
\and
Instituto de F\'{\i}sica de Cantabria (CSIC-Universidad de Cantabria), Avda. de los Castros s/n, Santander, Spain\\
\and
Istituto di Fisica del Plasma, CNR-ENEA-EURATOM Association, Via R. Cozzi 53, Milano, Italy\\
\and
Jet Propulsion Laboratory, California Institute of Technology, 4800 Oak Grove Drive, Pasadena, California, U.S.A.\\
\and
Jodrell Bank Centre for Astrophysics, Alan Turing Building, School of Physics and Astronomy, The University of Manchester, Oxford Road, Manchester, M13 9PL, U.K.\\
\and
Kavli Institute for Cosmology Cambridge, Madingley Road, Cambridge, CB3 0HA, U.K.\\
\and
LAL, Universit\'{e} Paris-Sud, CNRS/IN2P3, Orsay, France\\
\and
LERMA, CNRS, Observatoire de Paris, 61 Avenue de l'Observatoire, Paris, France\\
\and
Laboratoire AIM, IRFU/Service d'Astrophysique - CEA/DSM - CNRS - Universit\'{e} Paris Diderot, B\^{a}t. 709, CEA-Saclay, F-91191 Gif-sur-Yvette Cedex, France\\
\and
Laboratoire Traitement et Communication de l'Information, CNRS (UMR 5141) and T\'{e}l\'{e}com ParisTech, 46 rue Barrault F-75634 Paris Cedex 13, France\\
\and
Laboratoire de Physique Subatomique et de Cosmologie, Universit\'{e} Joseph Fourier Grenoble I, CNRS/IN2P3, Institut National Polytechnique de Grenoble, 53 rue des Martyrs, 38026 Grenoble cedex, France\\
\and
Laboratoire de Physique Th\'{e}orique, Universit\'{e} Paris-Sud 11 \& CNRS, B\^{a}timent 210, 91405 Orsay, France\\
\and
Lawrence Berkeley National Laboratory, Berkeley, California, U.S.A.\\
\and
Max-Planck-Institut f\"{u}r Astrophysik, Karl-Schwarzschild-Str. 1, 85741 Garching, Germany\\
\and
McGill Physics, Ernest Rutherford Physics Building, McGill University, 3600 rue University, Montr\'{e}al, QC, H3A 2T8, Canada\\
\and
MilliLab, VTT Technical Research Centre of Finland, Tietotie 3, Espoo, Finland\\
\and
Niels Bohr Institute, Blegdamsvej 17, Copenhagen, Denmark\\
\and
Observational Cosmology, Mail Stop 367-17, California Institute of Technology, Pasadena, CA, 91125, U.S.A.\\
\and
SB-ITP-LPPC, EPFL, CH-1015, Lausanne, Switzerland\\
\and
SISSA, Astrophysics Sector, via Bonomea 265, 34136, Trieste, Italy\\
\and
School of Physics and Astronomy, Cardiff University, Queens Buildings, The Parade, Cardiff, CF24 3AA, U.K.\\
\and
School of Physics and Astronomy, University of Nottingham, Nottingham NG7 2RD, U.K.\\
\and
Space Sciences Laboratory, University of California, Berkeley, California, U.S.A.\\
\and
Special Astrophysical Observatory, Russian Academy of Sciences, Nizhnij Arkhyz, Zelenchukskiy region, Karachai-Cherkessian Republic, 369167, Russia\\
\and
Stanford University, Dept of Physics, Varian Physics Bldg, 382 Via Pueblo Mall, Stanford, California, U.S.A.\\
\and
Sub-Department of Astrophysics, University of Oxford, Keble Road, Oxford OX1 3RH, U.K.\\
\and
Theory Division, PH-TH, CERN, CH-1211, Geneva 23, Switzerland\\
\and
UPMC Univ Paris 06, UMR7095, 98 bis Boulevard Arago, F-75014, Paris, France\\
\and
Universit\'{e} de Toulouse, UPS-OMP, IRAP, F-31028 Toulouse cedex 4, France\\
\and
University of Granada, Departamento de F\'{\i}sica Te\'{o}rica y del Cosmos, Facultad de Ciencias, Granada, Spain\\
% ...
}

\abstract{We present the current estimate of instrumental and systematic effect uncertainties for the \Planck-Low Frequency Instrument relevant to the first release of the \Planck\ cosmological results. We give an overview of the main effects and of the tools and methods applied to assess residuals in maps and power spectra. We also present an overall budget of {known} systematic effect uncertainties, which are dominated sidelobe straylight pick-up and imperfect calibration. However, even these two effects are at least two orders of magnitude {weaker} than the cosmic microwave background (CMB) fluctuations as measured in terms of the angular temperature power spectrum. 
A residual signal above the noise level is present in the multipole range $\ell<20$, most notably at 30\,GHz, and is likely caused by residual Galactic straylight contamination. Current analysis aims to further reduce the level of spurious signals in the data and to improve the systematic effects modelling, in particular with respect to straylight and calibration uncertainties.}

\maketitle
\tableofcontents

\section{Introduction}
\label{sec_introduction}

  This paper, one of a set \all2103resultspapers associated with the 2013 release of data from the \Planck\footnote{\Planck\ (\url{http://www.esa.int/Planck}) is a project of the European Space Agency (ESA) with instruments provided by two scientific consortia funded by ESA member states (in particular the lead countries France and Italy), with contributions from NASA (USA) and telescope reflectors provided by a collaboration between ESA and a scientific consortium led and funded by Denmark.} mission \citep{planck2013-p01}, describes the \Planck-LFI instrument systematic effects and their related uncertainties in CMB temperature maps and power spectra. {Systematic effects in \Planck-HFI data are discussed in \citet{planck2013-p03} and \citet{planck2013-p03e}}.

The LFI implements a pseudo-correlation differential design similar to \textit{WMAP} \citep{jarosik2003a,jarosik2003b} to suppress 1/$f$ amplifier gain and noise fluctuations \citep{seiffert2002,mennella2003,bersanelli2010} as well as correlated effects from thermal and electrical variations affecting both the sky signal and reference loads. The reference signal is provided by stable 4.5\,K blackbodies thermally and mechanically connected to the external structure of the High Frequency Instrument (HFI) 4\,K box \citep{valenziano2009,lamarre2010}. The offset between the sky and reference signals, of the order of 1\,--\,2\,K, is balanced in software during data processing on the ground \citep{mennella2003, planck2011-1.6}. The differenced time streams are characterised by 1/$f$ noise knee frequencies in the range 10\,--\,100\,mHz \citep{mennella2010,planck2011-1.4}, leaving residual correlated low-frequency fluctuations in gain and signal that are removed during calibration and map-making.

The LFI is also an excellent polarimeter, with very low systematic effects. Depolarisation by the optics and by imperfections in the orthomode transducers, separating the orthogonal linear polarisations, has been accurately measured on the ground and is almost negligible \citep{leahy2010}. 

Asymmetrical bandpass response in the two radiometers is the main source of $I\rightarrow(Q,U)$ leakage in the foreground-dominated sky regions, especially at low frequencies. Although accurate knowledge of the bandpass response allows us, in principle, {to correct} for this effect during data analysis, the ground bandpass measurements were not accurate enough to maintain this residual below 1\% \citep{zonca2009}. For this reason the spurious polarisation from bandpass mismatch was estimated and removed using flight data, as described in \citet{planck2013-p02}.

Optical effects arise mainly from Galactic and CMB dipole pick-up caused by primary and secondary mirror spillovers \citep{tauber2010b, sandri2010}. This is relevant especially for polarisation measurements at 30\,GHz, where Galactic emissions are stronger. 

In this paper we provide a {preliminary} overview of the instrument systematic effects and the uncertainties they cause on CMB temperature maps and power spectra (see Sect.~\ref{sec_summary_table}). In Sect.~\ref{sec_overview} we outline and discuss the known instrumental effects, separating them into two broad categories: (i) effects that do not depend on the sky signal and impact the radiometric measurements as an additive spurious fluctuation or a gain variation, and (ii) effects that {do} depend on the sky signal, i.e., on its amplitude and/or on the scanned sky region. Some of these effects are removed in the data processing pipeline according to algorithms described in \citet{planck2013-p02}. The assessment of the residual uncertainty, discussed in Sect.~\ref{sec_assessment}, was performed according to two different strategies. Null tests were the primary tool to check for systematic effect residuals exceeding the white noise level. We also assessed their impact on radiometric time streams, even if below the white noise limit, by exploiting in-flight housekeeping and scientific data. 

Some of the effects discussed in this paper are also relevant for calibration, and are discussed in detail in \citet{planck2013-p02b}. In this case we provide here only a brief discussion of the most relevant points and results, deferring to the dedicated paper any further details.

Throughout this paper we follow the naming convention described in Appendix A of \citet{mennella2010} and also available on-line on the Explanatory Supplement \citep{planck2013-p28}.

\section{Summary of uncertainties due to systematic effects}
\label{sec_summary_table}
  
  In this section we provide a top-level overview of the uncertainties due to systematic effects in the \Planck-LFI CMB temperature maps and power spectra. Table~\ref{tab_list_systematic_effects} provides a list of these effects, with short descriptions of their cause, strategies for their removal and references to sections and/or papers where more information can be found. {This section also provides a summary of the main results of our analysis, as detailed in Sect.~\ref{sec_assessment} and corresponding subsections.}

\begin{table*}[tmb]     
\begingroup
\newdimen\tblskip \tblskip=5pt
\caption{List of known instrumental systematic effects in \Planck-LFI}                         
\label{tab_list_systematic_effects}
\nointerlineskip
\vskip -3mm
\footnotesize
\setbox\tablebox=\vbox{
   \newdimen\digitwidth 
   \setbox0=\hbox{\rm 0} 
   \digitwidth=\wd0 
   \catcode`*=\active 
   \def*{\kern\digitwidth}
   \newdimen\signwidth 
   \setbox0=\hbox{+} 
   \signwidth=\wd0 
   \catcode`!=\active 
   \def!{\kern\signwidth}

{
\halign{ 
#\leaderfil\tabskip=1em& 
#\leaderfil&
#\leaderfil&
#\hfil\tabskip=0pt 
 \cr      
\noalign{\doubleline}
\omit\hfil{\bf Effect}\hfil&
\omit\hfil{\bf Source}\hfil&
\omit\hfil{\bf Control/Removal}\hfil&
\omit\hfil{\bf Reference}\hfil\cr      
\noalign{\vskip -3pt}
\noalign{\vskip 5pt\hrule\vskip 3pt}
\noalign{\vskip 10pt}
\multispan4\hskip 6cm{\bf Effects independent of sky signal}\hfil\cr 
\noalign{\vskip 6pt}
White noise correlation& 
Phase switch imbalance& 
Diode {weighting}& 
\ref{sec_correlated_one_over_f}\cr
\noalign{\vskip 6pt}
1/$f$ noise&
RF amplifiers&
Pseudo-correlation and destriping&
\ref{sec_correlated_one_over_f}\cr
\noalign{\vskip 6pt}
Bias fluctuations&
RF amplifiers, back-end electronics&
Pseudo-correlation and destriping&
\ref{sec_bias_fluctuations}\cr
\noalign{\vskip 6pt}
Thermal fluctuations&
4\,K, 20\,K and 300\,K thermal stages&
Calibration, destriping&
\ref{sec_thermal_effects},  \ref{sec_assessment_thermal}\cr
\noalign{\vskip 6pt}
1\,Hz spikes&
Back-end electronics&
Template fitting and removal&
\ref{sec_spikes},  \ref{sec_assessment_spikes}\cr 
\noalign{\vskip 10pt}
\multispan4\hskip 5.8cm{\bf Effects dependent on the sky signal}\hfil\cr 
\noalign{\vskip 6pt}
{Main beam ellipticity}& {Main beams}&
{Accounted for in window function} &
{\citet{planck2013-p02d}}\cr
\noalign{\vskip 6pt}
{Intermediate sidelobes}& {Optical response at angles}&
{Masking of Galaxy and point} &
{Not treated in this release}\cr
\omit{pickup}\hfil&\omit $<5^\circ$ {from the main beam}\hfil&\omit{sources}\hfil&\cr
\noalign{\vskip 6pt}
{Far} sidelobes pickup& Main and sub-reflector spillovers&
Model sidelobes removed from timelines &
\ref{sec_overview_sidelobes}, \ref{sec_assessment_sidelobes}\cr
\omit&\omit&\omit(not implemented in this release)\hfil&\cr
\noalign{\vskip 6pt}
Bandpass asymmetries&
Differential orthomode transducer&
Spurious polarisation removal&
\citet{planck2013-p02}\cr
\omit&\omit and receiver bandpass response\hfil&\omit&\cr
\noalign{\vskip 6pt}
Analogue-to-digital&
Back-end analogue-to-digital&
Template fitting and removal&
\ref{sec_adc},  \ref{sec_assessment_adc}\cr
\omit converter non linearity\hfil&\omit  converter\hfill&\omit&\citet{planck2013-p02}\cr
\noalign{\vskip 6pt}
Imperfect photometric&
Sidelobe pickup, radiometer noise&Calibration using the 4\,K reference&\ref{sec_overview_calibration}, \ref{sec_assessment_calibration},\cr
\omit calibration\hfil&\omit temperature changes and other\hfil&\omit load voltage output\hfil& \citet{planck2013-p02b} 
\cr
\omit&\omit non-idealities\hfil&\omit&\cr
\noalign{\vskip 6pt}
Pointing&
Uncertainties in pointing reconstru-&
Negligible impact on temperature&
\ref{sec_pointing_effects}, \ref{sec_assessment_pointing}\cr
\omit&\omit ction, thermal changes affecting\hfil &\omit anisotropy measurements\hfil&\cr
\omit&\omit focal plane geometry\hfil&\omit&\cr
\noalign{\vskip 5pt\hrule\vskip 3pt}
}
}}
\endPlancktablewide       
\endgroup
\end{table*}

The impact of 1/$f$ noise has been assessed using ``half-ring'' noise maps (see Sect.~\ref{sec_half_ring_null_tests}) normalized to the white noise estimate at each pixel obtained from the white noise covariance matrix, so that a perfectly white noise map would be Gaussian and isotropic with unit variance. Deviations from unity trace the contribution of residual $1/f$ noise in the final maps, which ranges from $0.06\%$ at 70\,GHz to $2\%$ at 30\,GHz, as detailed in Sect.~12.2 of \citet{planck2013-p02}.

Pixel uncertainties due to other systematic effects have been calculated on simulated maps degraded to $N_{\rm side}=128$ at 30 and 44\,GHz and $N_{\rm side}=256$ at 70\,GHz in order to approximate the optical beam size.

In Table~\ref{tab_summary_systematic_effects_maps} we list the r.m.s. and the difference between the 99\% and the 1\% quantiles in the pixel value distributions. For simplicity we refer to this difference as the peak-to-peak (p-p) difference, although it neglects outliers but effectively approximates the peak-to-peak variation of the effect on the map.


\begin{table}[tmb]               
\begingroup
\newdimen\tblskip \tblskip=5pt
\caption{Summary of systematic effects uncertainties on maps$^a$ in \muKCMB.}            
\label{tab_summary_systematic_effects_maps}                     
\nointerlineskip
\vskip -3mm
\footnotesize
\setbox\tablebox=\vbox{
   \newdimen\digitwidth 
   \setbox0=\hbox{\rm 0} 
   \digitwidth=\wd0 
   \catcode`*=\active 
   \def*{\kern\digitwidth}
   \newdimen\signwidth 
   \setbox0=\hbox{+} 
   \signwidth=\wd0 
   \catcode`!=\active 
   \def!{\kern\signwidth}
{\tabskip=0pt
\halign{ 
\hbox to 1.3in{#\leaderfil}\tabskip=0em& 
\hfil#\hfil\tabskip=.6em& 
\hfil#\hfil& 
\hfil#\hfil& 
\hfil#\hfil& 
\hfil#\hfil& 
\hfil#\hfil\tabskip=0pt 
 \cr                       
\noalign{\doubleline}
\omit&
\multispan2\hfil $30$\,GHz\hfil& 
\multispan2\hfil $44$\,GHz\hfil&
\multispan2\hfil $70$\,GHz\hfil\cr   
\noalign{\vskip -3pt}
\omit&\multispan6\hrulefill\cr
\noalign{\vskip 2pt}
\omit& 
\hfil p-p\hfil&\hfil rms\hfil&\hfil p-p\hfil&\hfil rms\hfil&\hfil p-p\hfil&\hfil rms\hfil\cr
\noalign{\vskip 4pt}
      Bias fluctuations& *0.08& *0.01& *0.10& *0.02& *0.23&*0.06\cr
\noalign{\vskip 4pt}
      Thermal fluctuations& *0.61& *0.11& *0.40& *0.08& *1.17& *0.20\cr
\noalign{\vskip 4pt}
      1-Hz spikes& *0.87& *0.17& *0.14& *0.03& *0.60& *0.12\cr
\noalign{\vskip 4pt}
      Sidelobes pickup& 18.95& *4.53& *1.92& *0.57& *6.39& *1.91\cr
\noalign{\vskip 4pt}
      ADC non-linearity& *3.87& *1.01& *0.89& *0.19& *0.92& *0.19\cr
\noalign{\vskip 4pt}
      {Calibration}& *4.33& *1.16& *4.74& *0.97& *6.51& *1.10\cr
\noalign{\vskip 10pt}
      Total\tablefootmark{b}& 21.02& *4.83& *5.61& *1.13& *7.87& *2.00\cr
\noalign{\vskip 5pt\hrule\vskip 3pt}
}
}}
\endPlancktable          
\tablenote a Calculated on a pixel size approximately equal to the
average beam FWHM.\par
\tablenote b The total has been computed on maps resulting from the sum of individual systematic effect maps.\par
\endgroup
\end{table}

\begin{figure*}
  \begin{tabular}{c c c}
   
    \multicolumn{3}{c}{\includegraphics[width=176mm]{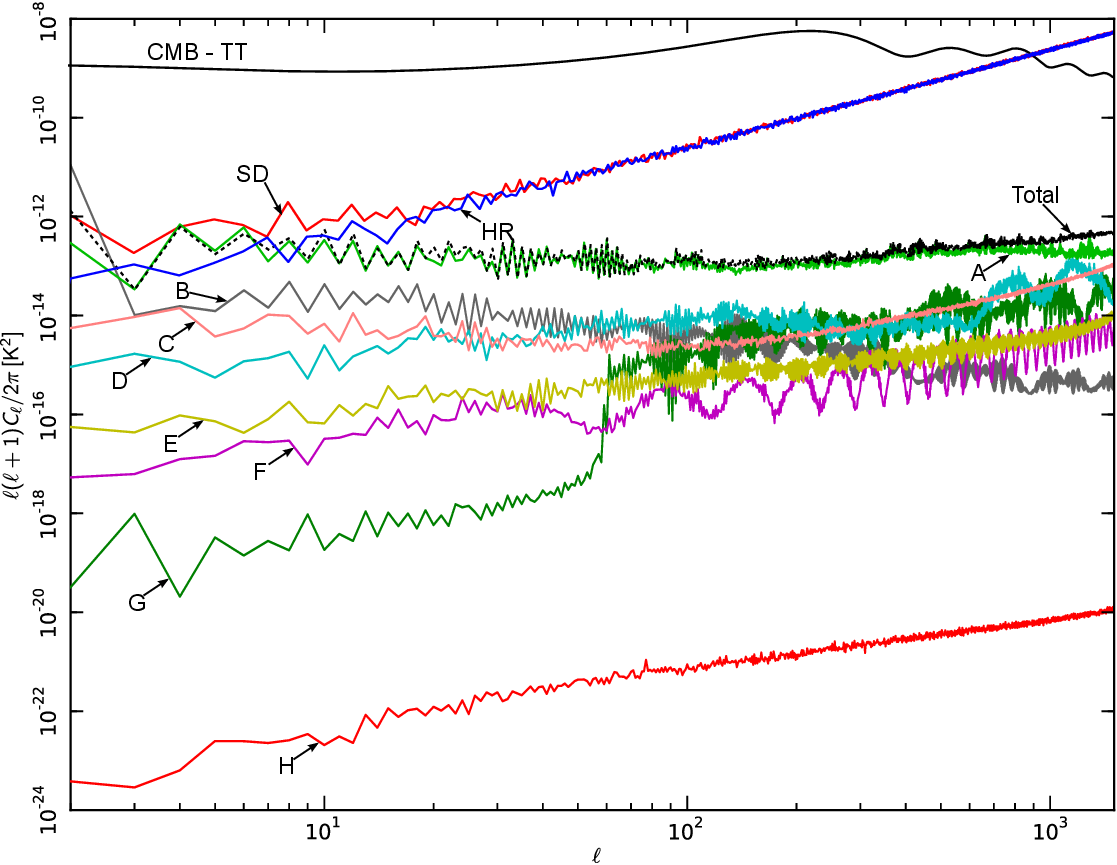}}\\
     & &\\
    \includegraphics[width=64.mm]{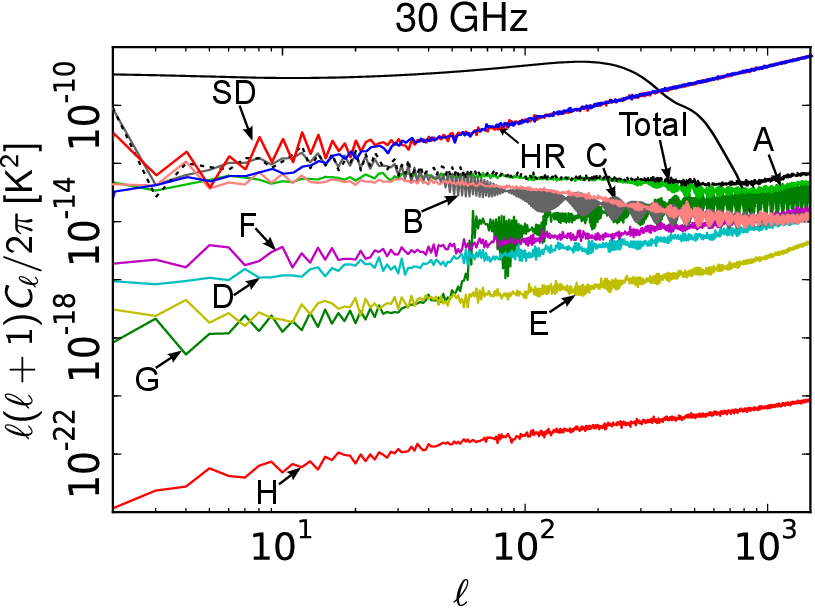}&
    \includegraphics[width=54.85mm]{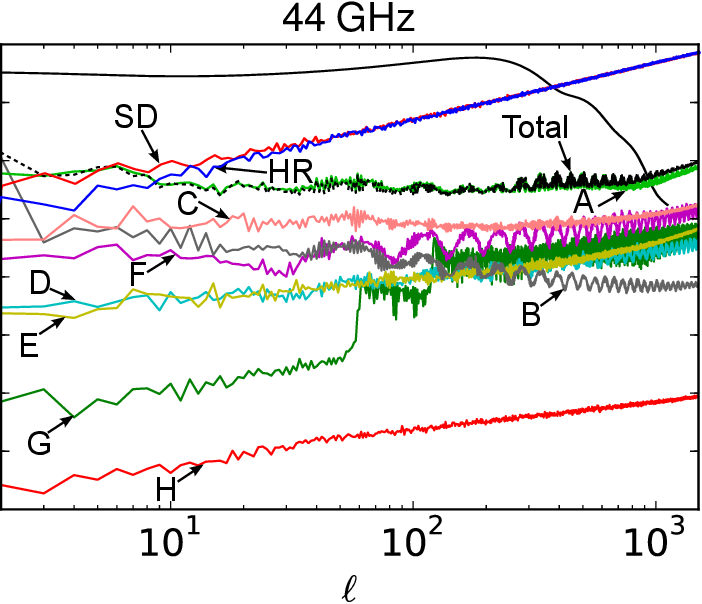}&
    \includegraphics[width=54.85mm]{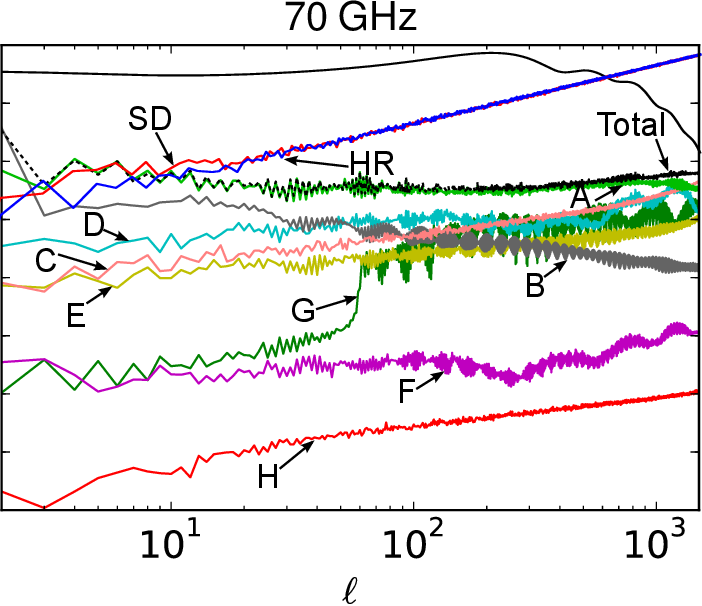}\\
    \multicolumn{3}{c}{\includegraphics[width=176mm]{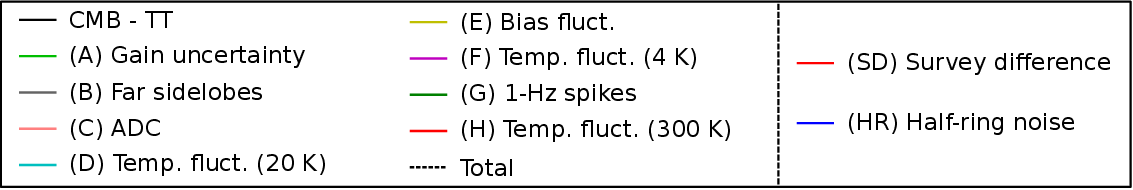}}\\
     & &\\

  \end{tabular}
  
  \caption{Angular power spectra of the various systematic effects compared to the \Planck\ temperature anisotropy spectrum. The black dashed curve, representing the total contribution, has been derived from a map where all the systematic effects have been summed. \textit{Top panel}: power spectra obtained from frequency independent maps resulting from the weighted-average of individual systematic effect frequency maps. \textit{Bottom panel}: contributions of systematic effects from individual frequency maps. The CMB curve corresponds to the \Planck\ best-fit model presented in \citet{planck2013-p08}. {In the bottom panels the CMB spectrum has been filtered by the beam window function for each frequency.}}
  \label{fig_systematic_effects_power_spectrum}
\end{figure*}

Angular power spectra have been obtained from full resolution ($N_{\rm side}=1024$) systematic effect maps at each frequency using the \texttt{HEALPix Anafast} routine \citep{gorski2005}. We have then evaluated the propagation of the various effects in the final CMB map by assuming a simple internal linear combination component separation, as explained in Sect.~\ref{sec_syseffect_compsep}. In Fig.~\ref{fig_systematic_effects_power_spectrum} we show how the power spectra of the various effects compare with the \Planck\ temperature spectrum, with the noise level coming from the half-ring difference maps (see Sect.~\ref{sec_half_ring_null_tests}) and with the residual map obtained from a difference map between survey 1 and survey 2\footnote{Time periods relative to individual surveys are defined in Table 11 of \citet{planck2013-p02}} (see Sect.~\ref{sec_survey_difference_null_tests}). The large plot in the top panel shows the power spectra obtained from frequency-independent maps resulting from the weighted-average of frequency maps using the weights specified in Sect.~\ref{sec_syseffect_compsep}. Spectra in the three small plots in the lower panel, instead, show contributions of systematic effects from individual frequency maps. 

{Our analysis is based on a combined assessment of known and unknown systematic effects via simulations and null-maps. It is worth underlining that some effects could be undetected in difference maps, although none of these effects are likely to affect significantly the results of our analysis, as discussed in detail in Sect.~\ref{sec_insensitive_null_test_effects}}.  Our assessment shows that the global impact of systematic effect uncertainties is at least two order of magnitudes less than the CMB power spectrum, and demonstrates the robustness of \Planck-LFI temperature anisotropy measurements. Comparison between the total simulated systematic effects and the residual signal obtained by differencing survey 1 and survey 2 maps highlights an excess signal in the multipole range $\ell \lesssim 20$ that is not completely accounted for in our simulations. This excess comes mainly from the 30\,GHz channel and is likely to be caused by Galactic emissions picked up by beam sidelobes. {Also the 44\,GHz and 70\,GHz channels show residuals at low multipoles, although smaller than at 30\,GHz.} Understanding this excess and further reducing the level of residual systematic uncertainties is the primary goal of our current analysis to obtain polarisation measurements with a level of purity comparable to what has been achieved for temperature anisotropies.

\section{Overview of LFI systematic effects}
\label{sec_overview}

  Known systematic effects in the \Planck-LFI data can be divided into two broad categories: effects independent of the sky signal, which can be considered as additive or multiplicative spurious contributions to the measured timelines, and effects which are dependent on the sky and that cannot be considered independently of the observational strategy. 

{These effects can generate correlations in the data, and should be removed from timelines before noise is assessed and maps are generated. For this release, based on temperature data only, we have removed from timelines three of these effects: diode-diode correlations (Sect.~\ref{sec_correlated_one_over_f}), ADC non linearity (Sections~\ref{sec_adc} and \ref{sec_assessment_adc}) and 1-Hz frequency spikes (Sections~\ref{sec_spikes} and \ref{sec_assessment_spikes}). The remaining effects have been treated as noise, and their effect assessed via the noise covariance matrices \citep{planck2013-p02} and half-ring difference maps \citep[see][ and Sect~\ref{sec_assessment_null_tests}]{planck2013-p02}. The future release will include a deeper assessment and removal of instrumental effects to match the required accuracy for polarisation.}

\subsection{Effects independent of sky signal}
\label{sec_additive_effects}

  \subsubsection{Noise correlations and 1/$f$ noise}
  \label{sec_correlated_one_over_f}
  
    Each \Planck-LFI receiver is a pseudo-correlation system viewing a scalar feed directed through the telescope at the sky, together with a reference cold load thermally stable near 4\,K. Non-white noise from the cold front-end amplifiers is reduced via the correlation, while fluctuations in the later stages of the receiver are minimized by modulating a phase switch in the correlation section at 8192\,Hz. The LFI receiver design, construction, ground performance and initial flight performance have been extensively documented \citep{bersanelli2010, mennella2010, planck2011-1.4}.

The noise properties of the receivers play an important role in downstream data analysis. In particular, we need good estimates of the white noise level, long term stability (1/$f$-type noise) and any correlated noise components.

The receiver architecture is symmetric, with two complementary detector diodes as output for each receiver channel. As described in \citet{seiffert2002} and \citet{planck2011-1.4} imperfect matching of components {limits} isolation between the complementary diodes of a receiver between $-10$ and $-15$\,dB. This imperfect isolation leads to a small anti-correlated component in the white noise that is cancelled by a weighted average of the time ordered data from the two diodes of each receiver as the first step of analysis. This avoids the complication of tracking the anti-correlated white noise throughout the analysis. 

We treat the combined diode data as the raw data, and calibration, noise estimation, map-making etc. are performed on these combined data.  The weights were determined from some initial estimates of the calibrated noise for each detector, and are kept fixed for the entire mission.

\begin{figure}
  \begin{center}
    \includegraphics[width=\columnwidth]{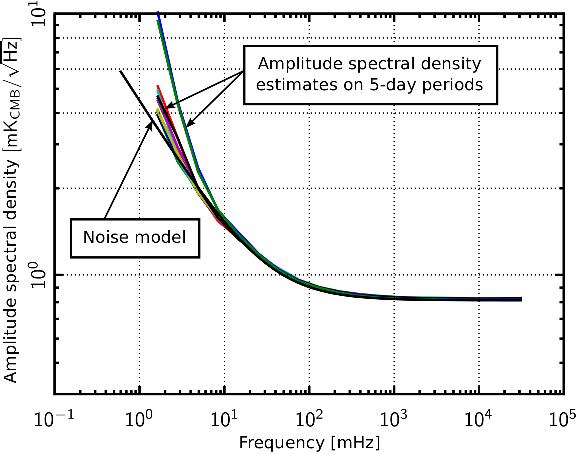}
    \caption{
      Amplitude spectral density estimates on 5-day time periods (coloured lines) compared to the nominal mission noise model for a representative 70 GHz radiometer (\texttt{LFI23M}).
    }
    \label{asd_sample_23m}
  \end{center}
\end{figure}

\begin{figure}
  \begin{center}
    \includegraphics[width=\columnwidth]{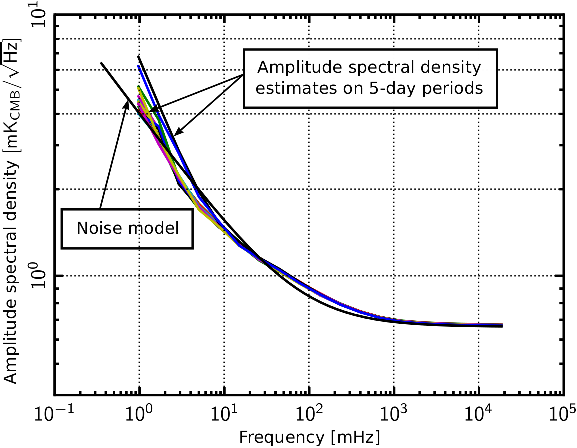}
    \caption{
      Amplitude spectral density estimates on 5-day time periods (coloured lines) compared with the nominal mission noise model (black line) for one 44 GHz radiometer (\texttt{LFI24S}).
    }
    \label{asd_sample_24s}
  \end{center}
\end{figure}

Noise parameters were reported in \citet{planck2011-1.4}. A longer data set, some thermal instabilities in the instrument (particularly during survey 3), and refinements of the data analysis (map making and noise covariance matrix) all require a more detailed look at the long term evolution of the noise characteristics of the receivers.

The noise power spectral density $P(f)$ of the receivers is generally well described by

\begin{equation}
    \label{e:PSD}
    P(f)=\sigma^{2}\left[1+\left(\frac{f}{f_{\mathrm{k}}}\right)^{\alpha}\right],
\end{equation}
where $\sigma$ characterizes the white noise component, the knee frequency, $f_{\mathrm{k}}$, denotes the frequency where white noise and $1/f$ contribute equally in power to the total noise, and $\alpha$ characterizes the slope of the power spectrum for frequencies $f<f_{\mathrm{k}}$. {In the following, low frequency power-law noise will referred to as $1/f$ noise, regardless of its slope, $\alpha$.}

We estimate the signal-subtracted noise power spectrum of each receiver on 5-day time periods. Except for specific, mostly well understood events, shorter time scale noise estimation does not produce any evident trends. For nearly all the radiometers our noise model is a very good approximation of the power spectrum. We plot a representative comparison in Fig.~\ref{asd_sample_23m}. A few channels show features not well captured by this simple model; the worst is displayed in Fig.~\ref{asd_sample_24s}.

\begin{figure}
  \begin{center}
    \includegraphics[width=\columnwidth]{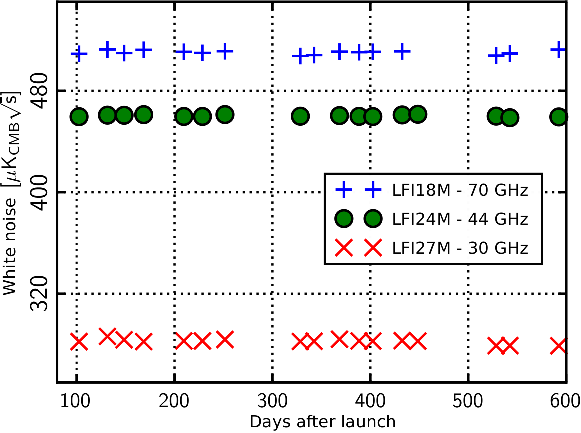}
    \caption{
      Fitted white noise parameters over the nominal survey for representative radiometers at 30, 44 and 70 GHz. Values are estimated on 5-day sections of data.
    }
    \label{wnlevel}
  \end{center}
\end{figure}

\begin{figure}
  \begin{center}
    \includegraphics[width=\columnwidth]{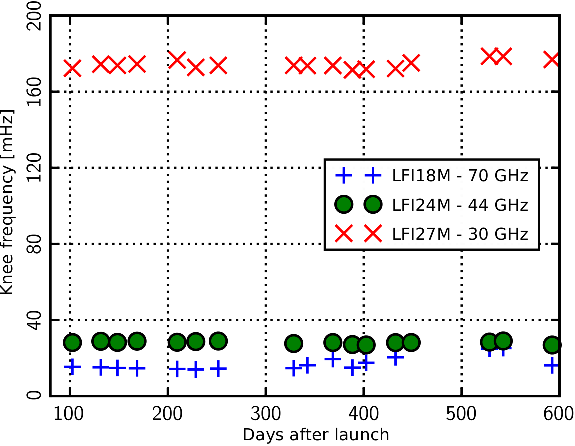}
    \caption{
      Fitted knee frequencies over the nominal survey for representative radiometers at 30, 44 and 70 GHz. Values are estimated on 5-day sections of data.
    }
    \label{fik}
  \end{center}
\end{figure}

\begin{figure}
  \begin{center}
    \includegraphics[width=\columnwidth]{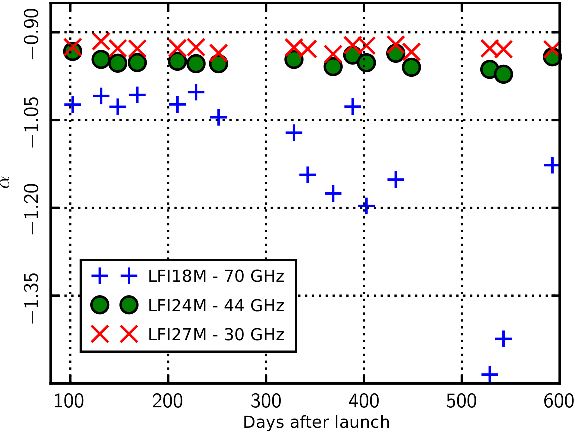}
    \caption{
      Fitted power-law slopes for low frequency noise. Here we note significant instability {after day 300}. This is due to substantially greater thermal instability of the 20\,K stage just before and after switch over between the two sorption coolers{, which occurred at day 460}. 
    }
    \label{alpha}
  \end{center}
\end{figure}

Over the course of the nominal mission, the noise is well fit by the model, with the exception of the early parts of sky survey 3. During this time, thermal instabilities brought on by the switch-over from the nominal to the redundant sorption cooler cause poor fits and some changes in the parameters. In Figs.~\ref{wnlevel} through \ref{alpha} we show the behaviour of the three noise parameters in Eq.~(\ref{e:PSD}) estimated on 5-day sections of data over the nominal time period. White noise and knee frequency are stable, while the slope starts increasing in absolute value after day 300, as a result of larger temperature fluctuations in the 20\,K focal plane. The jump in slope after day 500 is correlated with the sorption cooler switch-over (see Sect.~\ref{sec_thermal_effects} for further details).

  \subsubsection{Thermal effects}
  \label{sec_thermal_effects}

\begin{figure}[h!]
  \includegraphics[width=8.8cm]{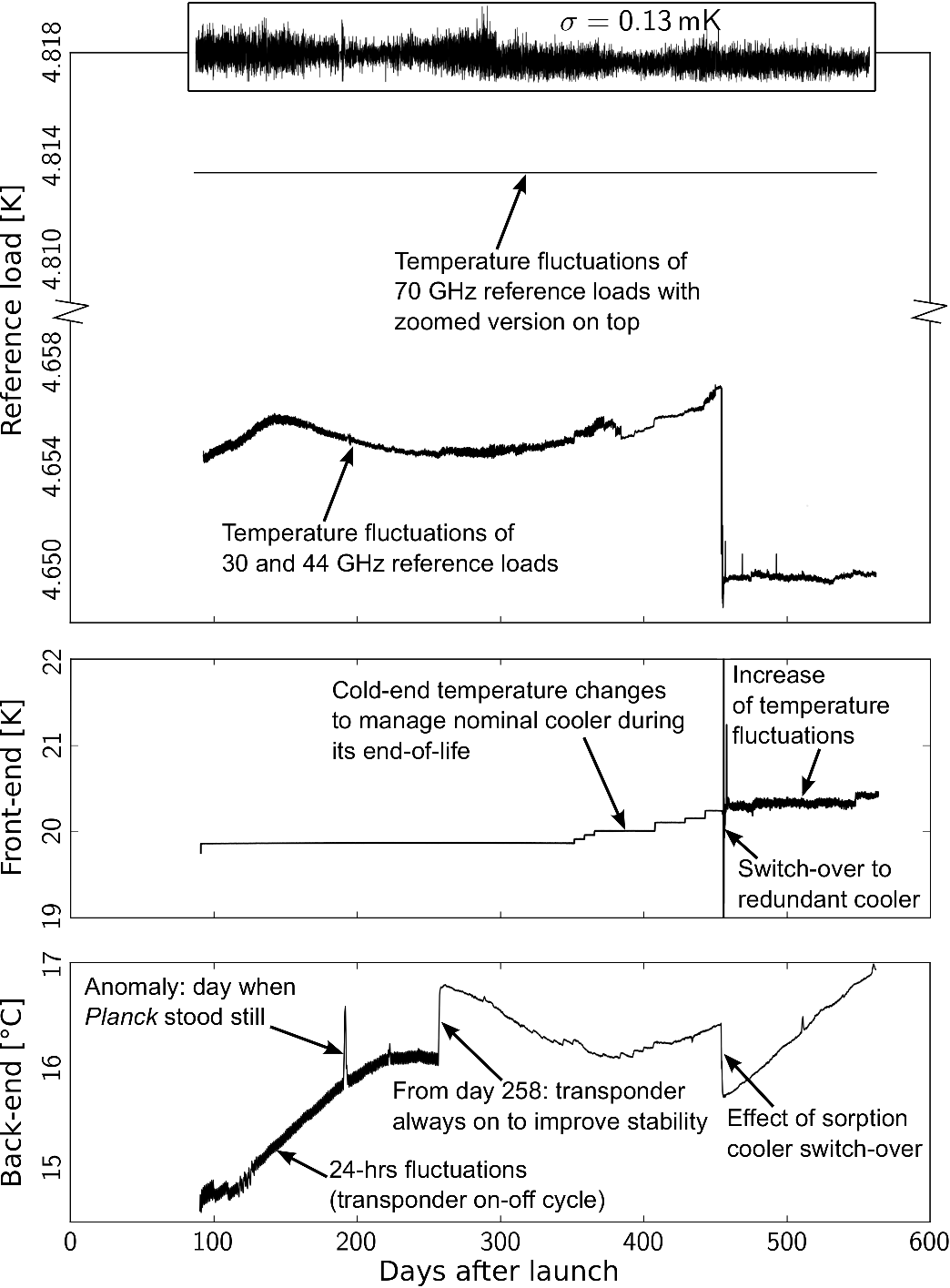}
  \caption{Main temperatures in \Planck-LFI. From top to bottom: 70\,GHz reference loads, 30 and 44\,GHz reference loads, 20\,K focal plane (sensor placed on feed horn flange of \texttt{LFI28}) and 300\,K back-end (sensor placed on the back-end electronics box). A brief description of the main operational events affecting the thermal behaviour is provided in each panel.}
  \label{fig_temperatures}
\end{figure}

    The LFI is susceptible to temperature fluctuations in the 300\,K back-end modules, in the 4\,K reference loads and in the 20\,K focal plane. Figure~\ref{fig_temperatures} provides an overview of the main temperatures during the period between day 91 (the start of nominal operations) and 563 after launch. 

The two topmost plots show the reference load temperatures at the level of the 70\,GHz and 30\,--\,44\,GHz channels, respectively. The temperature of the 70\,GHz reference loads is actively controlled by a proportional-integral-derivative (PID) system and is very stable ($\delta T_{\rm rms}\sim 0.13$\,mK, see the zoomed plot in the inset). Reference loads of the 30 and 44\,GHz channels, instead, do not benefit from active thermal control. Their temperature is consequently more unstable and susceptible to major system-level events like, for example, the switch over to the redundant sorption cooler.

The third plot from the top of Fig.~\ref{fig_temperatures} shows the 20\,K LFI focal plane temperature measured by a sensor placed on the feed horn flange of the \texttt{LFI28} receiver. The temperature during the first sky survey was very stable, with a $\delta T_{\rm rms}\lesssim 1$\,mK. Towards the end of the first year of operations the sorption cooler performance started to degrade and its stability was maintained with a series of controlled temperature changes. The switch over to the redundant cooler was performed on August, 11$^\mathrm{th}$ 2010, leaving a clear signature on all the main LFI temperatures. After this operation the level of temperature fluctuations in the focal plane increased unexpectedly, and this was later understood to be the effect of liquid hydrogen that was still present in the cold-end of the nominal cooler, because the degraded compressor system was not able to absorb all the hydrogen that was present in the cooler line. Although this effect was later mitigated by a series of dedicated operations, most of the third sky survey suffered from a higher-than-nominal level of temperature variation.

The last plot shows the temperature of the 300\,K electronics box, measured by one of its temperature sensors. During the first sky survey the back-end temperature suffered from a daily fluctuation caused by the satellite transponder that was switched on daily during contact with the ground station. After day 258 the system was left continuously on, and the modulation disappeared. This operation caused an increase of the absolute temperature level. The second temperature change occurred in correspondence to the sorption cooler switch-over operation. The plot also shows a yearly temperature modulation due to the satellite rotation around the Sun and a temperature spike at day 191 after launch. This was caused by an operational anomaly that led the satellite to fail to re-point for an entire day with a corresponding temperature increase of the warm units.

More details about the thermal stability performance of \Planck\ can be found in \citet{planck2011-1.3}, while the susceptibility of the LFI to temperature variations is discussed in \citet{terenzi2009b}.

  \subsubsection{Bias fluctuations}
  \label{sec_bias_fluctuations}
  
          The signal detected by the radiometers can vary because of fluctuations in the front-end and back-end amplifier bias voltages. In the LFI these fluctuations occurred according to two time scales:
    
    \begin{itemize}
     \item slow electric drifts, due to thermal changes in the power supply, in the RF amplifiers, and in the detector diodes;
     \item fast and sudden electric instabilities, arising in the warm electronics or from electromagnetic interference effects, and affecting both the cold amplifiers and the warm detector diodes. 
    \end{itemize}

    The effect of slow drifts is suppressed by the pseudo-correlation architecture of the differential radiometers. Fast electric changes produce quasi-random fluctuations and abrupt steep drops or jumps in the signal. If jumps are caused by instabilities in the front-end bias voltage then the effect involves the output voltage of both diodes in the radiometer. When the jumps occur in the back-end detector diodes (so-called ``popcorn noise'') they impact only the output voltage of the corresponding diode and affect sky and reference load samples. In both cases the differenced signal is largely immune from these effects.

  \subsubsection{1-Hz spikes}
  \label{sec_spikes}
  
      This effect is caused by pickup from the housekeeping electronics clock that occurs after the detector diodes and before the analogue-to-digital converter (ADC) \citep{meinhold2009,mennella2010,planck2011-1.4}. This spurious signal is detected in the radiometer time-domain outputs as a 1\,s {rectangular} wave with a rising edge near 0.5\,s and a falling edge near 0.75\,s in on-board time. In the frequency domain it appears  at multiples of 1\,Hz. 

Frequency spikes are present at some level in the output from all detectors, but affect the 44\,GHz data most strongly because of the low voltage output and high post-detection gain values in that channel. For this reason spikes are removed from the 44\,GHz time-ordered data via template fitting, as described in \citet{planck2013-p02}.

\subsection{Effects dependent on sky signal}
\label{sec_effects_with_sky}

  \subsubsection{Sidelobe pick-up}
  \label{sec_overview_sidelobes}

\begin{figure}[h!]
  \includegraphics[width=8.8cm]{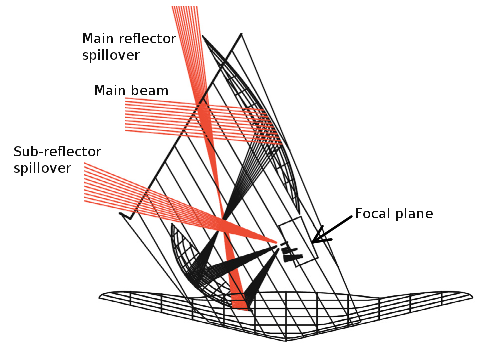}
  \caption{Main and sub-reflector spillover, and main beam directions in the \Planck\ telescope.}
  \label{fig_spillover}
\end{figure}

    Straylight contamination arises from the spurious signal pickup from the telescope far sidelobes. Main sources of straylight contamination are the Galaxy, especially at 30\,GHz, and the cosmological dipole, mainly detected in the directions of the main and sub-reflector spillover, as sketched in Fig.~\ref{fig_spillover}. In principle we should also include the straylight contribution from the orbital dipole, but its effect is a factor ten lower than the cosmic dipole, so that it can safely be neglected in this framework (but it has been considered in the calibration pipeline). 

{Intermediate sidelobes, i.e., the lobes in the pattern at angles less than 5$^\circ$ from the main beam, represent another source of systematic effects. The fraction of power intercepted by intermediate sidelobes ranges from 0.02\% to 0.08\% of the total beam power, which is about 10 times less than the fraction in far sidelobes (ranging from 0.18\% to 0.68\%). Their effect is therefore correspondingly smaller, of the order of $\sim1.5\,\mu$K on the maps. Moreover, because intermediate lobes involve sky regions very close to the main beam, their effect can be controlled by masking the Galaxy and point sources. In this paper we have therefore neglected the effect from intermediate sidelobes, which will be addressed in detail in a future paper dedicated to the analysis of the full mission dataset.}

Straylight impacts the measured signal in two ways: (i) through direct contamination and coupling with the main beam sky signal, and (ii) in the photometric calibration of the radiometer detected signal. In this paper we concentrate on the direct detection, while the impact on calibration and the adopted mitigation strategies are described in \citet{planck2013-p02b}.

Because of the beam orientation, the straylight fingerprint is different in odd surveys compared to even surveys. The Galaxy, for example, is detected by the sub-reflector spillover in the odd surveys and by the main-reflector spillover in the even surveys. Because the sub-reflector spillover points approximately in the main beam direction, the Galaxy straylight pattern is close to the Galactic plane. The main-reflector spillover, instead, points at about 85$^\circ$ from the main beam so that the Galaxy is re-imaged onto a ring (see figures in Sect.~\ref{sec_assessment_sidelobes}).

{Further details about the \Planck\ optical system are reported in \citet{tauber2010b}, while the LFI and HFI beams and window functions are provided in \citet{planck2013-p02d} and \citet{planck2013-p03c}, respectively.}

  \subsubsection{ADC non linearity}
  \label{sec_adc}
  
\begin{figure}[h!]
  \includegraphics[width=88mm]{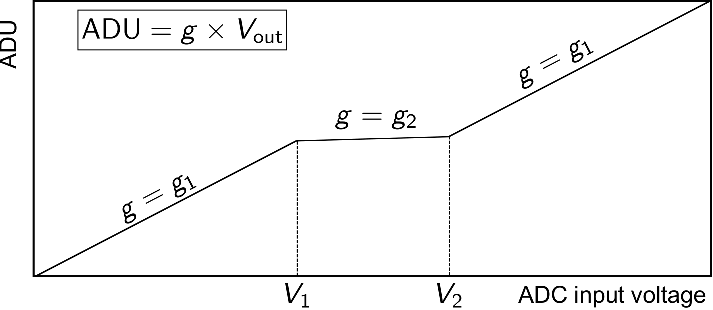}
  \caption{Schematic of the ADC non-linearity effect. For a small range of voltages the ADC response changes slope. }
  \label{fig_adc_concept}
\end{figure}

      The ADC linearity requires that the voltage step sizes between successive binary outputs are constant over the entire input dynamic range. If these steps are not constant (see the sketch in Fig.~\ref{fig_adc_concept}) we have a non-linearity in the ADC response that leads to calibration errors. A brief description of the mathematical model of this effect is provided in Appendix~\ref{app_adc_formalism}.

\begin{figure}[h!]
  \includegraphics[width=8.8cm]{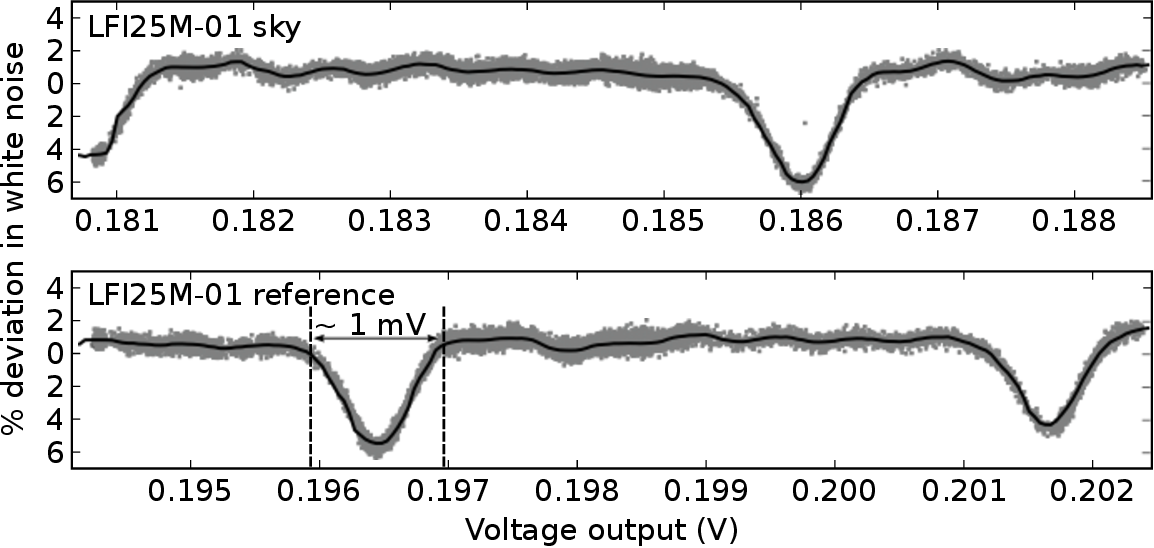}
  \caption{Percentage variation in the single detector white noise estimates with detector voltage.}
  \label{fig_adc_whitenoise_volts}
\end{figure}

{In case of linear response, the voltage output of a coherent receiver scales linearly with the white noise}. The typical fingerprint of ADC non-linearity is a variation of the detector voltage output white noise not paired by a detectable variation in the voltage level. This effect was observed in the LFI radiometer data for the first time in flight, where drops of a few percent were observed in the voltage white noise but not in the output level over periods of few weeks.  Fig.~\ref{fig_adc_whitenoise_volts} shows this effect as a plot of relative white noise variation versus the detector output voltage for one the the most affected radiometer channels (the 44\,GHz detector \texttt{LFI25M-01}).

The grey points represent an average over each pointing (about 40\,min) while the solid line has been obtained by further binning the data in 200 bins over the plotted range in order to reduce the scatter and show more detail. The figure shows that the typical amplitude of the region where the non-linearity occurs is of the order of 1\,mV, corresponding to about three bits in the ADC. The ADC effect is strongest (3 to 6\%) in the 44~GHz channels, because of their lower detector voltages. 

\begin{figure}[h!]
  \includegraphics[width=8.8cm]{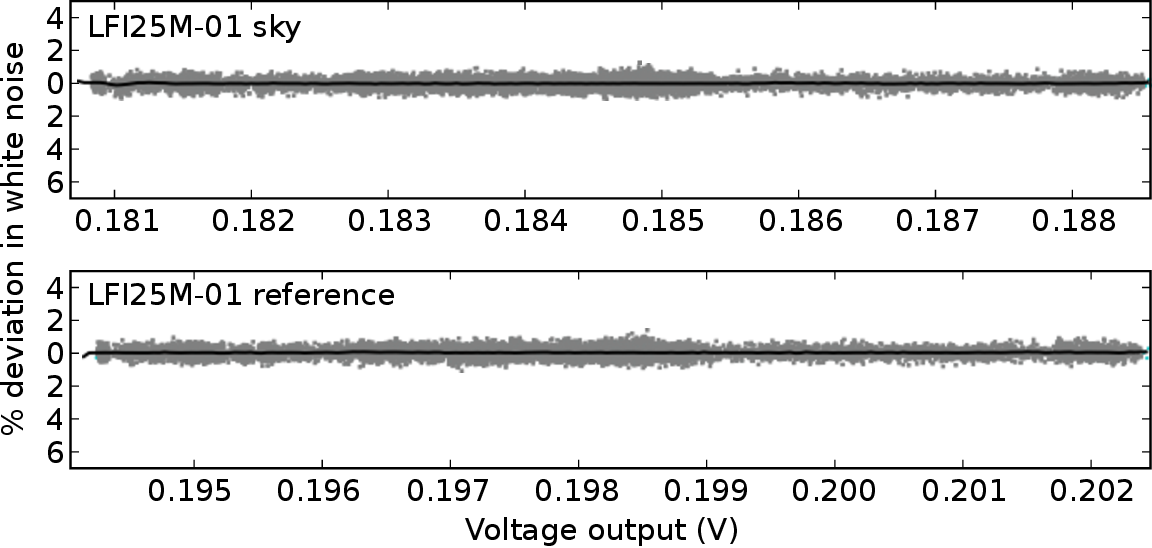}
  \caption{Same as in Fig.~\ref{fig_adc_whitenoise_volts} after correction of the ADC non-linearity effect.}
  \label{fig_adc_whitenoise_volts_corrected}
\end{figure}

The ADC non-linearity effect has been characterised from flight data and removed from the data streams according to the procedure described in \citet{planck2013-p02}. In Fig.~\ref{fig_adc_whitenoise_volts_corrected} we show the same data as in Fig.~\ref{fig_adc_whitenoise_volts} after the correction has been applied. The figure clearly shows that the anomalous white noise dips disappear after correction.

{In general we cannot exclude other causes of this anomalous scaling of voltage with noise. The ADC linearity tests performed before launch were not sensitive enough to highlight this effect and we could not perform post-launch tests on similar devices. On the other hand, the effect occurs repeatedly at specific values of the input ADC voltage and the the ADC non-linearity model applied to correct the data proved effective. These facts give us confidence that this hypothesis is sound.}

  \subsubsection{Imperfect photometric calibration}
  \label{sec_overview_calibration}
  
      An important set of systematic effects are those related to the photometric calibration of the radiometers. Such effects are discussed in \citet{planck2013-p02b}; here we will only provide the most important information to put the results of that paper within the context of this work.

There are three different kinds of systematic effects that can affect the calibration.
\begin{enumerate}
  \item \textit{Incorrect assumptions regarding the calibration signal}. In the case of LFI, the signal used for the calibration is the dipolar   field caused by the motion of the Solar System with respect to the CMB rest frame and by the motion of the spacecraft around the Sun.  We model the former using the values quoted by \citet{hinshaw2009} and the latter using the spacecraft's attitude information. Any error in the numbers would directly lead to an error in the calibration of \Planck-LFI data.
  \item \textit{Incorrect treatment of the calibration signal}. To actually use any previous knowledge of the CMB dipole, we need to convolve the signal with the beam response of the LFI radiometers. Any error in this step would produce a systematic effect in the map, not only because of the wrong shape expected for the calibration signal, but also because of the removal of the (wrong) dipole from the calibrated maps done by the \Planck-LFI pipeline \citep{planck2013-p02}. Possible types of errors include: 
  wrong convolution of the expected dipole with the radiometer beams, incorrect masking of the Galaxy when fitting the observed signal with the dipole, etc.
  \item \textit{Incorrect reconstruction of gain fluctuations}. Some of the algorithms we used in calibrating LFI data for this release use the radiometer equation and the recorded variations of the radiometers total-power output to track gain changes. In principle, any deviation in the behaviour of the radiometer from the {implemented model} 
  can induce systematic effects in the gain curves.
\end{enumerate}

{The calibration strategy and uncertainties for HFI are discussed in \citet{planck2013-p03f}.}

\subsection{Pointing effects}
\label{sec_pointing_effects}

  Pointing uncertainties are translated into uncertainties in pixel temperature measurements. If pointing uncertainties are not constant in time then the statistics of the sky anisotropy measurements is not preserved, with a consequent impact on power spectrum and cosmological parameters. For \Planck-LFI, pointing uncertainties arise from two main effects: 

\begin{enumerate}
  \item {\it Satellite pointing determination}. The \Planck\ Attitude Control Movement System guarantees a pointing accuracy of about $2\arcs$ \citep{planck2013-p01, pso2010-sience_operations}, which is well within scientific requirements. However, small non-idealities in the system and errors in the attitude reconstruction (caused, for example, by thermo-elastic effects) can affect the data.

  \item {\it Uncertainties in the focal plane geometry reconstruction}. The measurement of the \Planck-LFI focal plane geometry is based on the determination of the beam pointing with respect to the nominal line of sight exploiting Jupiter observations. The peak of each beam has been determined by fitting data with a bi-variate Gaussian function which may not be representative of the real beam centre.  
\end{enumerate}

\section{Assessing residual systematic effect uncertainties in maps and power spectra}
\label{sec_assessment}
  
  In this section we discuss the assessment of the impact of residual systematic effects on maps and power spectra. This assessment has been performed according to two strategies: by ``null maps'' obtained by differencing maps with the same sky signal, in order to highlight residuals, and by simulating known systematic effects in the timelines, exploiting a combination of flight data and measured instrumental properties.

\subsection{Null tests}
\label{sec_assessment_null_tests}

  We define a ``null test'' as any difference between two independent data sets which are anticipated to give nearly the same signal, on the assumption of perfect calibration, pointing reconstruction, and systematic effects removal. Null tests are a powerful means to assess the validity and self-consistency of \Planck\ data on various timescales and across different dimensions (detector, frequency, time), and to highlight systematic effects above the white noise level. 

The design of \Planck\ and its observing strategy provides a wide range of opportunities for null tests, with sensitivity to different systematic effects and implications for the scientific outputs. Although we refer to these tests as null tests, the results are generally not featureless. {Some of these features are caused by beam orientation and ellipticity that cause spurious effects in odd minus even survey difference maps in correspondence to point sources and in the galactic plane. The analysis of the effect of beam ellipticity on the CMB power spectrum is provided in \citet{planck2013-p02d} and will not be repeated here. A part of these residuals is caused by signal pickup from beam sidelobes. Our beam model captures, at least partly, these fingerprints. Future work will be aimed at exploiting this model to remove the sidelobe signal from the data. A fraction of this large-scale residual is not captured yet by our instrument model and will require further investigation to be understood and properly removed from the data.}

In the \Planck-LFI collaboration each internal data release is accompanied by a comprehensive set of null tests as a check of our processes and ongoing improvement in terms of systematic errors. In this section we report the results from the main tests supporting the systematic effect analysis for the first \Planck\ public data release. Unless otherwise noted, {the maps presented in this section} are masked to remove point sources and to include only pixels measured in both maps. Difference maps are divided by 2 to be statistically consistent with average maps, and are smoothed to 2$^\circ$ FWHM to enhance large scale features.

\subsubsection{{Systematic effects that are insensitive to null tests}}
\label{sec_insensitive_null_test_effects}

{Null maps are powerful means to understand residual systematic effects in the data, both of known and unknown origin, but they do not capture all possible effects. For example, fluctuations occurring on 20\,min time scale would be undetected in half-ring difference maps, and fast fluctuations (like 1-Hz spikes and short timescale temperature variations) and effects arising from near sidelobes would not be revealed by survey difference maps.  None of these effects, however, are likely to affect significantly the results of our analysis:} 

  
\begin{itemize}
  \item {Twenty-minute spurious fluctuations, if present, can be detected in power spectra calculated from time-ordered data (which are routinely calculated and assessed to derive noise properties). Their effect is strongly reduced by short (1-second) baseline destriping map-making. In-flight LFI noise properties have been presented and discussed in \citet{mennella2010}.}

  \item {Effects at 1-Hz have been assessed from in-flight timelines by stacking data from all the mission (for each detector) in 1-second time windows. This allowed us to produce time-domain templates of the 1-Hz spurious signal that have been removed from the data at 44 GHz, which is the channel most affected by this effect. }

  \item {Short timescale temperature fluctuations at the level of the radiometers and of the 4\,K reference loads are not expected as the LFI and HFI focal planes act as lowpass thermal filters. Measurements from the LFI and HFI temperature sensors confirm that only the slow temperature fluctuations propagate from the cooler cold ends to the focal plane detectors and reference loads. Thermal transfer functions have been derived both before launch \citep{terenzi2009a} and in-flight \citep{gregorio2013}. These have been used to produce the thermal systematic effect maps discussed in this paper. }

  \item {Near sidelobes can also produce spurious effects that are undetected in survey difference maps. The fraction of power intercepted by intermediate sidelobes ranges from 0.02\% to 0.08\% of the total beam power, about 10 times smaller than the fraction contained in far sidelobes (ranging from 0.18\% to 0.68\%). Their effect is therefore correspondingly smaller, of the order of $\sim$$1.5\,\mu$K in the maps. In this paper we therefore neglect the effect from intermediate sidelobes, but note that these will be addressed in detail in a future full-mission analysis paper.}
\end{itemize}

\subsubsection{Half-ring difference null tests}
\label{sec_half_ring_null_tests}

  Half-ring difference null tests, constructed by taking a weighted difference between the first and second halves of each pointing period, are useful to assess the data noise properties and systematic effects on time scales smaller than about 20\,min. Weights are calculated as explained in Sect.~9.2 of \citet{planck2013-p02}. 
  
\begin{figure}[h!]

  \includegraphics[width=88mm]{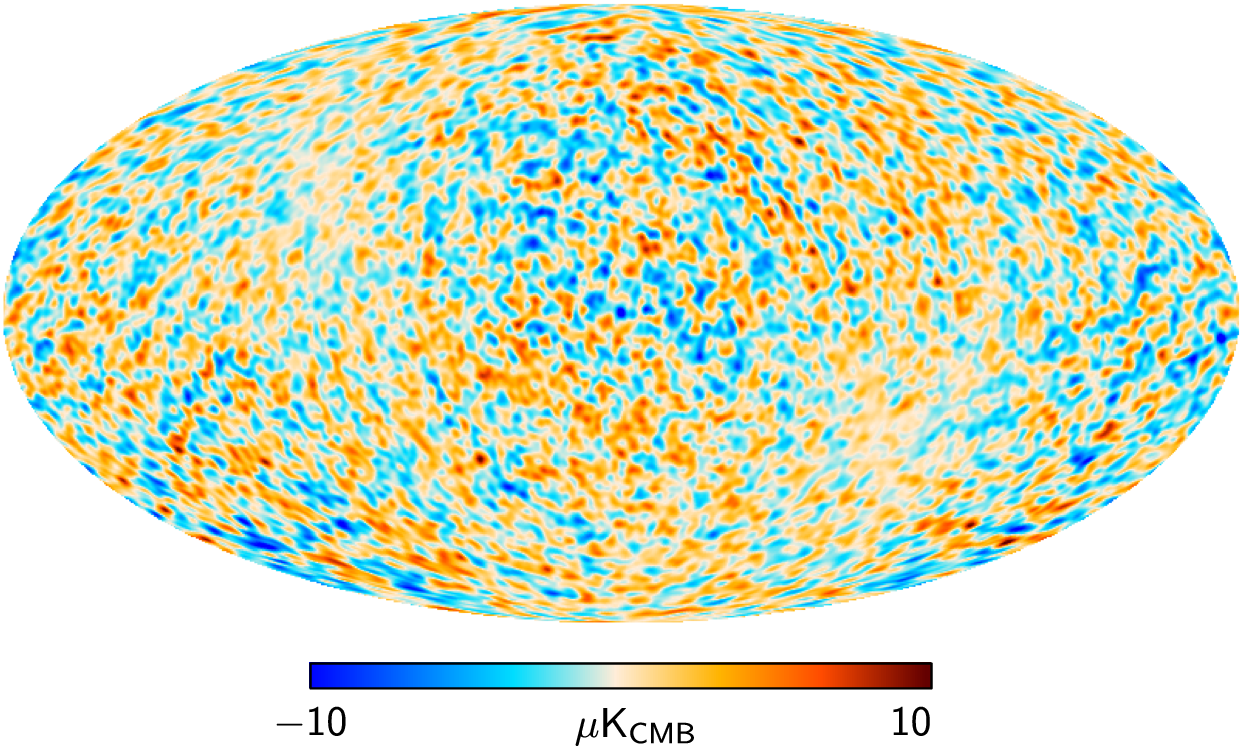}
  \includegraphics[width=88mm]{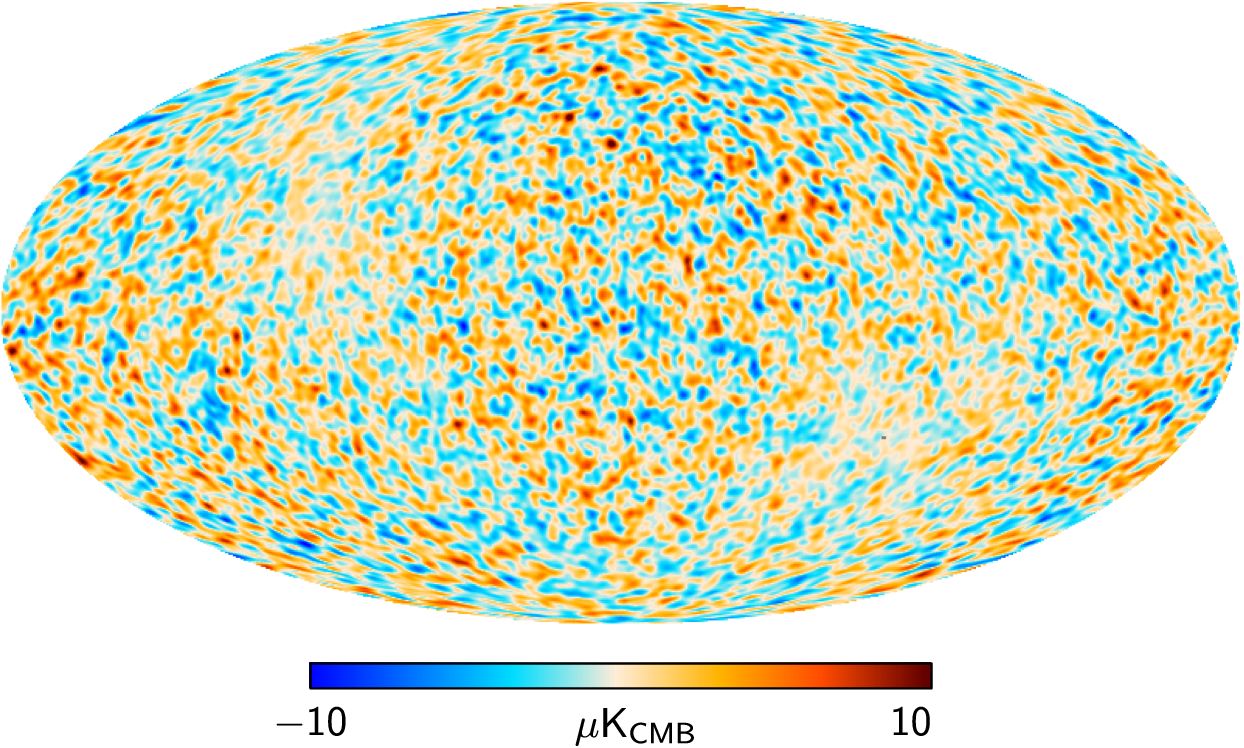}
  \includegraphics[width=88mm]{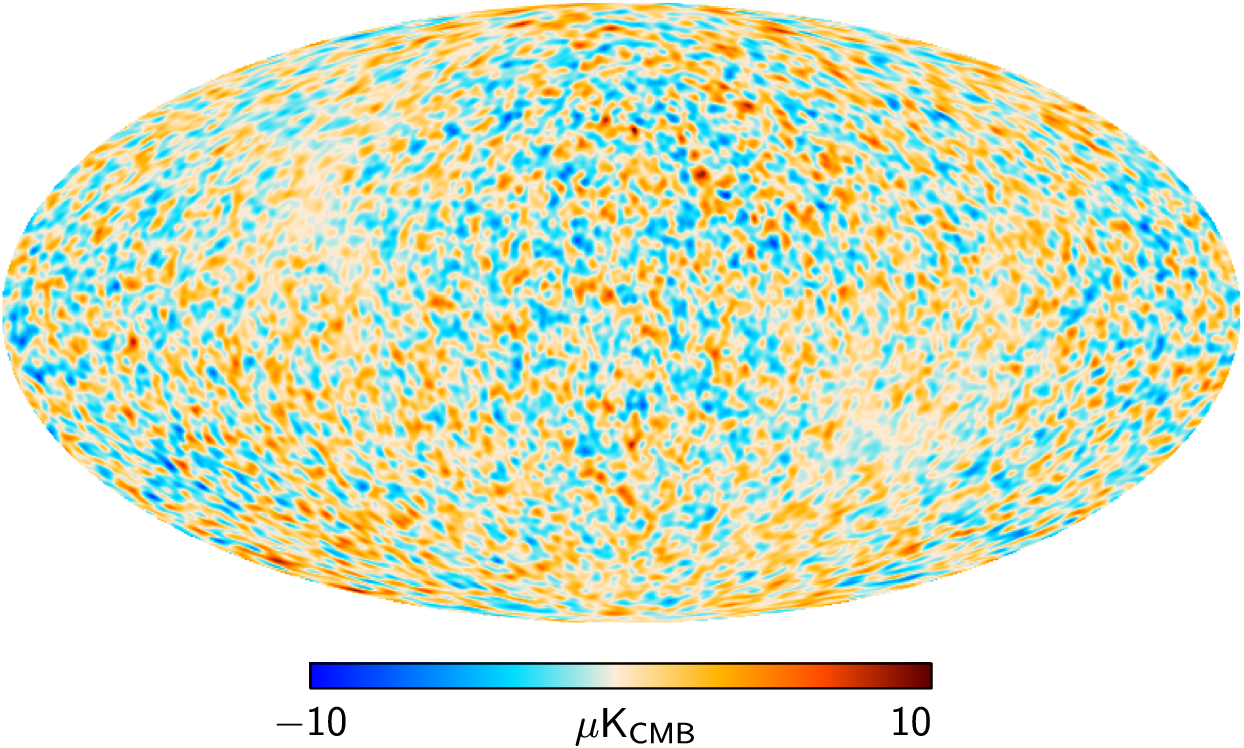}
  \caption{Half-ring difference maps: 30\,GHz (top), 44\,GHz (middle), and 70\,GHz (bottom).}

  \label{fig_half_ring_null_tests}

\end{figure}

  In Fig.~\ref{fig_half_ring_null_tests} we show the half-ring difference maps for the three LFI frequencies. A simple quantitative test was performed by dividing them pixel-by-pixel by the square root of the white noise covariance maps \citep{planck2013-p28} and checking the standard deviation of the resulting maps. We found this r.m.s. value to be very close to unity: 1.0211, 1.0089, and 1.0007 for 30, 44, and 70\,GHz, respectively. The deviation from unity is consistent with the different level of 1/$f$ noise in the three frequency channels \citep[see Tables 1 and 10 of][]{planck2013-p02}. {A more complete quantitative analysis of these maps including cross-spectra analysis is reported in the ``Data validation'' section of \citet{planck2013-p02}.}
  
\subsubsection{Survey difference null tests}
\label{sec_survey_difference_null_tests}

  Differences between single survey maps are useful to check for residual systematic effects at large angular scales.
  
\begin{figure*}

    \begin{center}
      \begin{tabular}{c c}

      \includegraphics[width=88mm]{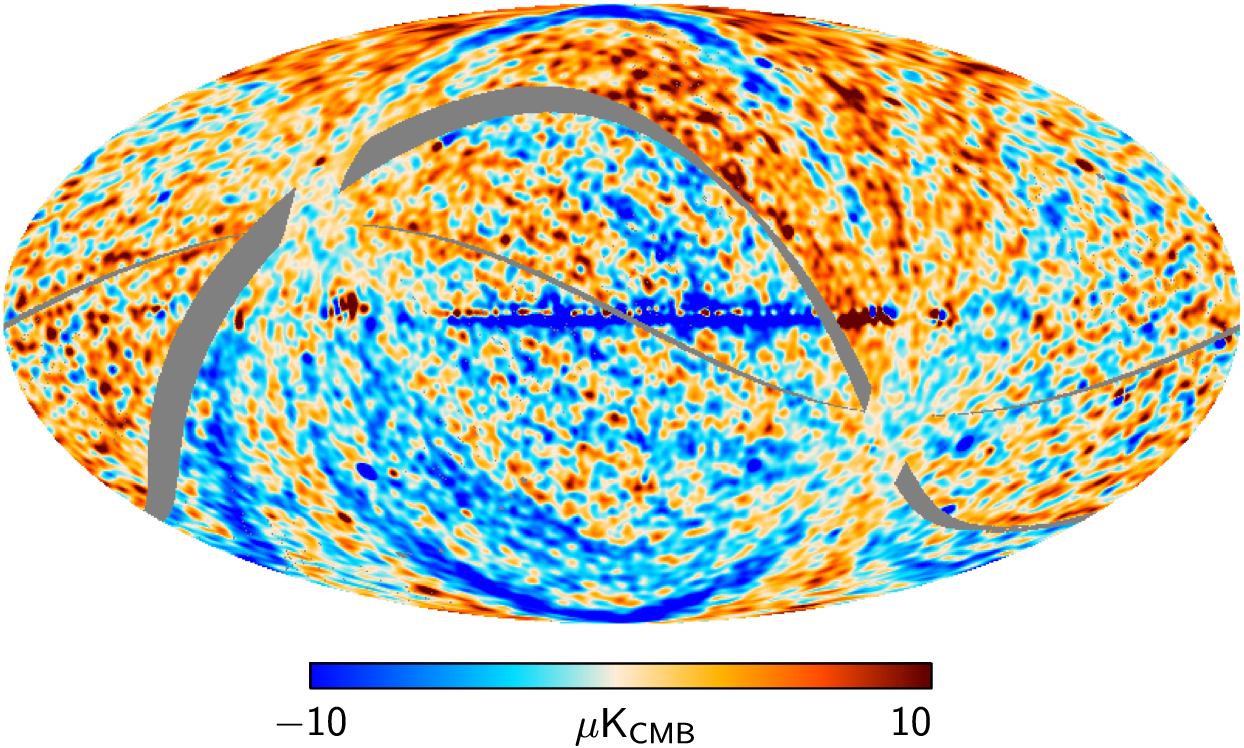} & \includegraphics[width=88mm]{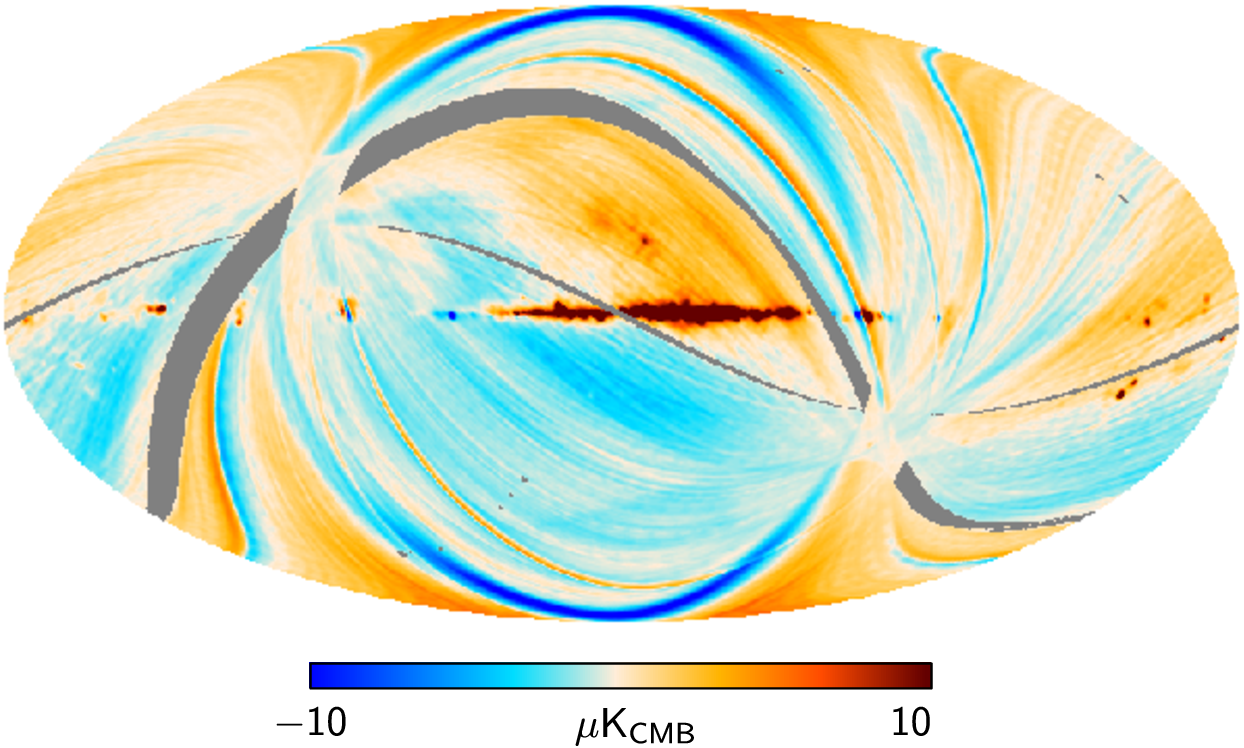}\\
      \includegraphics[width=88mm]{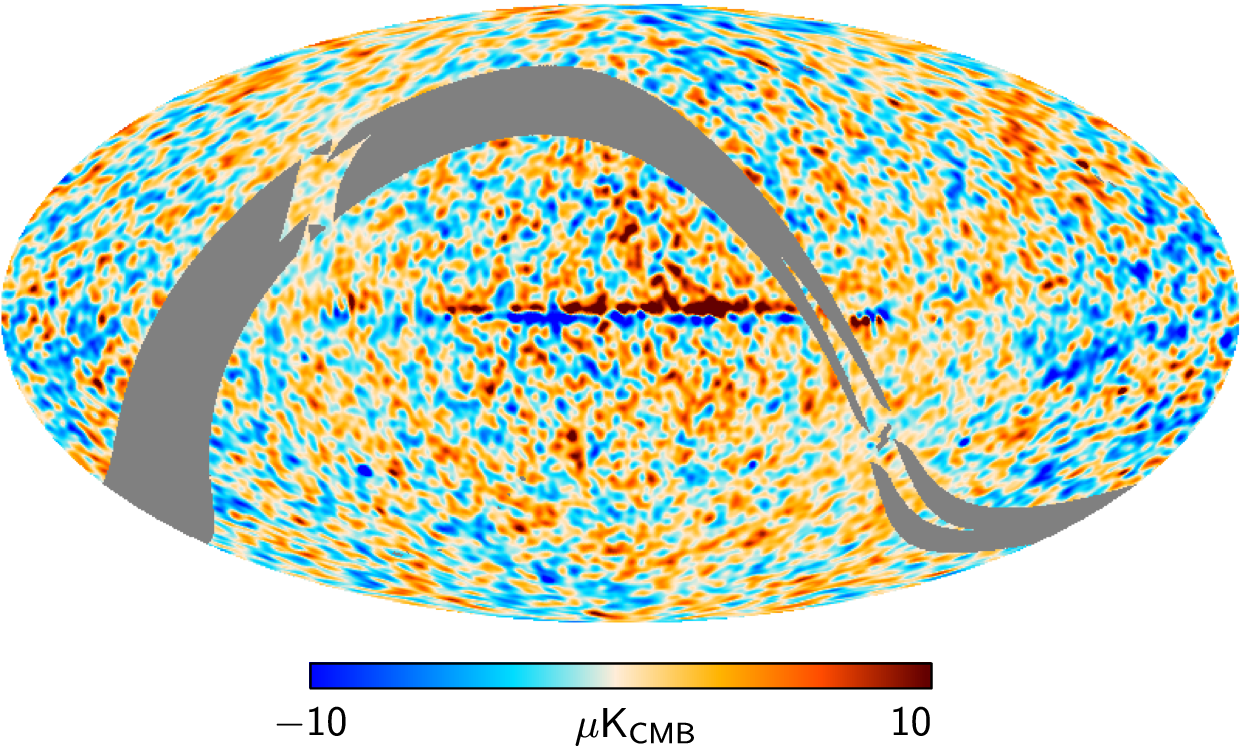} & \includegraphics[width=88mm]{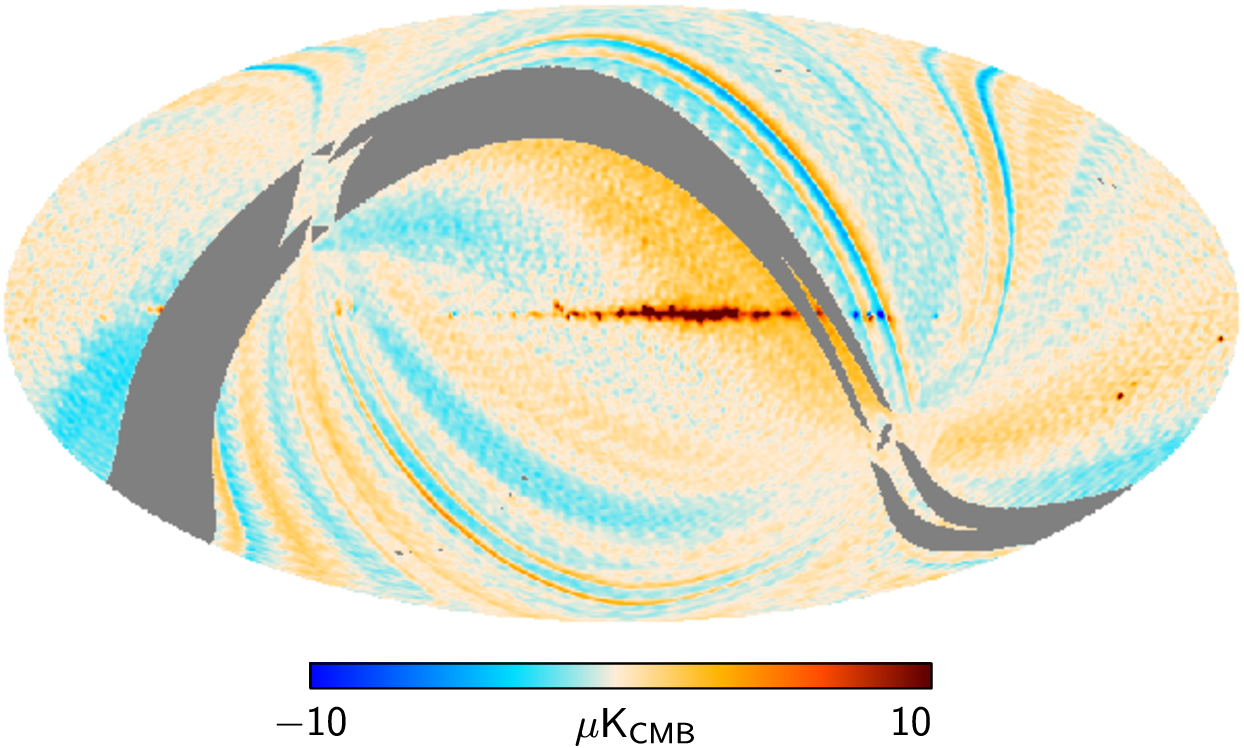}\\
      \includegraphics[width=88mm]{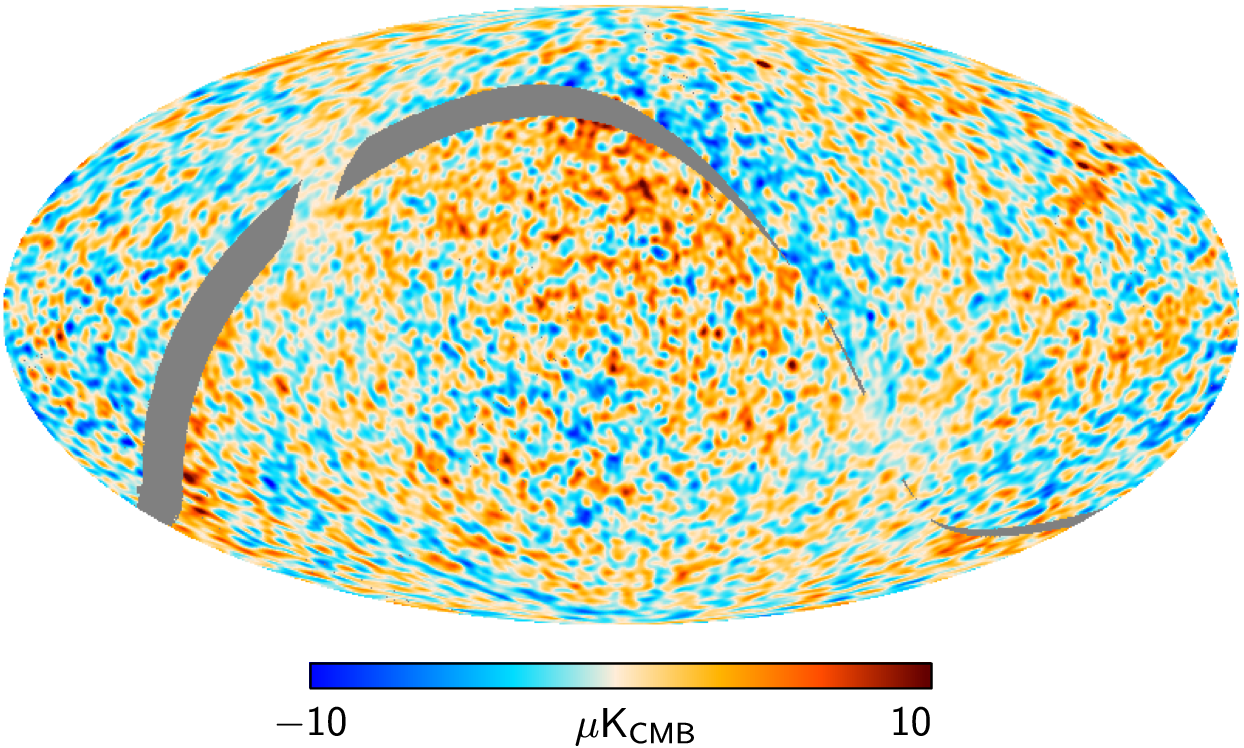} & \includegraphics[width=88mm]{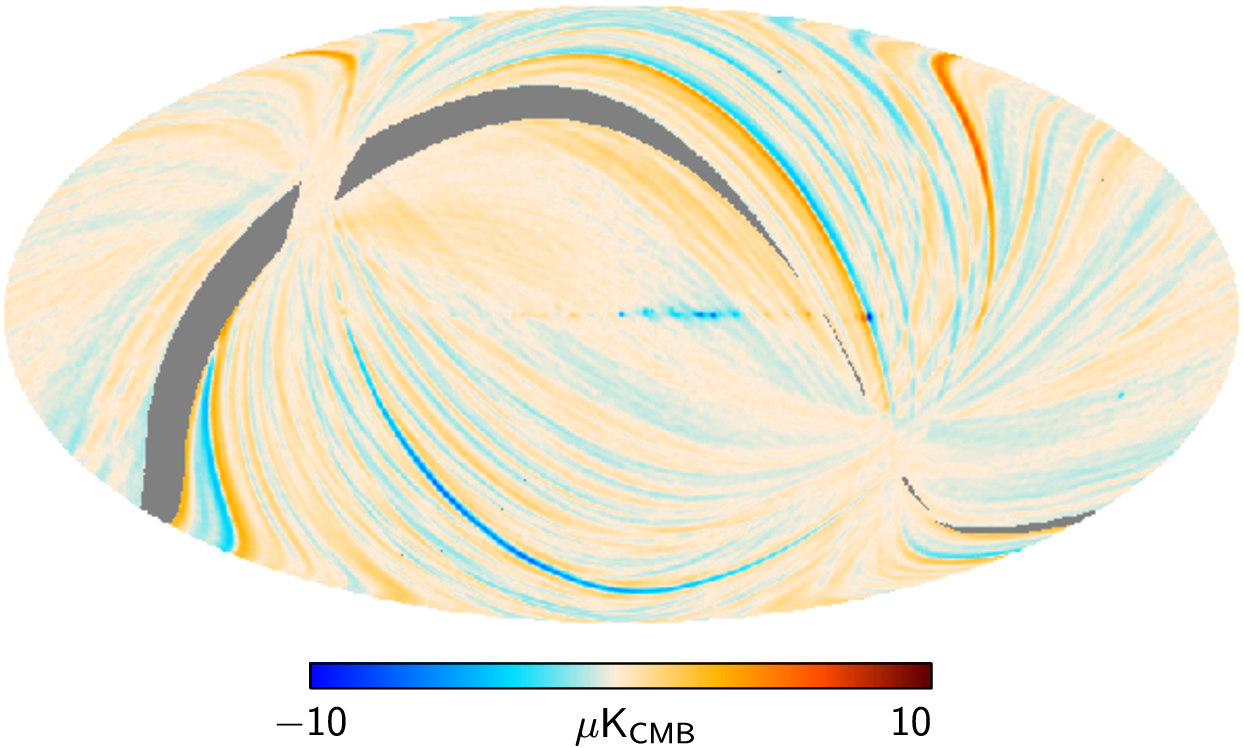}

      \end{tabular}
      \caption{{Survey 1 minus survey 2 difference maps calculated from actual measurements (left column) and from simulations (right column), for 30, 44 and 70\,GHz (top to bottom).}}

      \label{fig_odd-even_survey_diff}

    \end{center}

  \end{figure*}

  Difference maps of odd minus even surveys highlight effects arising from beam ellipticity and far sidelobes. {The left-hand column in} Fig.~\ref{fig_odd-even_survey_diff} shows the difference maps between surveys 1 and 2 {obtained from measured data at} the three LFI frequencies. The cosmic and orbital dipole signals are removed during calibration \citep[as discussed in][]{planck2013-p02b}, so the difference in orbital dipole signal between survey 1 and survey 2 is not visible. {These maps show large scale residuals above the noise floor, especially in the 30\,GHz channel, and in particular} far sidelobe pickup of the galactic plane in survey 2 is visible as a large blue ring. 

{For comparison, the right-hand column of Fig.~\ref{fig_odd-even_survey_diff} shows the same difference maps predicted by the systematics simulations discussed in Sections~\ref{sec_assessment_timeline_additive} and \ref{sec_assessment_effects_dependent_sky}. Clearly, our simulations reproduce patterns similar to those observed in the measured data, even if not every feature is exactly matched. The most notable example of the latter is the residual signal in the Galactic plane in the 30\,GHz map, which has an opposite sign in the simulations compared to the data, a discrepancy that has not yet been fully understood. One possibility is that the Galactic residual in the data null map may be dominated by beam ellipticity, an effect that was not accounted for in our simulations\footnote{Beam ellipticity is accounted for in the beam window function. For this reason we did not assess its impact on the final power spectra.}.}
  

\begin{figure}[h!]

    \begin{center}

      \includegraphics[width=88mm]{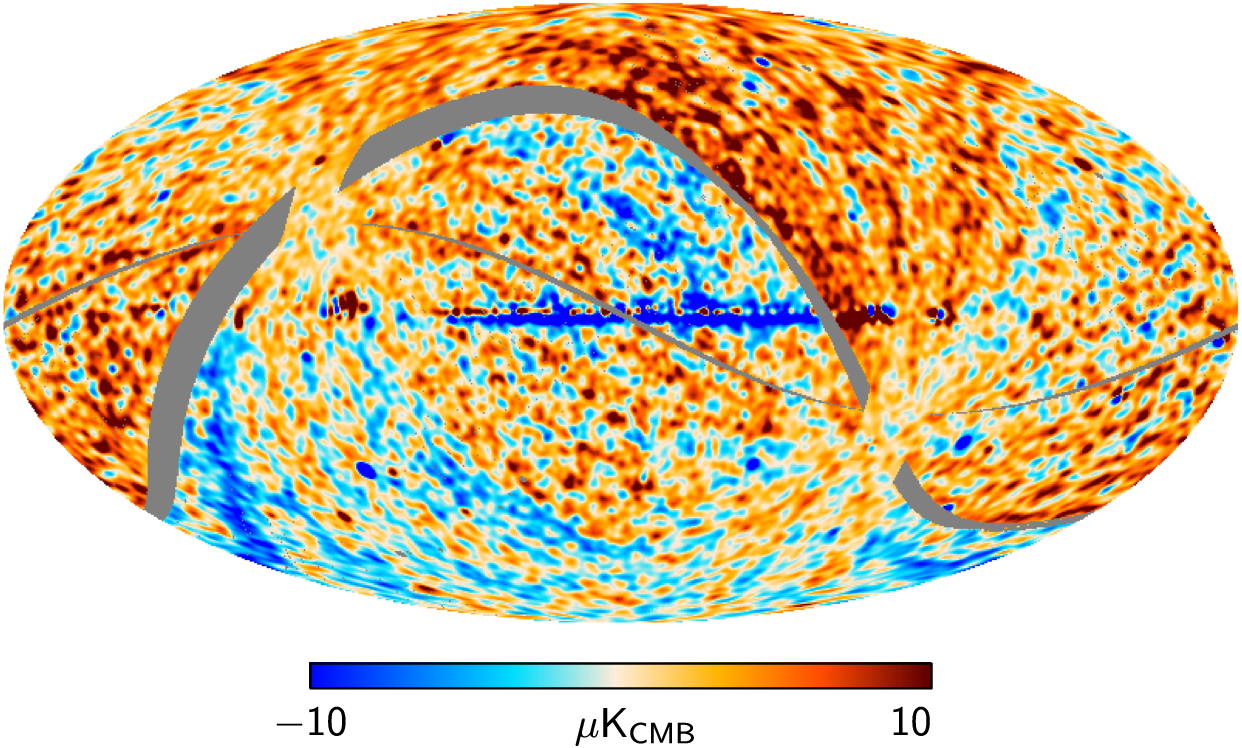}

      \caption{Survey $1 - 2$ difference map at 30\,GHz after subtracting a model of sidelobe contamination (see Fig.~\ref{fig_overview_straylight_30}). Note that the blue ring seen in Fig.~\ref{fig_odd-even_survey_diff} disappears, demonstrating both the origin of the structure and the accuracy of our model.}

      \label{fig_odd-even_survey_diff_no_sidelobes}

    \end{center}

  \end{figure}

{As a demonstration of the accuracy of our sidelobe model, we show in Fig.~\ref{fig_odd-even_survey_diff_no_sidelobes} the 30\,GHz survey 1 minus survey 2 difference map after removing the residual sidelobe signal as predicted by our model (see Fig.~\ref{fig_overview_straylight_30}). The blue ring structure disappears, confirming both the nature of this spurious feature in the map and that our sidelobe model captures a significant part of the large-scale residuals. Further investigation is needed to fully understand and remove the remaining level of spurious large-scale structures in our maps}.
  
\begin{figure}[h!]
  \begin{center}
      \includegraphics[width=88mm]{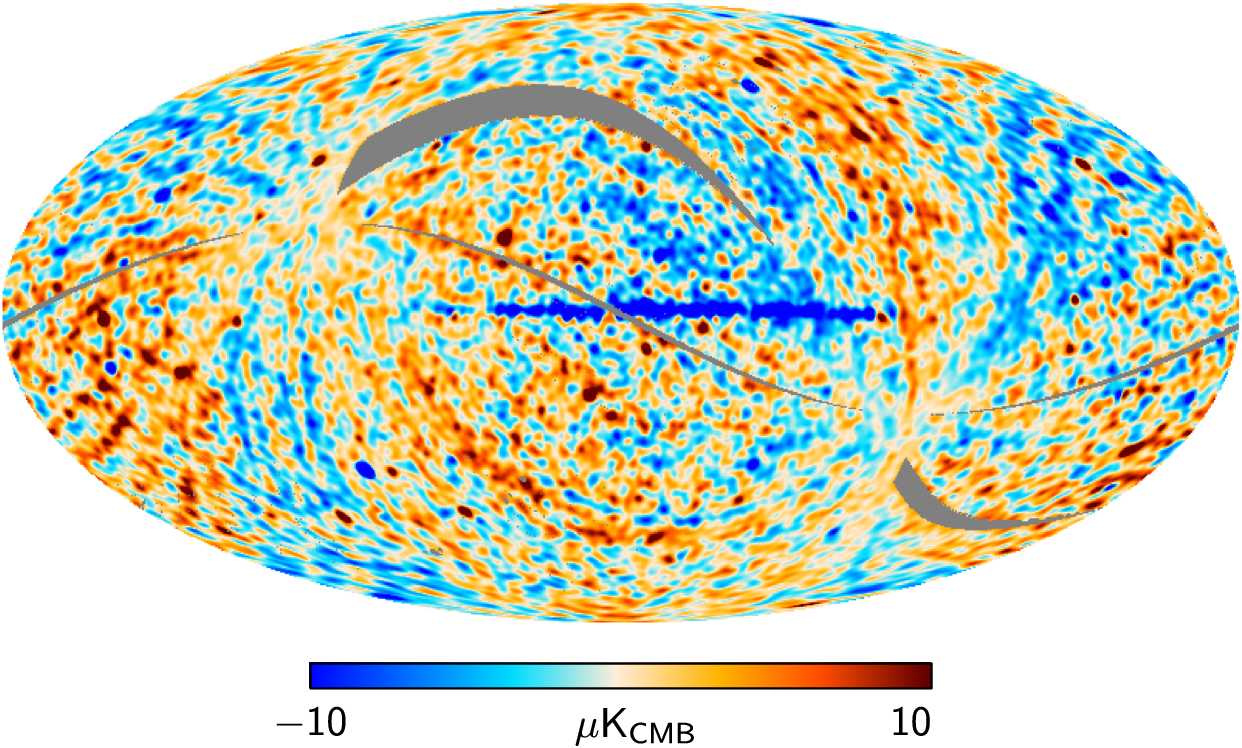}
      \includegraphics[width=88mm]{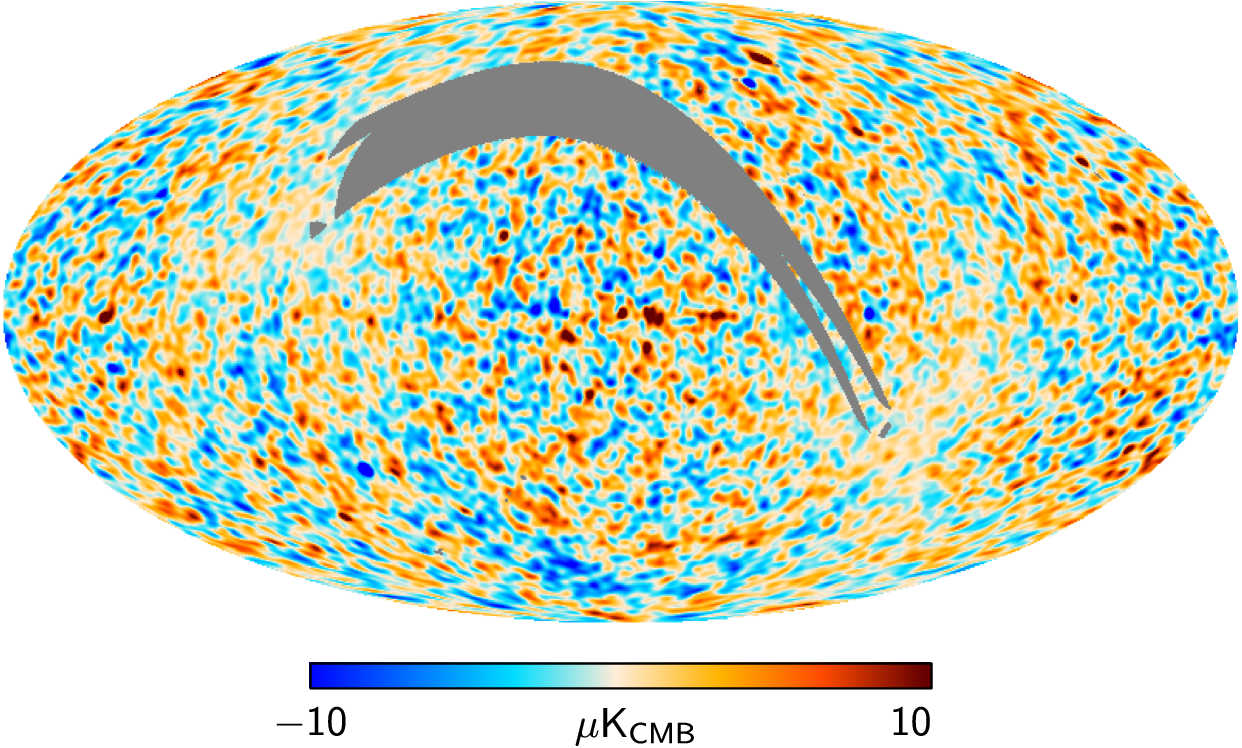}
      \includegraphics[width=88mm]{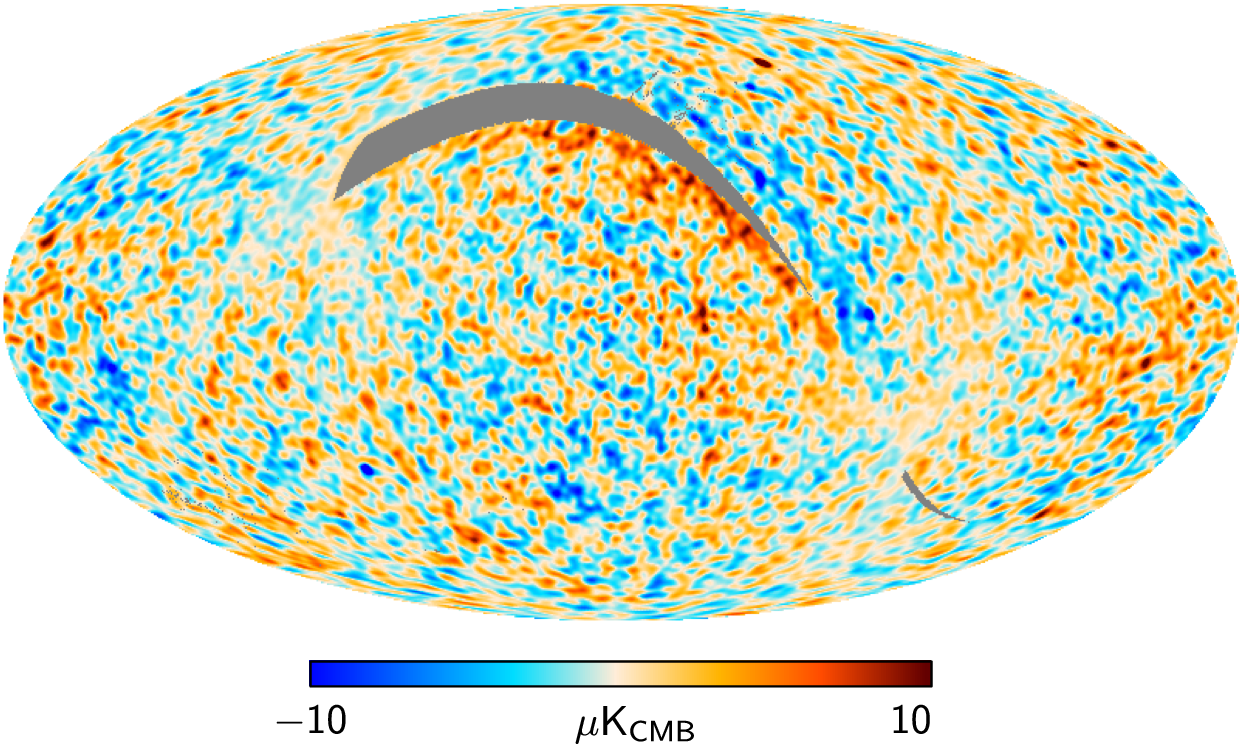}

    \caption{Survey 1 minus survey 3 difference maps: 30\,GHz (top), 44\,GHz (middle), and 70\,GHz (bottom).}

    \label{fig_odd-odd_survey_diff}

  \end{center}

\end{figure}

In Fig.~\ref{fig_odd-odd_survey_diff} we show survey 1 minus survey 3 null test maps. {These two surveys} cover the sky in nearly identical orientations, and would be consistent with noise if calibration and other systematics were perfectly controlled. {However, as seen in this figure, there are large-scale features also in these difference maps, and these are still under investigation. Because the first \Planck\ cosmological release is based only on data from the first two surveys, though, a detailed study of effects present in data beyond survey 2 is outside the scope of this paper, and will be discussed in the second data release.}

\begin{figure}[h!]

  \begin{center}

    \includegraphics[width=88mm]{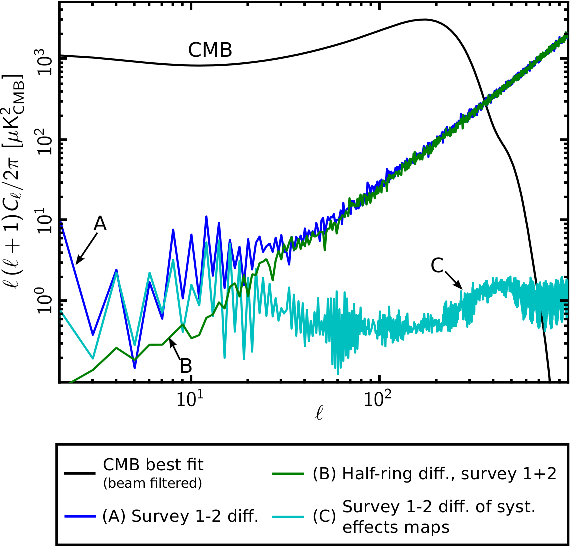}

    \caption{Angular power spectra for 30\,GHz null tests. Pseudo-spectra are calculated on 80\% of the sky, with the Galaxy and point sources masked. All spectra are corrected for sky fraction but not for beam smearing effects. {For comparison, we also show the simulated odd-even survey difference spectrum and the best-fit cosmological model spectrum {filtered by the beam window function}.  For $\ell\geq 30$ the spectrum of the survey 1 minus survey 2 map difference fully coincides with the half-ring difference spectrum calculated for the same time period, while for the 30~GHz channel there is a small sidelobe contribution at the {10}\,$\mu$K$^2$ level on larger scales.}}
    \label{null_spectra_30Ghz}

  \end{center}

\end{figure}

\begin{figure}[h!]

  \begin{center}

    \includegraphics[width=88mm]{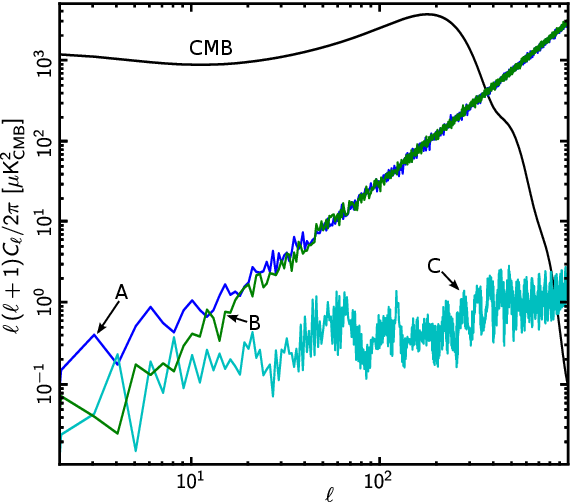}

    \caption{Angular power spectra for 44\,GHz null tests.  In this case he low-$\ell$ spectrum of the survey 1 minus survey 2 map difference is closer to that of the half-ring difference spectrum {than at 30~GHz}. Letters in the plot follow the same convention of the legend in Fig.~\ref{null_spectra_30Ghz}.}
    \label{null_spectra_44Ghz}

  \end{center}

\end{figure}

\begin{figure}[h!]

  \begin{center}

    \includegraphics[width=88mm]{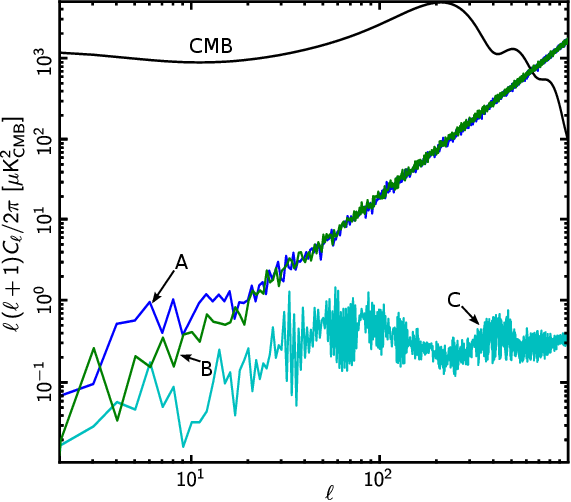}

    \caption{Angular power spectra for 70\,GHz null tests. {This channel has the smallest large-scale residuals among the three LFI channels.}  Letters in the plot follow the same convention of the legend in Fig.~\ref{null_spectra_30Ghz}. 
}
    \label{null_spectra_70Ghz}

  \end{center}

\end{figure}

Next, we quantify the impact of residuals seen in our difference maps through angular power spectrum analyses. {In Figs.~\ref{null_spectra_30Ghz} through \ref{null_spectra_70Ghz} we compare the pseudo-spectra of odd-even survey difference maps (left column of Fig.~\ref{fig_odd-even_survey_diff}) with the CMB spectrum filtered by the beam window function of each LFI channel; to spectra from simulated survey difference maps (right column of Fig.~\ref{fig_odd-even_survey_diff}); and to spectra from half-ring difference maps, which estimate the noise contribution. Note that the half-ring difference map noise is slightly lower than the survey difference map noise, simply because half-ring difference maps cover a larger time period than survey difference maps. For this reason we rescale the half-ring difference spectra by $\sqrt{t_{\rm HR}/t_{\rm SD}}$, where $t_{\rm HR}$ and $t_{\rm SD}$ are the average integration times in half-ring and survey difference maps, respectively.}  

{For multipoles $\ell\geq 30$ the survey difference power spectra closely match the instrumental noise. For $\ell<30$ there are additional residuals, especially at 30\,GHz, which are partially captured by our simulations. These residuals, however, are at least two orders of magnitudes below the CMB power spectrum. }  


\subsection{Assessment of timeline-additive systematic effects}
\label{sec_assessment_timeline_additive}

  \subsubsection{Thermal effects}
  \label{sec_assessment_thermal}
  
    \label{sec_assessment_thermal_method}
\paragraph{Method.}

  Thermal systematic effects maps have been generated using a simulation strategy that combines in-flight temperature sensor measurements \citep{mennella2010}, thermal modelling of the propagation of temperature fluctuations \citep{tomasi2010} and radiometric transfer functions measured during ground tests \citep{terenzi2009b}. Here we sketch the procedure used to combine these data into systematic effect maps.
  
  For each temperature effect and for each receiver detector diode we choose the most representative sensor, generally the closest to the receiver. Housekeeping data are low-pass filtered to remove high frequency sensor noise, Fourier-filtered to obtain the estimated temperature fluctuation at the receiver location, and then multiplied by the radiometric transfer function to obtain the simulated antenna temperature fluctuation on the undifferenced sky and reference load channels. 
  
  For each pointing period the average measured sky and reference load voltages are added to the two antenna temperature fluctuation data streams. After a weighted average of the two detector data values of each radiometer, we take the sky-load difference using the gain modulation factor, $r$, and multiply the resulting stream by the photometric constant, $G$. The weights, $r$, and $G$, are the same as used in the nominal pipeline to produce sky maps for that radiometer.

  After oversampling to the receiver sampling frequency using linear interpolation, we use these data to build maps accounting for the same in-flight pointings and map making procedure used to produce the final scientific products.
  
\paragraph{Results.}
\label{sec_assessment_thermal_results}
  
\begin{figure}[h!]
    \includegraphics[width=8.8cm]{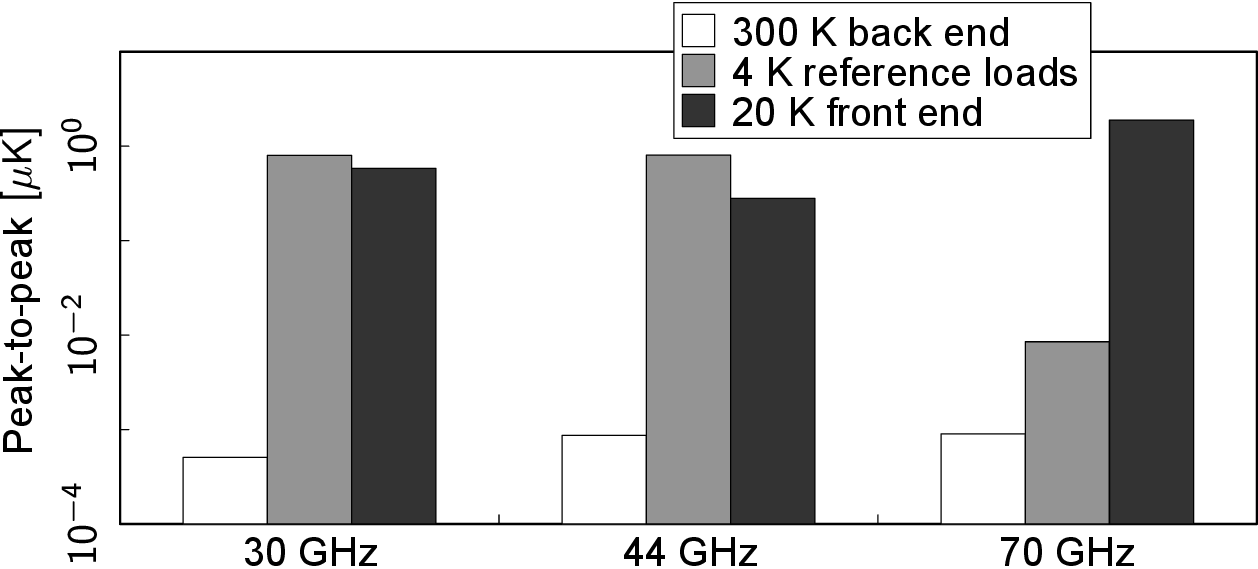}
    \caption{Peak-to-peak thermal effects in maps. Notice the logarithmic scale on the ordinate axis.}
    \label{fig_p2p_effect_thermal}
  \end{figure}

  In Fig.~\ref{fig_p2p_effect_thermal} we show the peak-to-peak amplitude of the various effects on final maps. Back-end temperature fluctuations have a sub-$\mu$K effect on maps. This low level can be understood if we consider that these fluctuations impact sky and reference load signals symmetrically and are effectively suppressed in the differential measurement. Furthermore the residual present in the data is a purely multiplicative effect, so it is essentially calibrated out through our gain model \citep{planck2013-p02b}. 
  
  Temperature variations in the 4\,K reference loads couple with the radiometric output as an asymmetric additive spurious signal. In this case the relative calibration model provides no benefit, leaving a residual of about 1\,$\mu$K peak-to-peak at 30 and 44\,GHz. At 70\,GHz, this effect is largely suppressed by the active thermal control system present on the HFI focal plane, close to the reference loads of this frequency channel. 

Front-end 20\,K temperature variations couple with the radiometric measurements through both gain and noise temperature fluctuations. For this reason the effect can only be partially calibrated out. Moreover, the asymmetry of the receiver chain before the orthomode transducer is such that the suppression provided by the sky-load differencing is not optimal. The residual effect is similar for the three frequency channels, and of the order of $1\,\mu$K peak-to-peak. 

\begin{figure}[h!]
    \includegraphics[width=8.8cm]{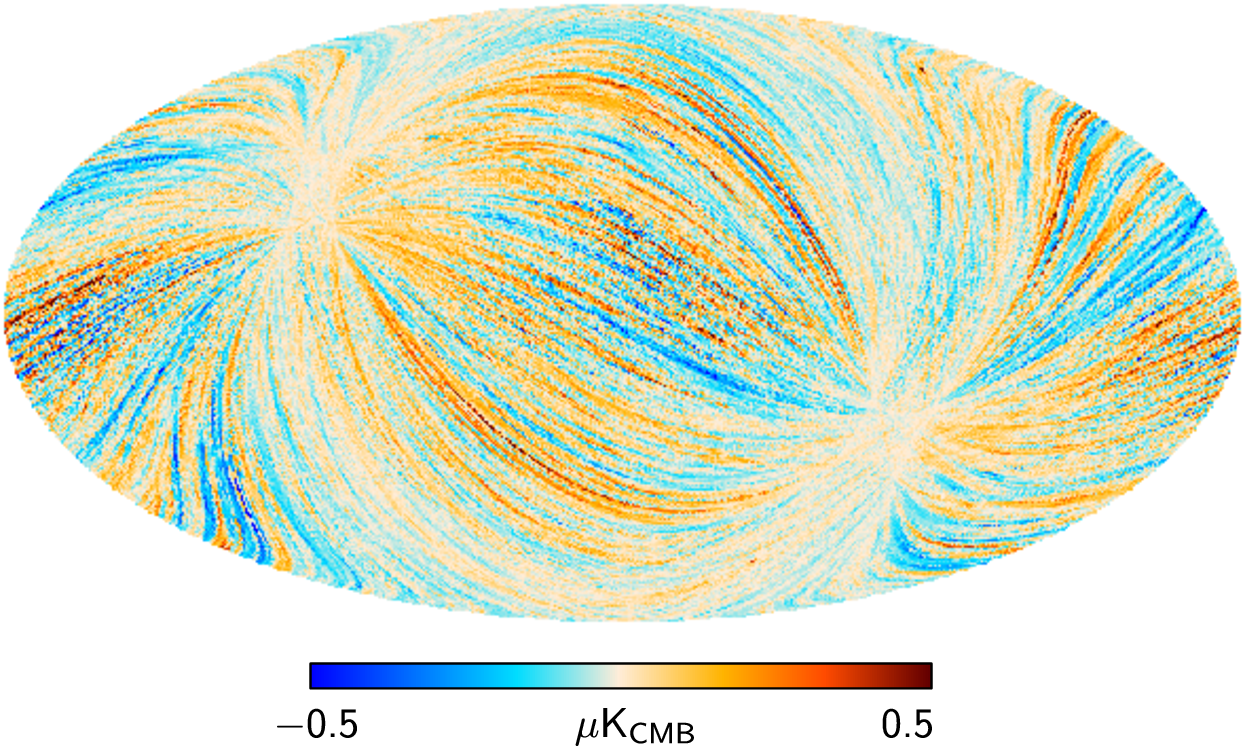} \\ 
    \\
    \includegraphics[width=8.8cm]{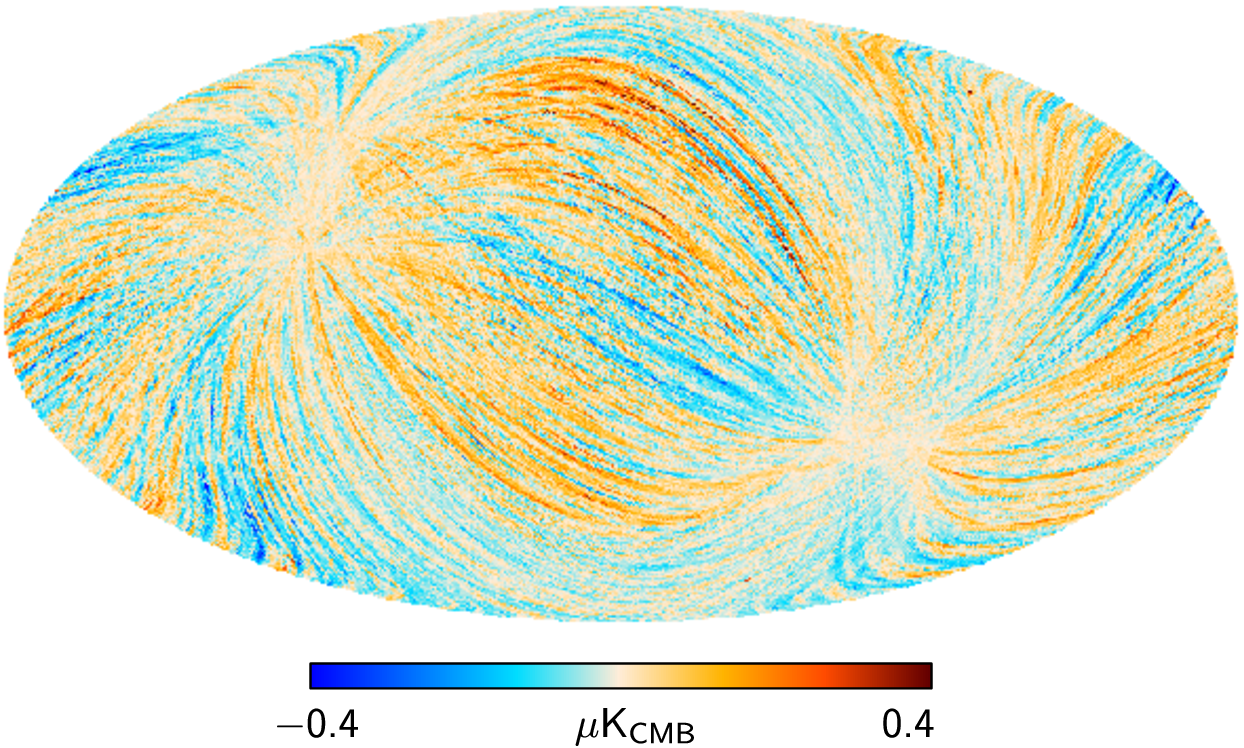} \\
    \\
    \includegraphics[width=8.8cm]{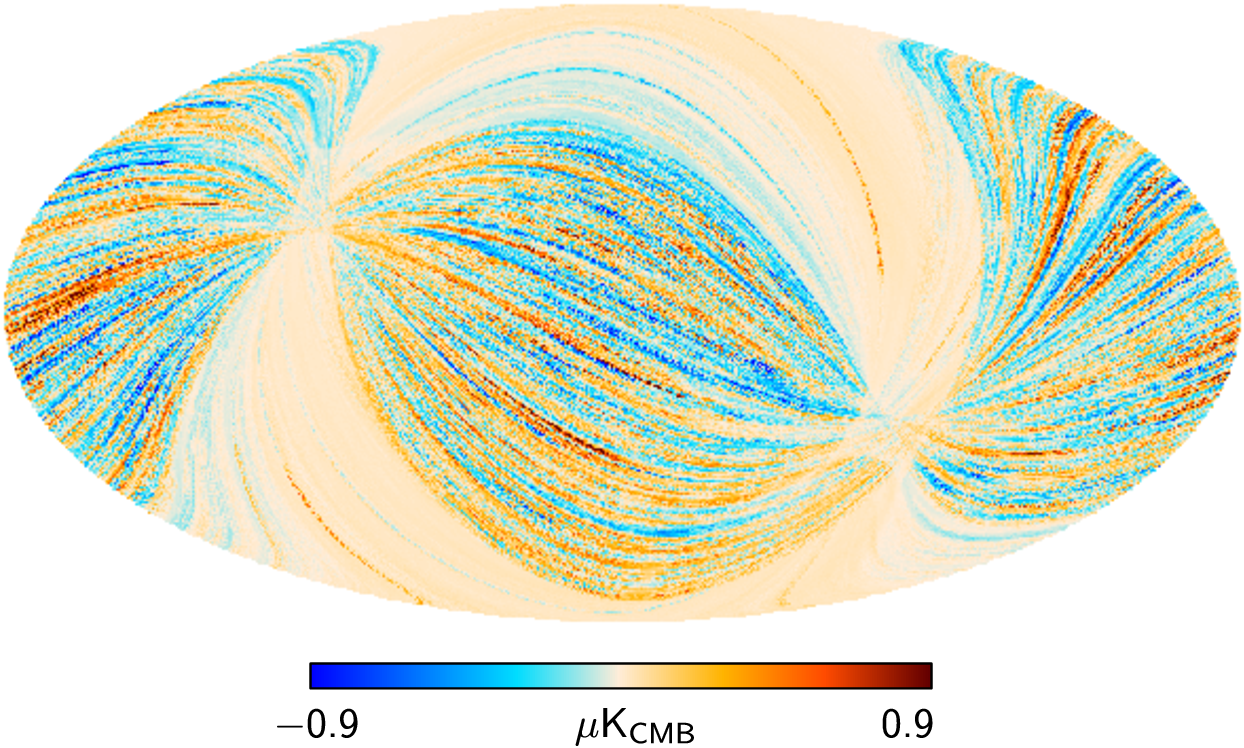}  
    \caption{Maps of combined thermal effects at 30\,GHz (top), 44\,GHz (middle), and 70\,GHz (bottom).}
    \label{fig_global_thermal_maps}
  \end{figure}

  Maps of the combined thermal effects at the three LFI frequency channels are shown in Fig.~\ref{fig_global_thermal_maps}.
  
   \subsubsection{Bias fluctuations}
   \label{sec_assessment_bias}

    \paragraph{Method.} The effect of bias fluctuations in the front-end amplifiers has been computed for maps and power spectra using the measured drain currents and a linear transfer function that links the drain currents of the two amplifiers to the radiometric output in antenna temperature. Since we are interested in assessing purely electrical instabilities, we correct the drain current housekeeping data for variations induced by temperature changes in the 20\,K and 300\,K temperature stages, i.e.,

\begin{equation}
  I_{\rm drain}^{\rm corr}(t) = I_{\rm drain}(t) - \alpha_{20\,\mathrm{K}} \delta T_{20\,\mathrm{K}}(t) - \alpha_{300\,\mathrm{K}} \delta T_{300\,\mathrm{K}}(t),
  \label{eq_id_thermal_correction}
\end{equation}
where $\delta T_{20\,\mathrm{K}}(t)$ and $\delta T_{300\,\mathrm{K}}(t)$ are temperature variations on the 20\,K and 300\,K temperature units, respectively, and $\alpha_{20\,\mathrm{K}}$ and $\alpha_{300\,\mathrm{K}}$ are the corresponding drain current thermal susceptibility coefficients, calculated using an iterative linear fitting process. First we calculate the coefficients of the susceptibility to back-end temperature fluctuations, exploiting the temperature change induced by the change in the transponder state, which occurred at day 258 (see Fig.~\ref{fig_temperatures}), and then we determine the coefficients of the susceptibility to front-end temperature fluctuations using data from a temperature susceptibility test run at the end of the in-flight calibration phase. The iterative process is closed by re-calculating back-end thermal coefficients after correcting drain currents for front-end temperature fluctuations.

Following thermal correction we correlate drain current changes with antenna temperature variations in sky and reference load samples. We recall here that the LFI receiver architecture implies that the signal characteristics at each detector depend on both radiometer front-end amplifiers \citep{bersanelli2010}. Thus, for each detector diode we first calculate the weight, $w$, providing the maximum correlation between the output voltage and the linear combination of the drain currents of the two radiometer amplifiers, then a linear fit between this combination and the voltage output provided the required transfer function. Mathematically the relationship between the corrected drain current fluctuations, $\delta I_{\rm drain}^{\rm corr}$, and the voltage output variations, $\delta V_{\rm sky(ref)}$, reads
\begin{equation}
  \delta V_{\rm sky(ref)}(t) = \alpha_{\rm sky(ref)}\left[ w\, \delta I_{\rm drain,1}^{\rm corr}(t) + (k-w)\, \delta I_{\rm drain,2}^{\rm corr}(t)\right],
  \label{eq_id_etf}
\end{equation}
where $k$ is a constant, $w$ is the weight and $\alpha_{\rm sky(ref)}$ is the slope of the linear fit between the weighted combination of the two drain currents and the sky (reference load) voltage outputs. The time-ordered data obtained by Eq.~(\ref{eq_id_etf}) are finally projected onto the sky using flight pointings.

\begin{figure}[h!]
    \includegraphics[width=8.8cm]{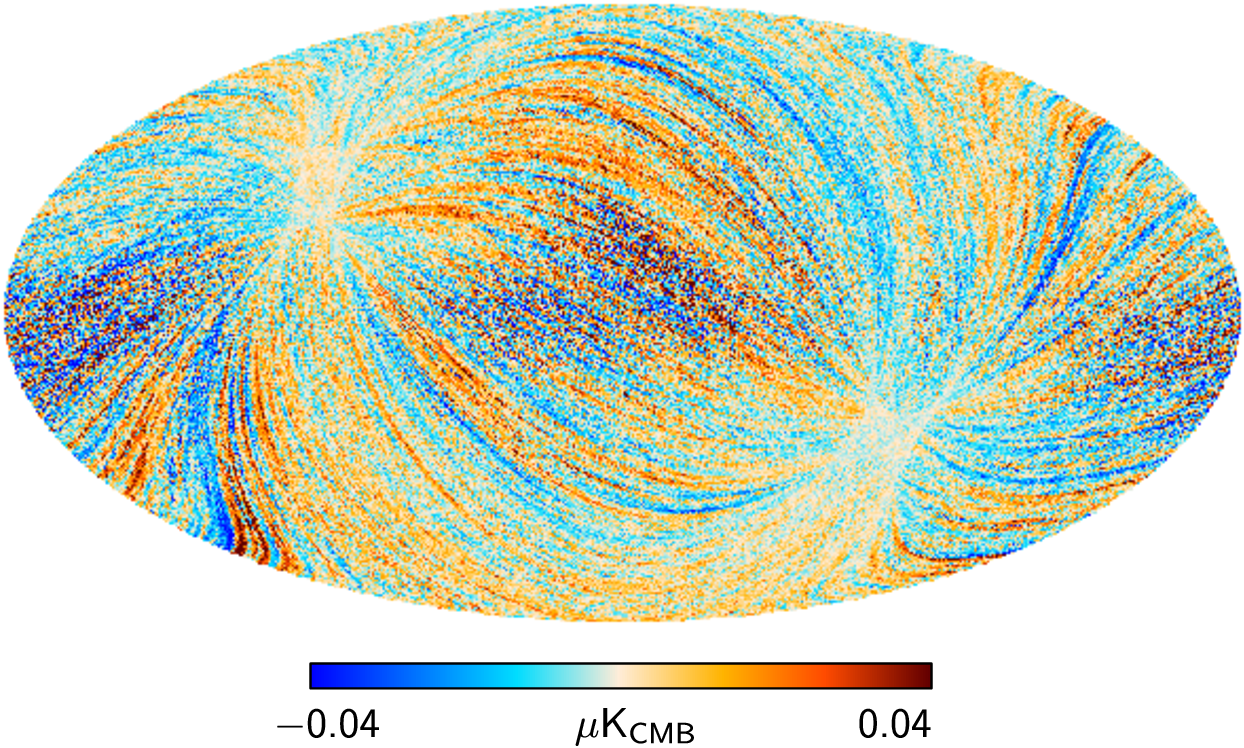} \\ 
    \\
    \includegraphics[width=8.8cm]{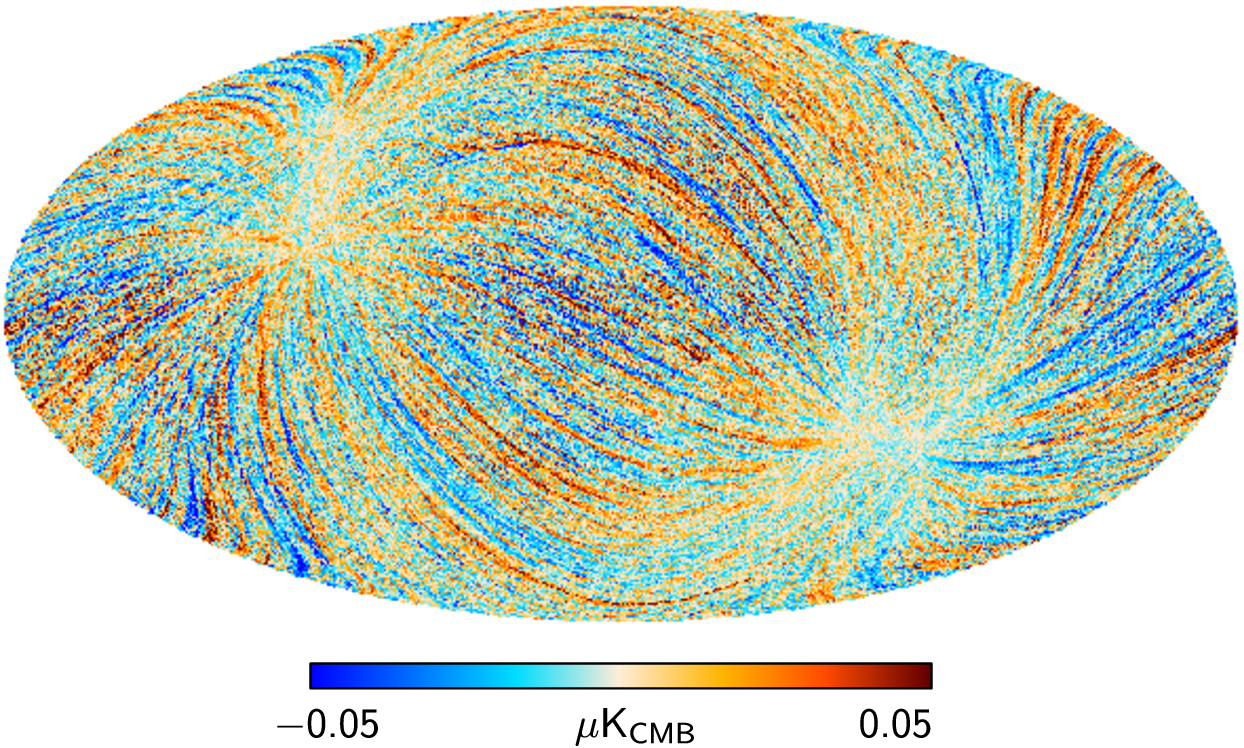} \\
    \\
    \includegraphics[width=8.8cm]{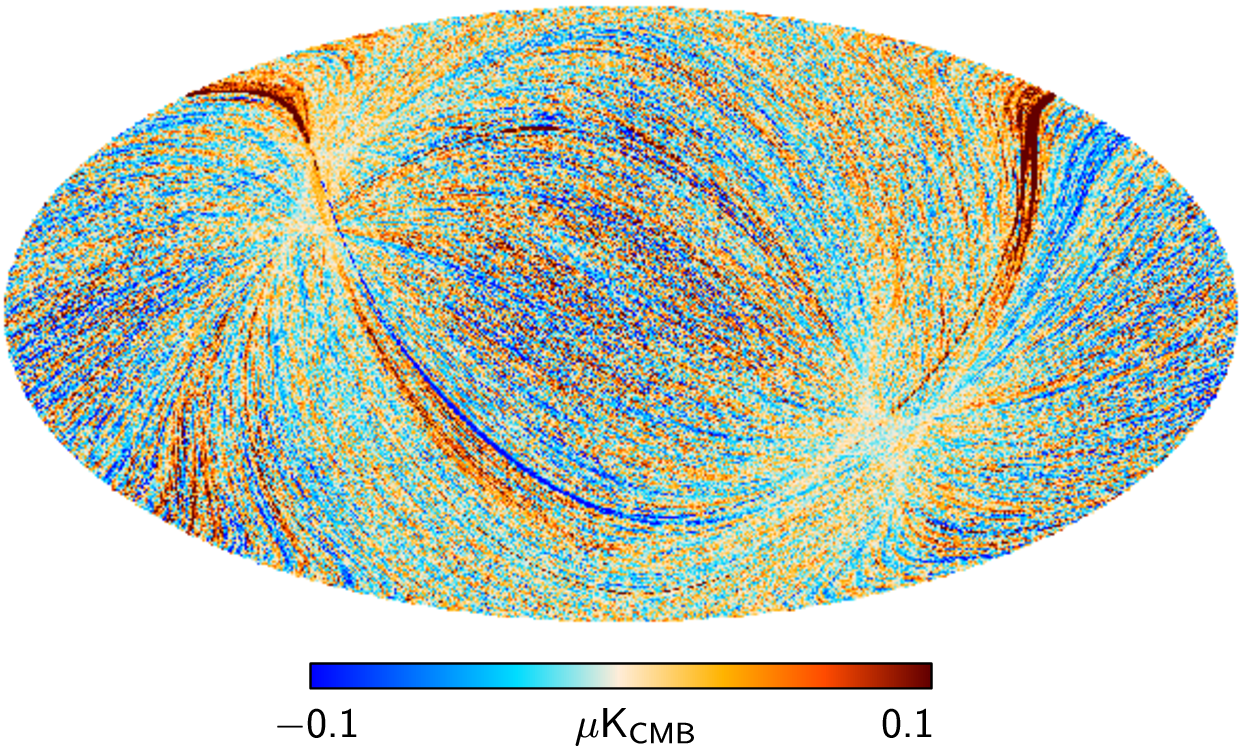}  
    \caption{Maps of the systematic effect from drain current fluctuations at 30\,GHz (top), 44\,GHz (middle) and 70\,GHz (bottom).}
    \label{fig_idrain_maps}
  \end{figure}

\paragraph{Results.} Maps of the residual effect arising from bias fluctuations are shown in Fig.~\ref{fig_idrain_maps}. This effect is smaller than 1\,$\mu$K peak-to-peak at all frequencies and presents little structure apart from a stripe in the 70\,GHz map, caused by a jump in the bias voltage occurring at day 258, following the change in transponder state. This jump affected in particular the 70\,GHz radiometers, leaving a small signature in the maps.

   \subsubsection{1-Hz spikes}
   \label{sec_assessment_spikes}
    
    \paragraph{Method.}

  Time ordered data containing the spike signal are generated using templates obtained from flight radiometric data. The details of this method are described in Sect.~7.1 of \citet{planck2011-1.4}, and will not be repeated here.
  
\paragraph{Results.}
\label{sec_assessment_spikes_results}

\begin{figure}[h!]
    \includegraphics[width=8.8cm]{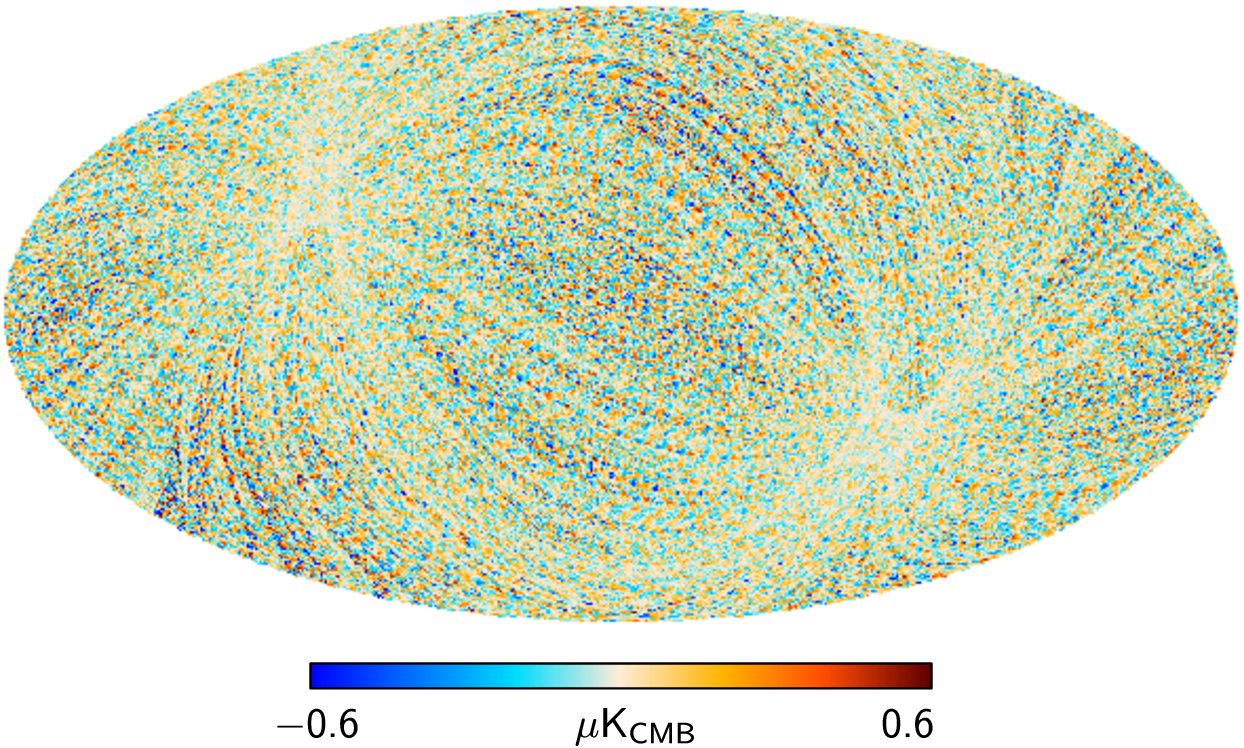} \\ 
    \\
    \includegraphics[width=8.8cm]{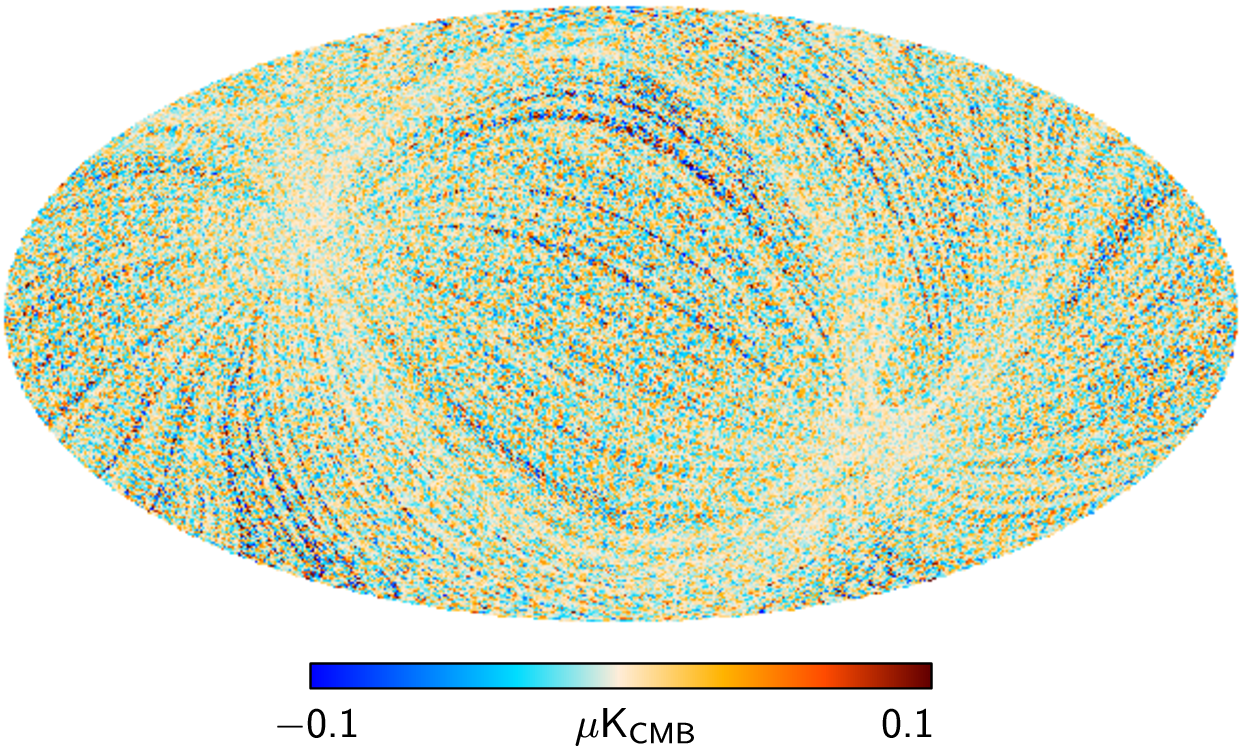} \\
    \\
    \includegraphics[width=8.8cm]{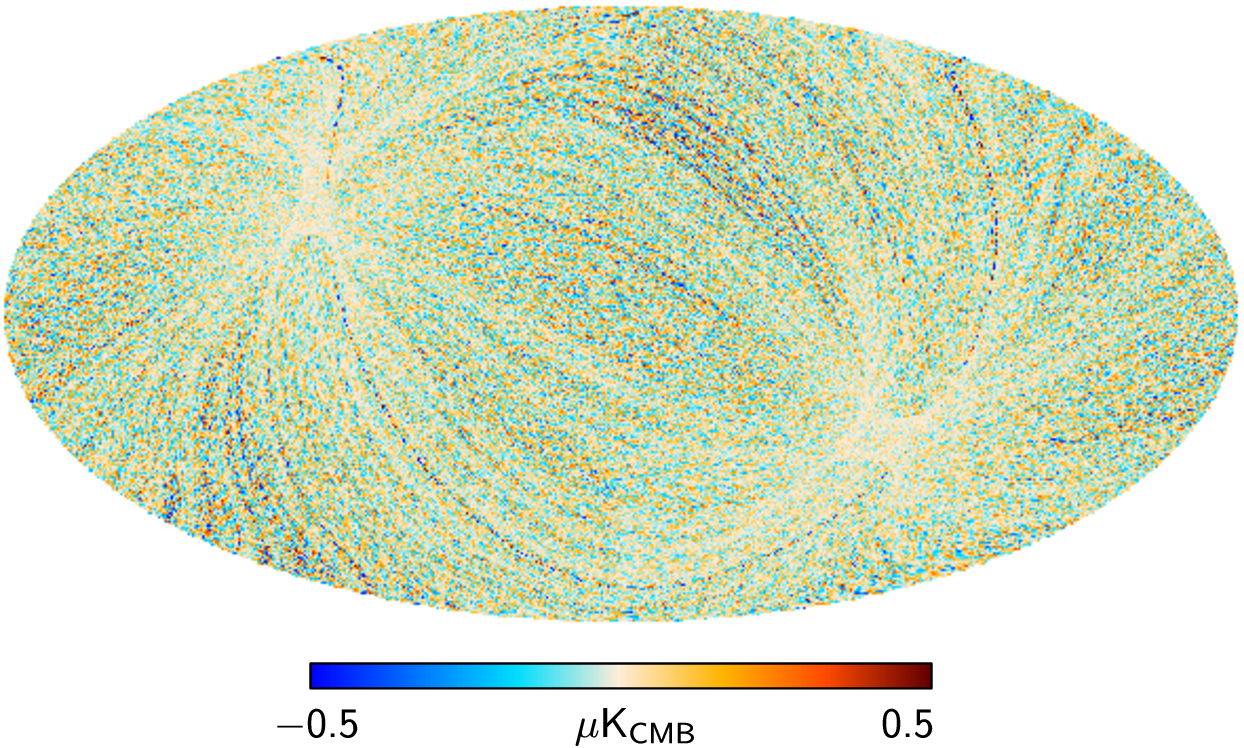}  
    \caption{Maps of 1\,Hz spikes at 30\,GHz (top), 44\,GHz (middle) and 70\,GHz (bottom). The map at 44\,GHz represents the residual after the spike signal has been removed from the time-ordered data, while maps at 30 and 70\,GHz represent the spike signal with no removal applied.}
    \label{fig_spike_maps}
  \end{figure}

  In Fig.~\ref{fig_spike_maps} we show maps of the spike systematic effect at the three LFI frequencies. Because spikes are removed from the 44\,GHz channel, but not from 30 and 70\,GHz, the corresponding maps represent the residual effect after removal at 44\,GHz, and the spike effect with no removal applied at 30 and 70\,GHz. In all the three channels the r.m.s. effect is at the sub-$\mu$K level.

\subsection{Assessment of effects dependent on the sky}
\label{sec_assessment_effects_dependent_sky}

  \subsubsection{Far sidelobes}
  \label{sec_assessment_sidelobes}

    \paragraph{Method.}
\label{sec_assessment_sidelobes_method}

\begin{figure}[h!]
    \includegraphics[width=8.8cm]{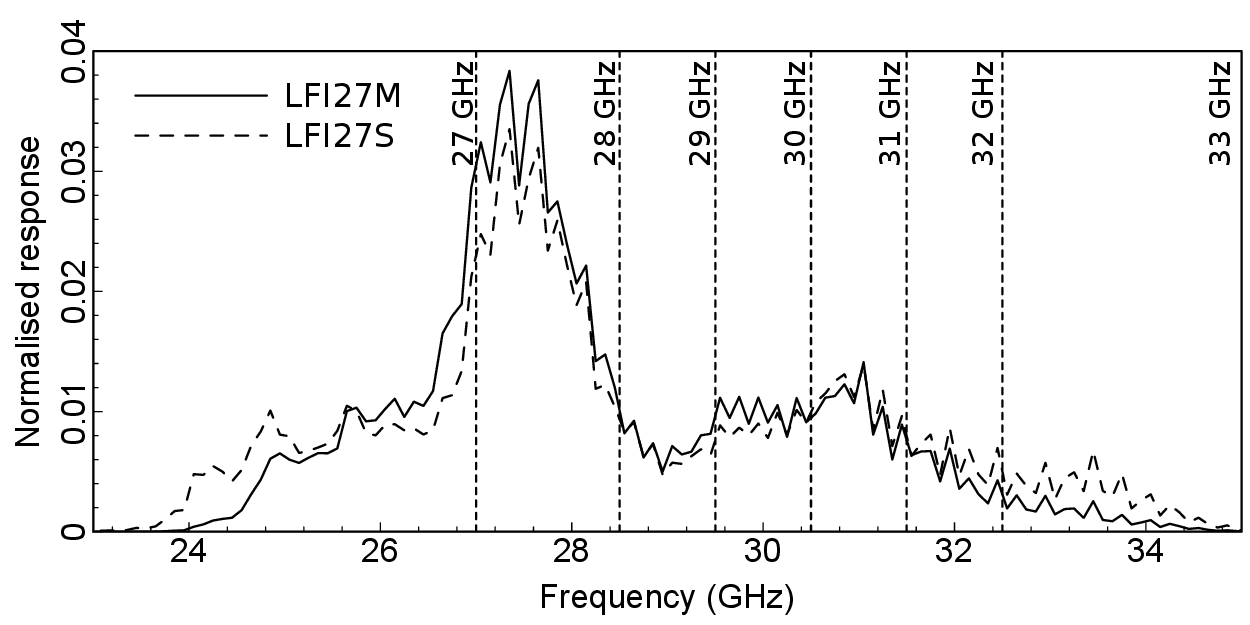}
    \caption{Bandpass response of the two radiometers of the \texttt{LFI27} receiver. The figure shows the seven frequency intervals and the corresponding frequencies at which sidelobes have been simulated. For each interval the weight is the integral of the bandpass response curve.}
    \label{fig_sidelobes_bandpass_response}
  \end{figure}

  The external straylight contamination is evaluated with simulations in which the sky model includes the diffuse Galactic emission and the dipole, the two most important sources of external straylight contamination. At 30\,GHz, the straylight assessment includes the beam frequency dependence and the receiver in-band response \citep[see][]{zonca2009,planck2013-p03d} by dividing the bandpass response into discrete frequency intervals. For each frequency interval a weight factor is calculated as the integral of the bandpass response over the interval itself. In Fig.~\ref{fig_sidelobes_bandpass_response} we show, as an example, the bandpass response of the \texttt{LFI27} receiver (main and side arms) and the seven frequency intervals considered in the simulations. The weights correspond to the integral of the bandpass response curve over the frequency interval. In parallel, far sidelobes are computed using the \texttt{GRASP MrGTD}\footnote{Multi-reflector geometrical theory of diffraction} software (\url{www.ticra.com}) at the frequencies indicated on the top of each slice reported in Fig.~\ref{fig_sidelobes_bandpass_response} (27, 28, 29, 30, 31, 32, and 33\,GHz). The optical model used in \texttt{GRASP} simulations is the ``RFTM'' reported in \citet{planck2013-p02d}.   For each frequency interval an observation of the sky model is simulated for all 30 GHz detectors using the beam sidelobes and the real sky pointings, neglecting beam smearing effects and weighting the data stream with the above mentioned weight factors. Finally we run the \texttt{Madam} map-making code to generate maps from simulated data streams.

\paragraph{Results.}
\label{sec_assessment_sidelobes_results}
  
\begin{figure}[h!]
    \begin{tabular}{cc}
      \centering
	\includegraphics[width=4.25cm]{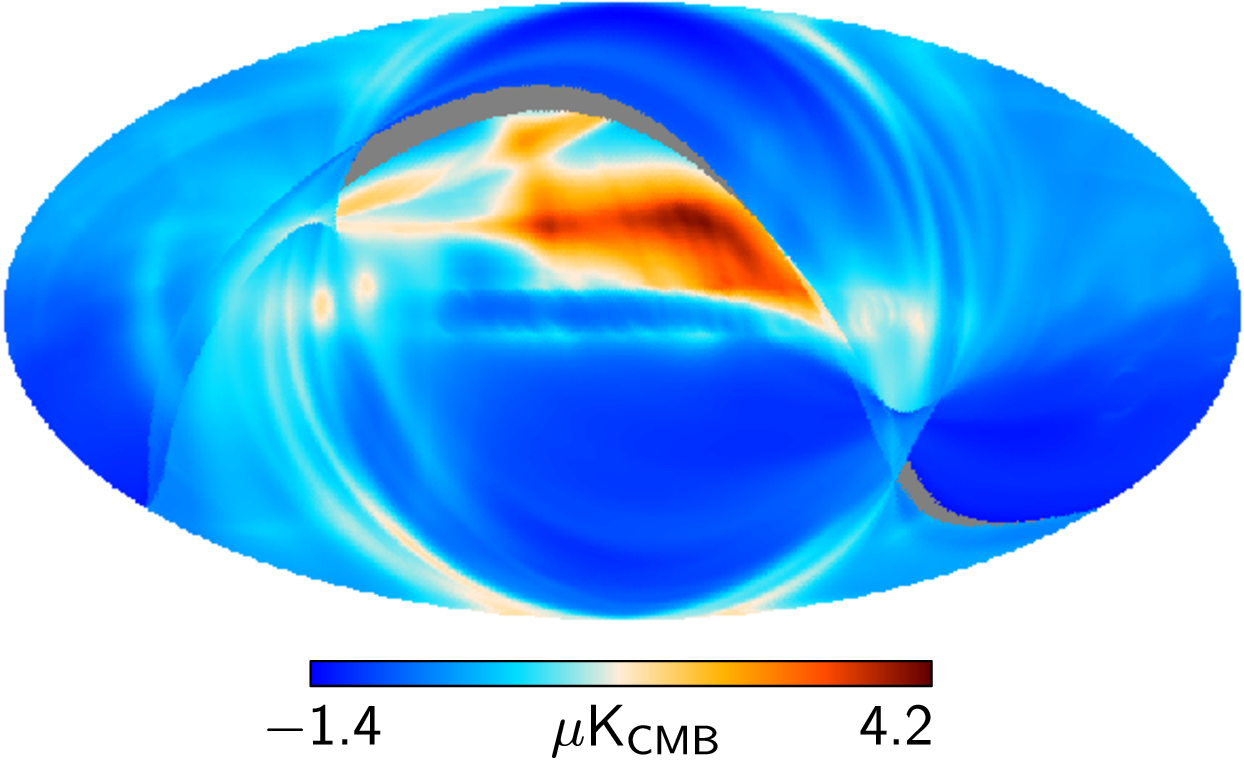} & 
	\includegraphics[width=4.25cm]{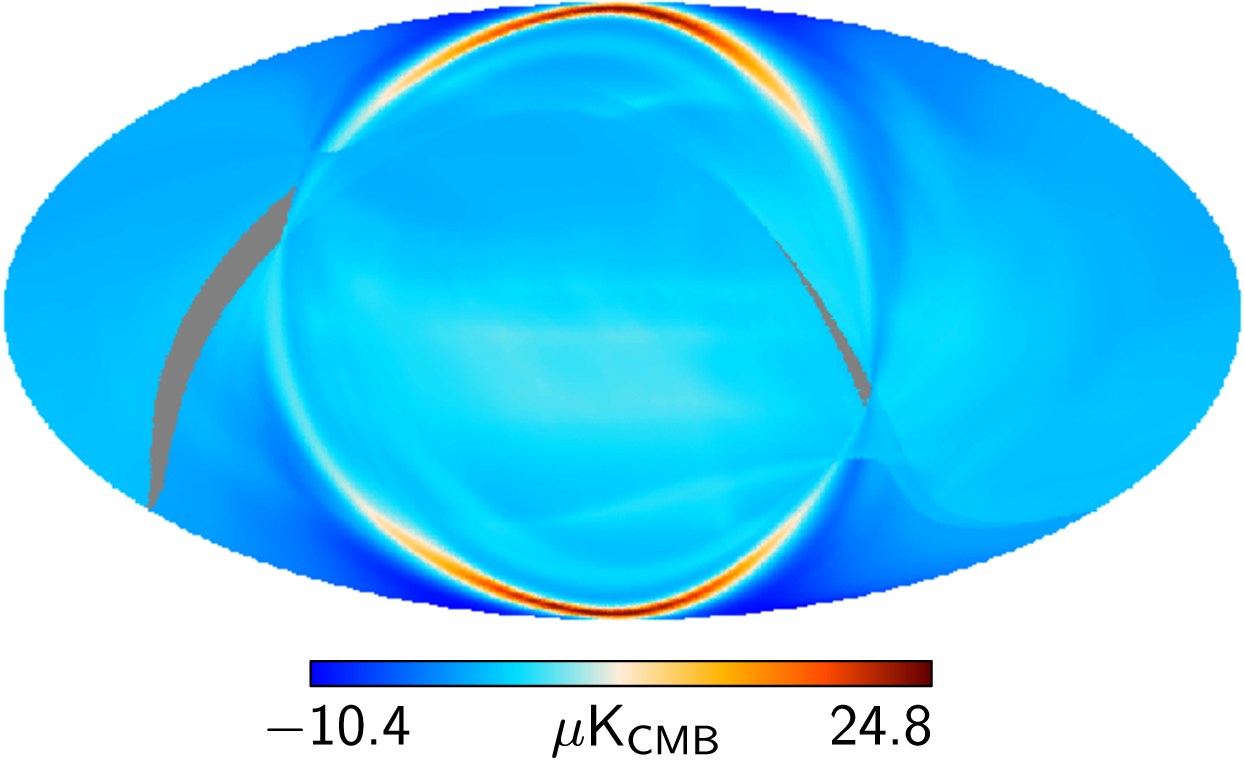} \\
	\includegraphics[width=4.25cm]{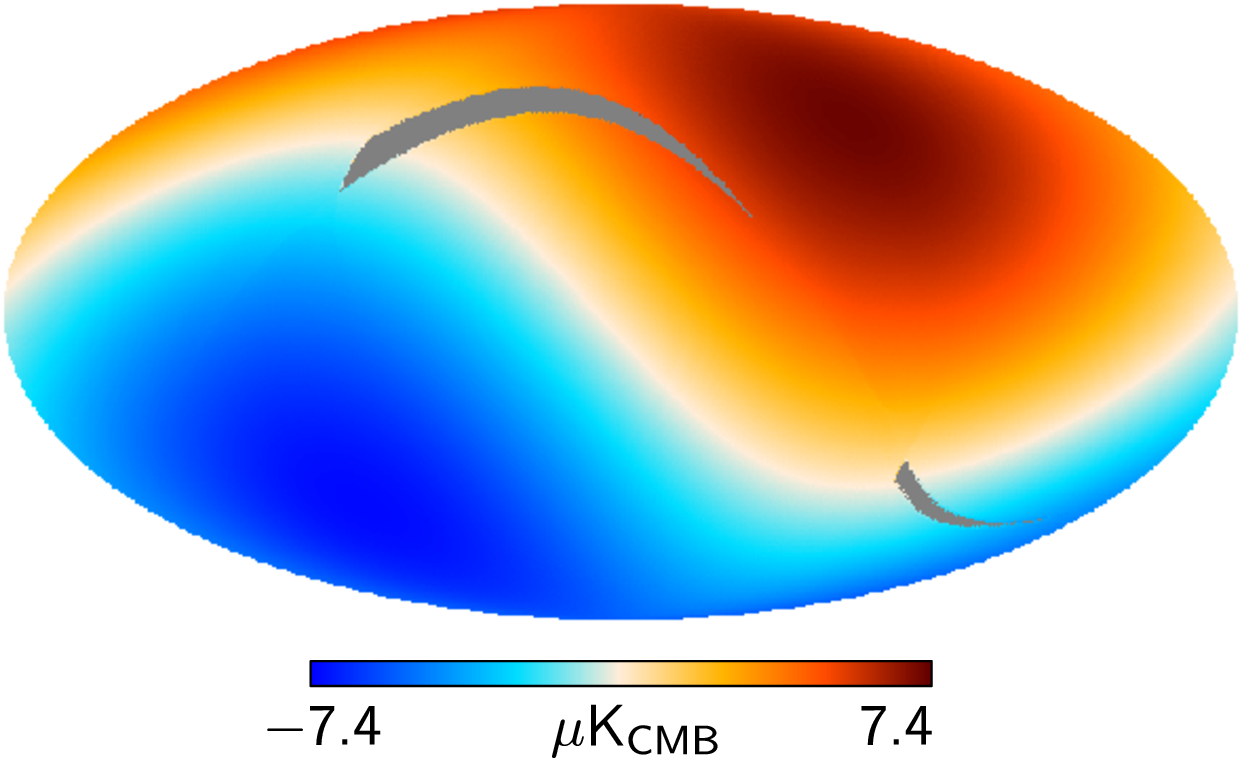} & 
	\includegraphics[width=4.25cm]{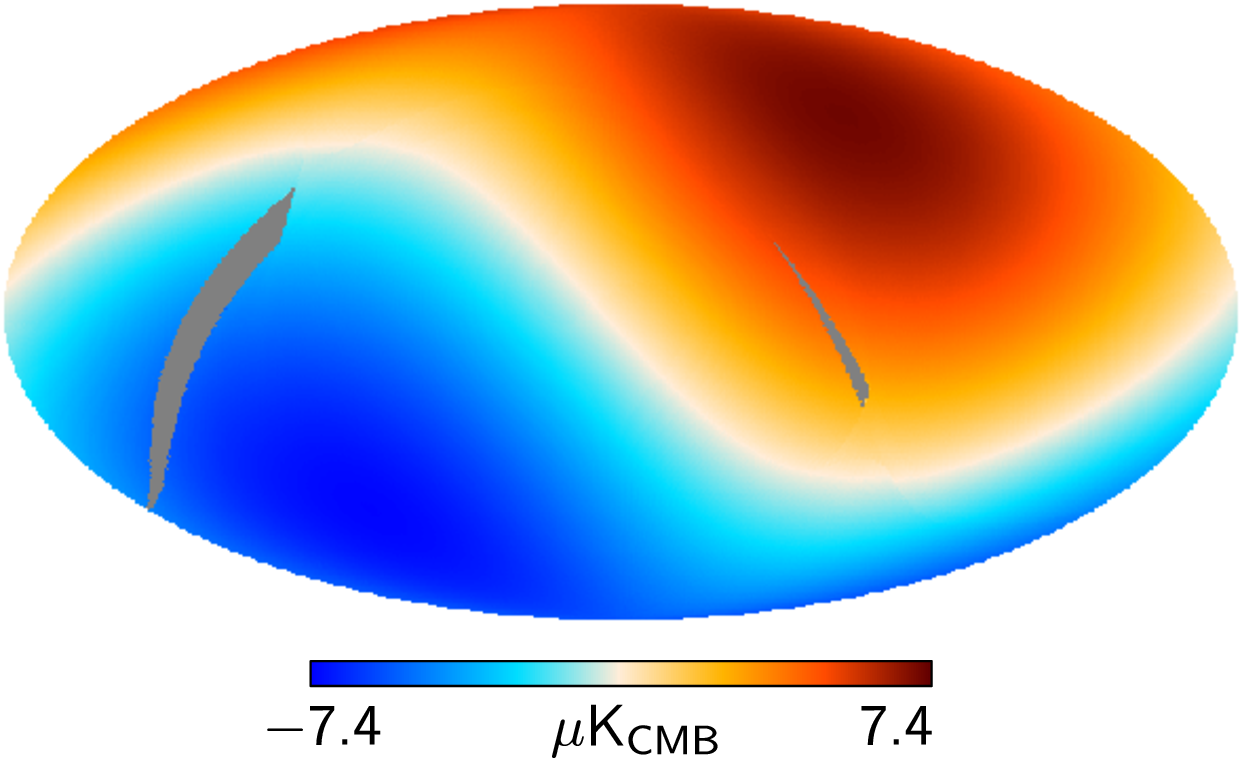} 
      \end{tabular}
      \caption{Sidelobe fingerprint in the 30~GHz channel due to Galactic foregrounds (top row) and cosmological dipole (bottom row) for surveys 1 (left) and 2 (right).}
      \label{fig_overview_straylight_30}
  \end{figure}

\begin{figure}[h!]
    \begin{tabular}{cc}
      \centering
	\includegraphics[width=4.25cm]{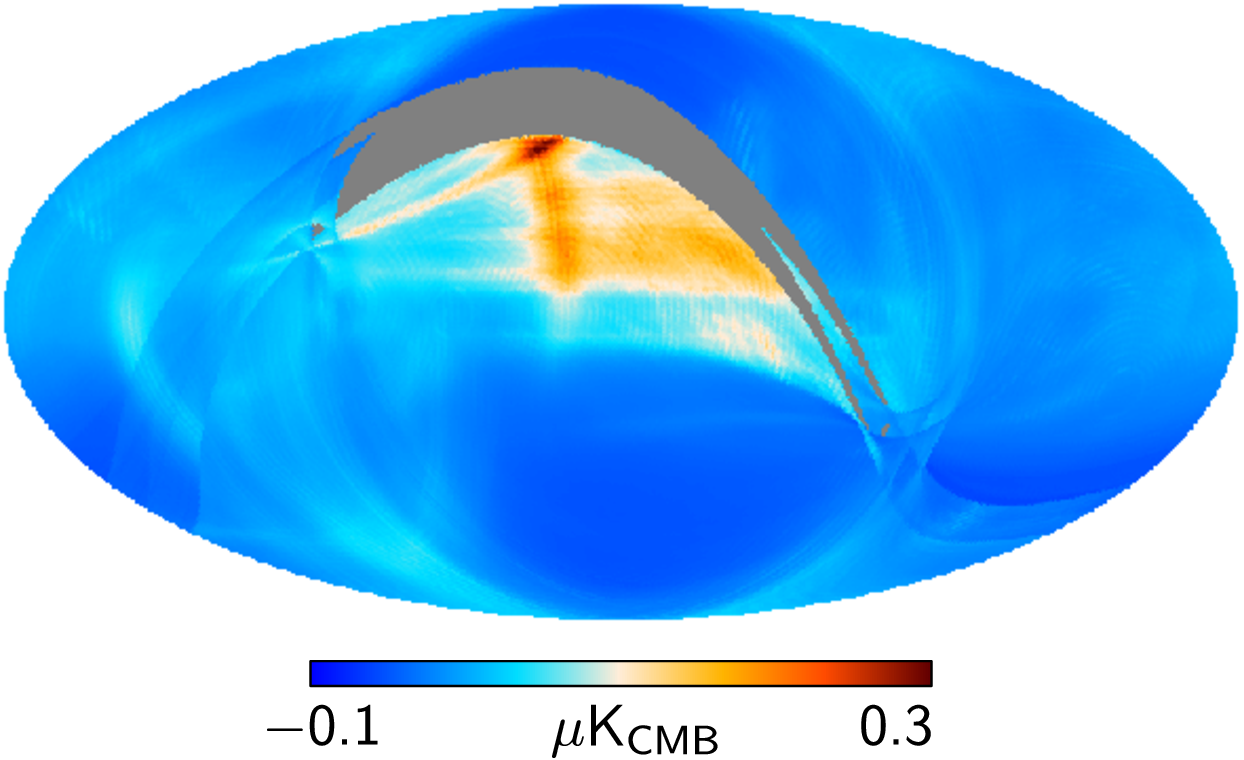} & 
	\includegraphics[width=4.25cm]{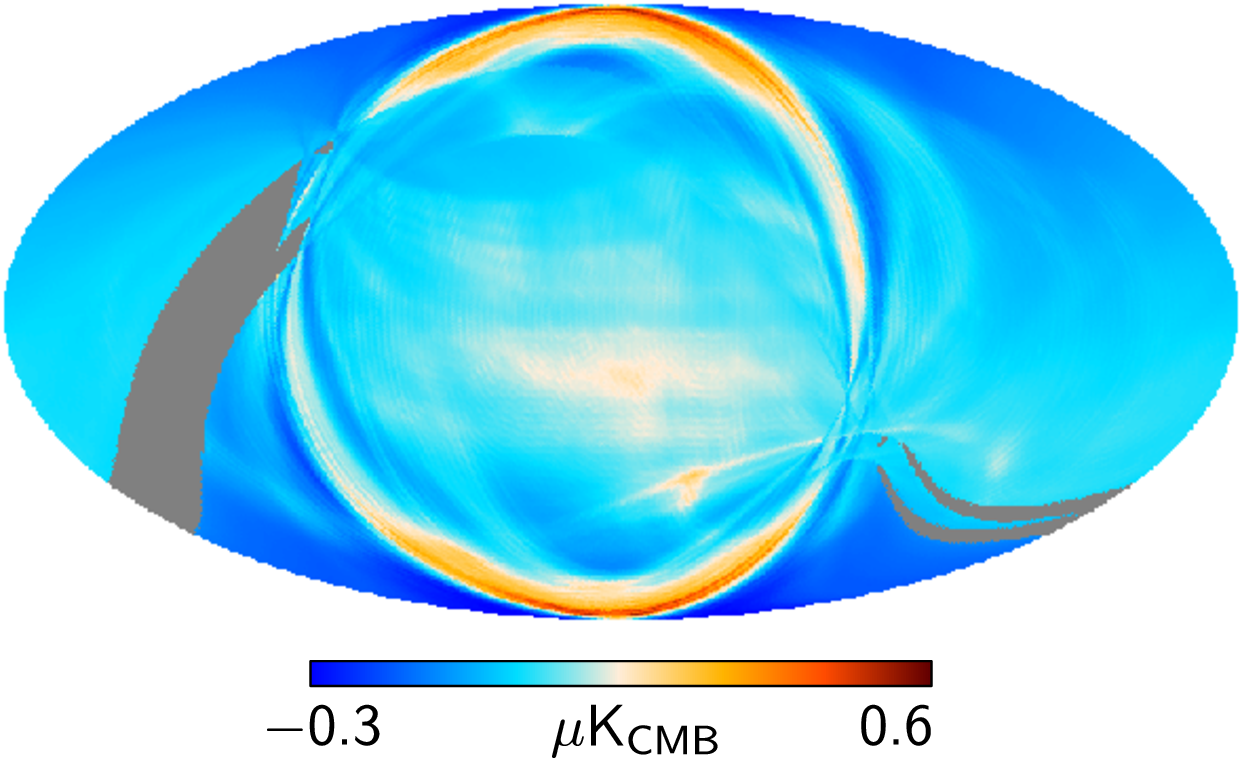} \\
	\includegraphics[width=4.25cm]{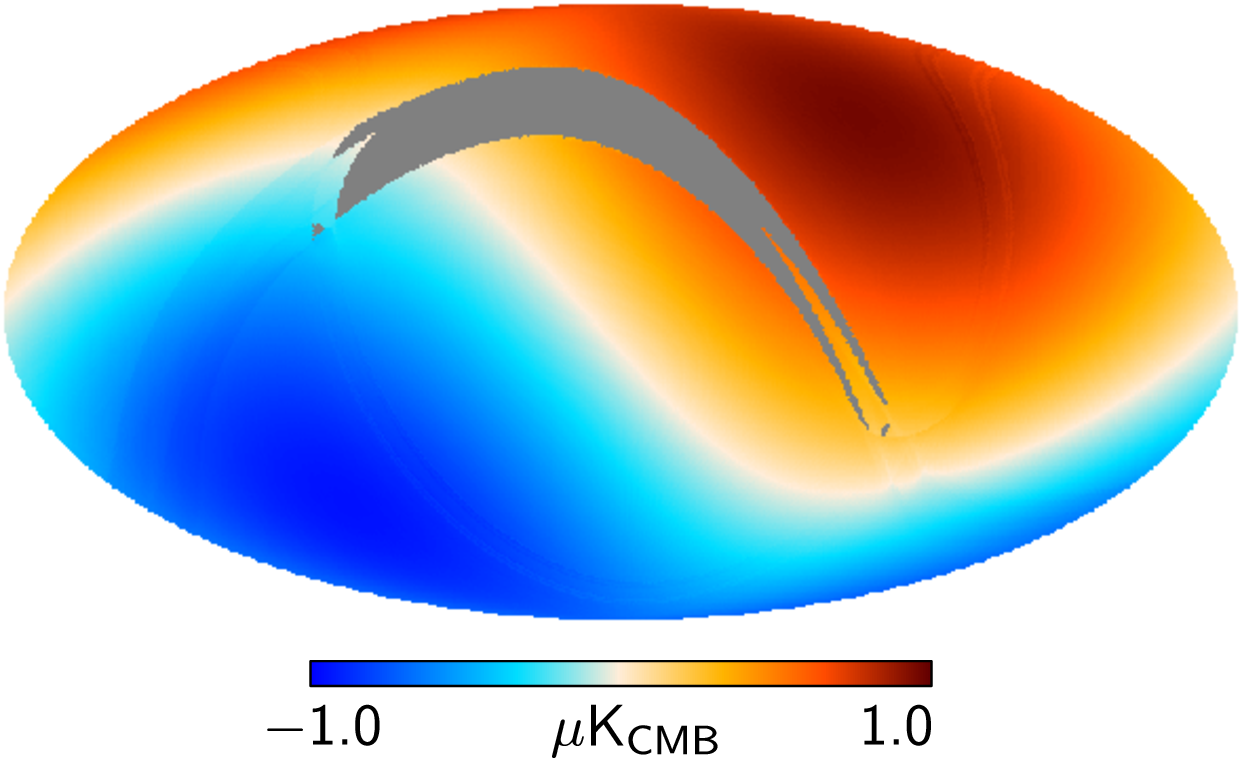} & 
	\includegraphics[width=4.25cm]{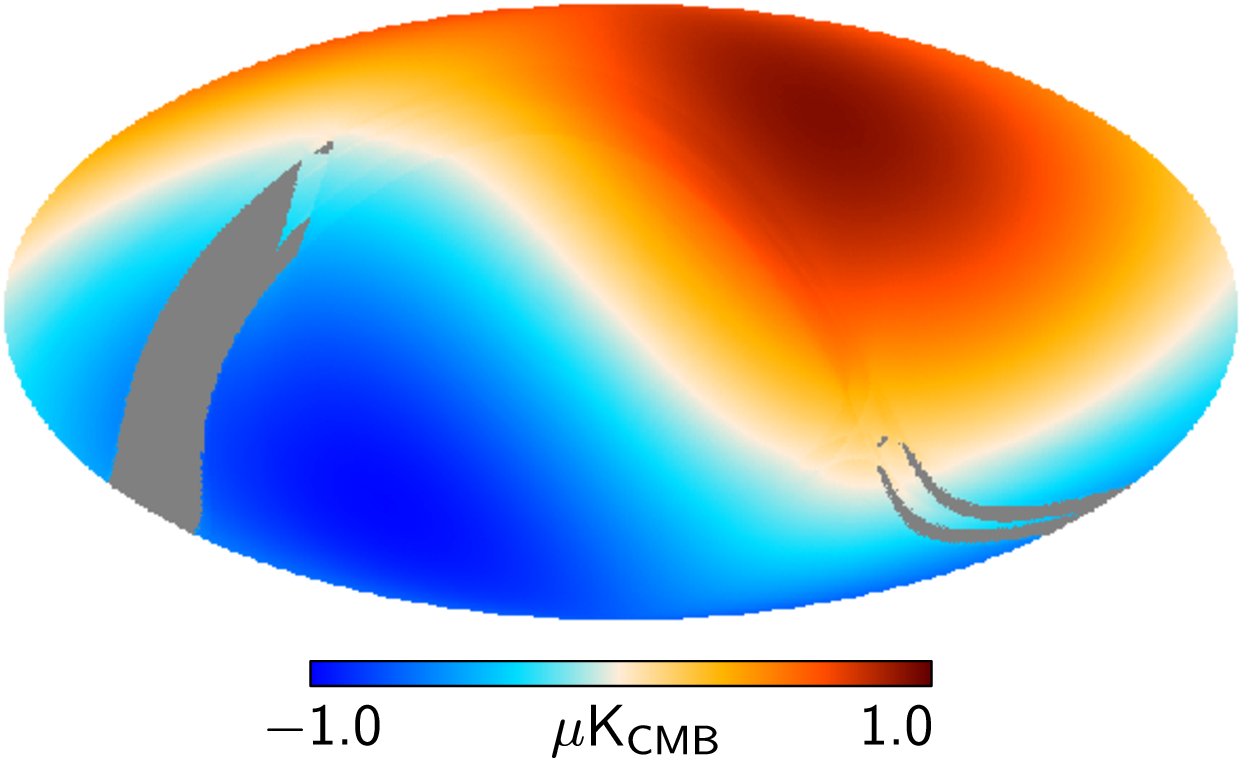} 
      \end{tabular}
      \caption{Sidelobe fingerprint in the 44~GHz channel due to Galactic foregrounds (top row) and cosmological dipole (bottom row) for surveys 1 (left) and 2 (right).}
      \label{fig_overview_straylight_44}
  \end{figure}

\begin{figure}[h!]
    \begin{tabular}{cc}
      \centering
	\includegraphics[width=4.25cm]{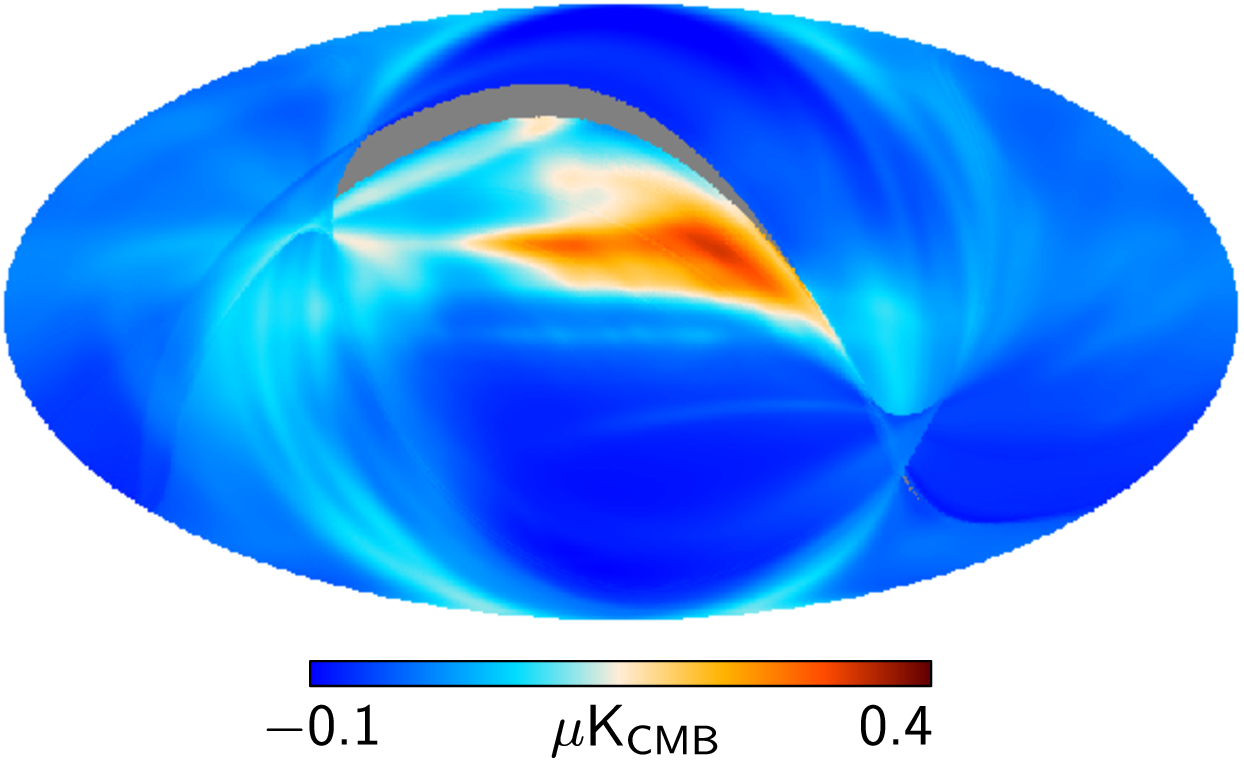} & 
	\includegraphics[width=4.25cm]{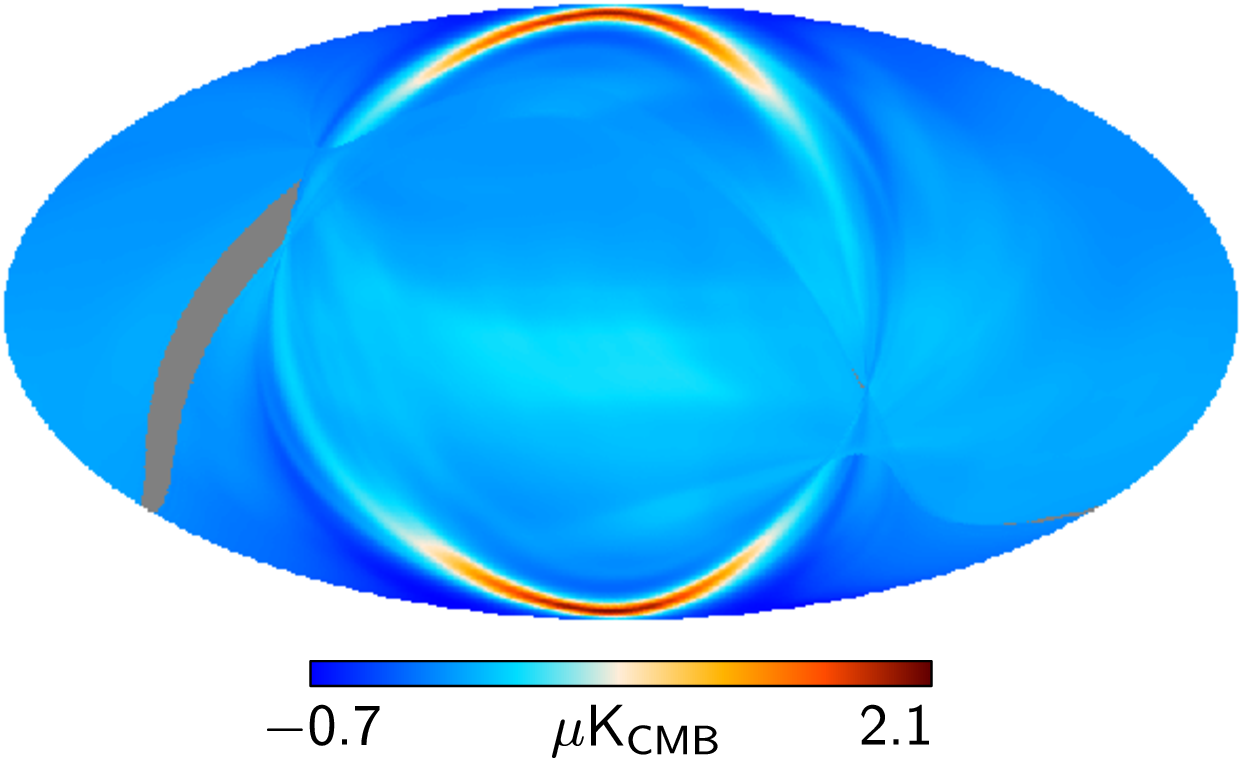} \\
	\includegraphics[width=4.25cm]{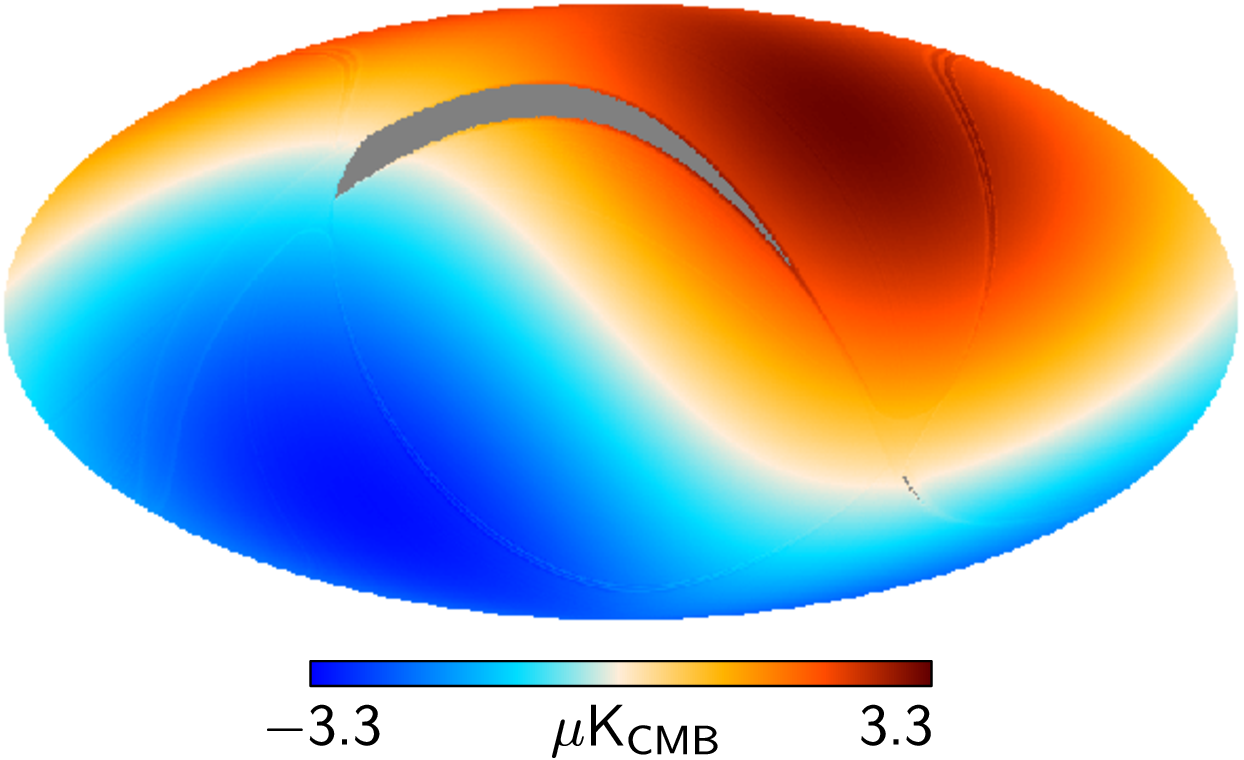} & 
	\includegraphics[width=4.25cm]{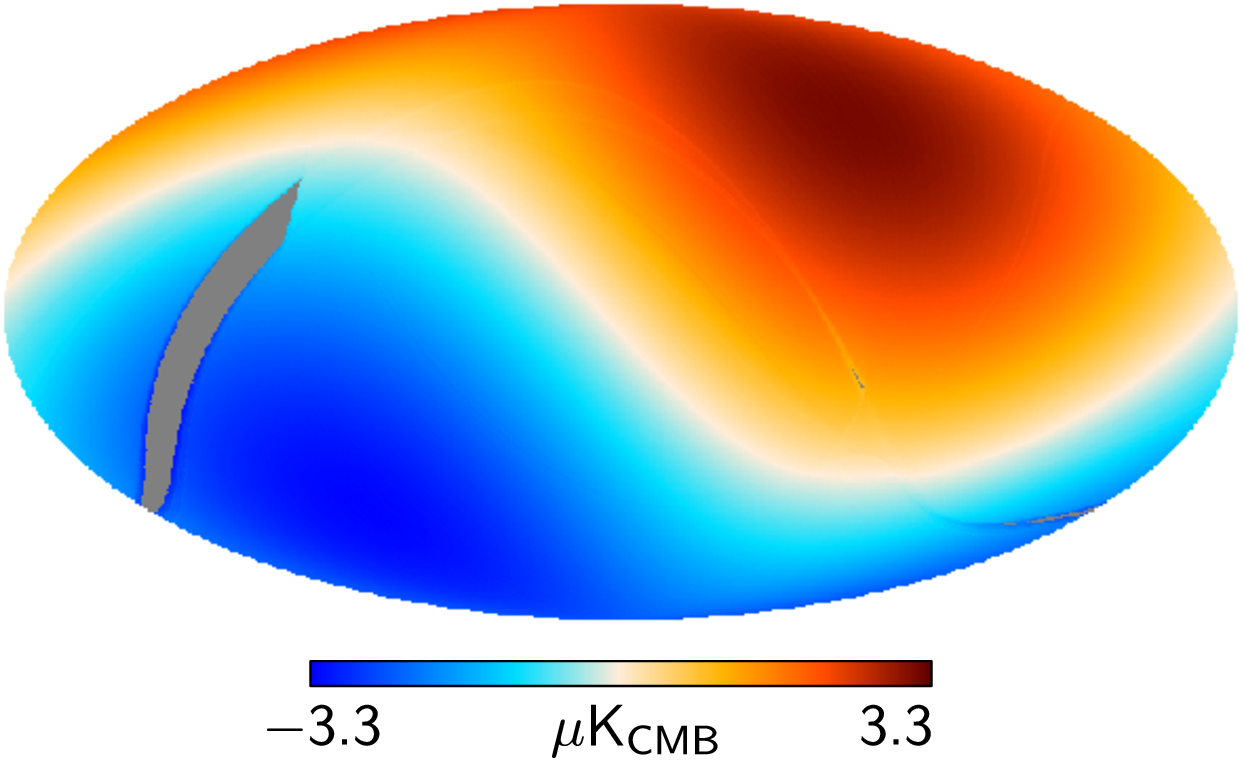} 
      \end{tabular}
      \caption{Sidelobe fingerprint in the 70~GHz channel due to Galactic foregrounds (top row) and cosmological dipole (bottom row) for surveys 1 (left) and 2 (right).}
      \label{fig_overview_straylight_70}
  \end{figure}

  In Figs.~\ref{fig_overview_straylight_30} through \ref{fig_overview_straylight_70} we show the simulated sidelobe fingerprint on the sky after the destriping process, for the odd (left side) and even (right side) surveys, respectively. These figures show that the straylight from the cosmological dipole is similar in the two surveys, while the Galaxy straylight is, as expected, larger in the second. The ring-shaped fingerprint in the second survey is also observed at the expected level in the real data by taking the difference of even minus odd survey maps (see Figs.~\ref{fig_odd-even_survey_diff} and \ref{fig_odd-even_survey_diff_no_sidelobes} in Sect.~\ref{sec_assessment_null_tests}), thus confirming the accuracy of our simulations.
  
  These results show that the most sensitive channel to straylight is 30\,GHz, followed in order by 70\,GHz and 44\,GHz. This is consistent with the telescope optical performance at the various frequencies. The primary mirror is strongly under-illuminated by the 44\,GHz horns, resulting in a low straylight sensitivity at the expense of a larger main beam, especially for the \texttt{LFI25} and \texttt{LFI26} horns. The 30 and 70\,GHz horns are characterised by similar illumination properties, so that their straylight susceptibilities are comparable, with a slightly better performance of the 70\,GHz horns with respect to the 30\,GHz ones. If we also take into account the larger sensitivity of the 30\,GHz channel to the Galactic signal it is apparent that this channel is, overall, the most susceptible to straylight contamination.
  
\begin{figure}[h!]
    \includegraphics[width=8.8cm]{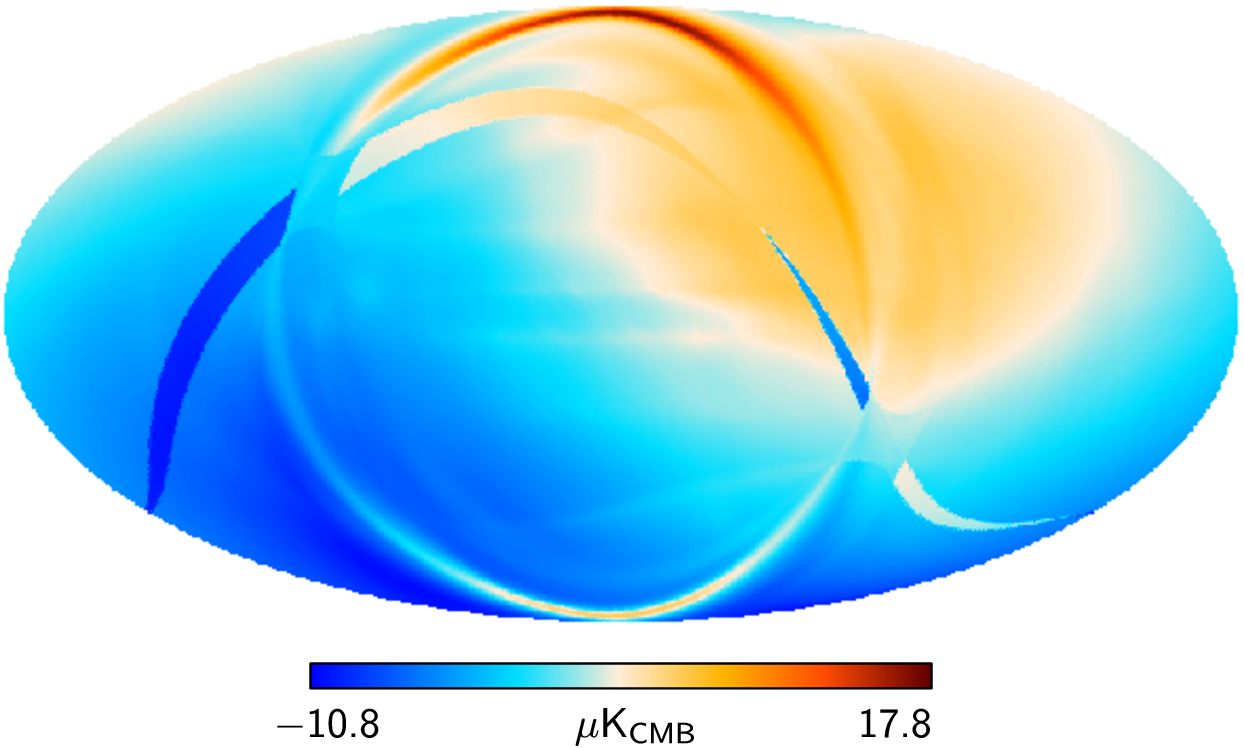} \\ 
    \\
    \includegraphics[width=8.8cm]{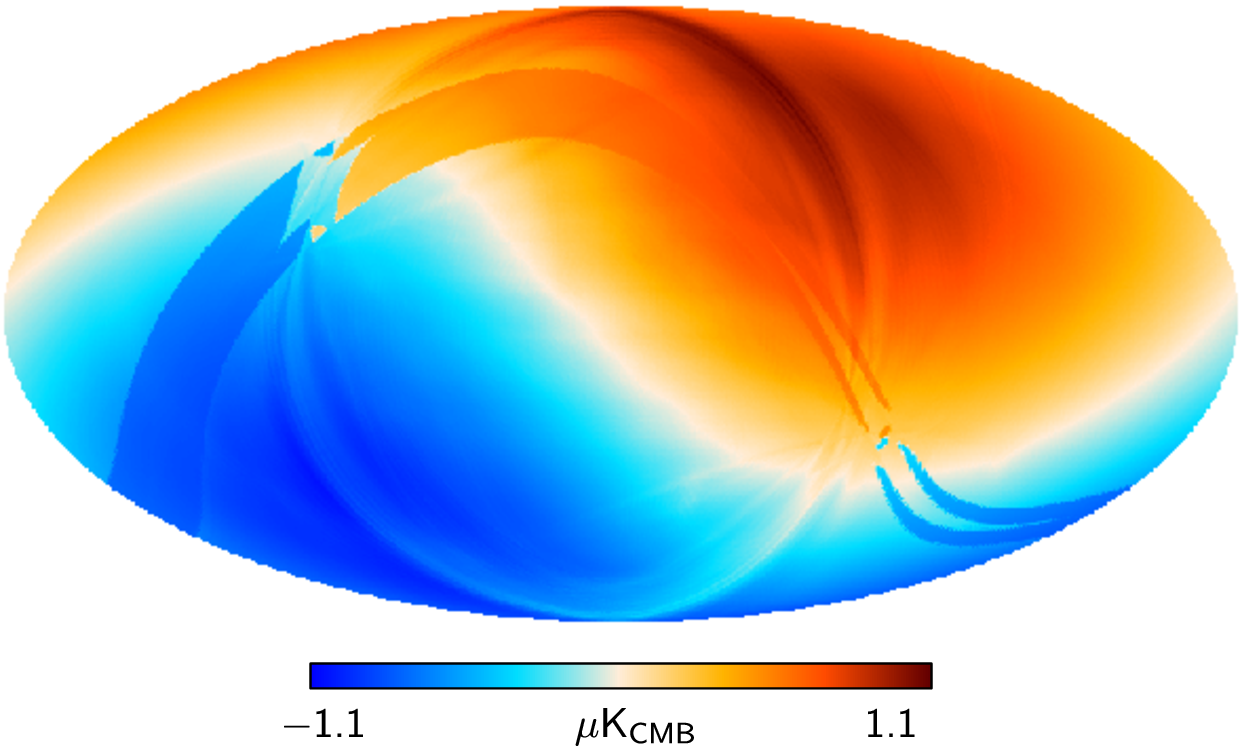} \\
    \\
    \includegraphics[width=8.8cm]{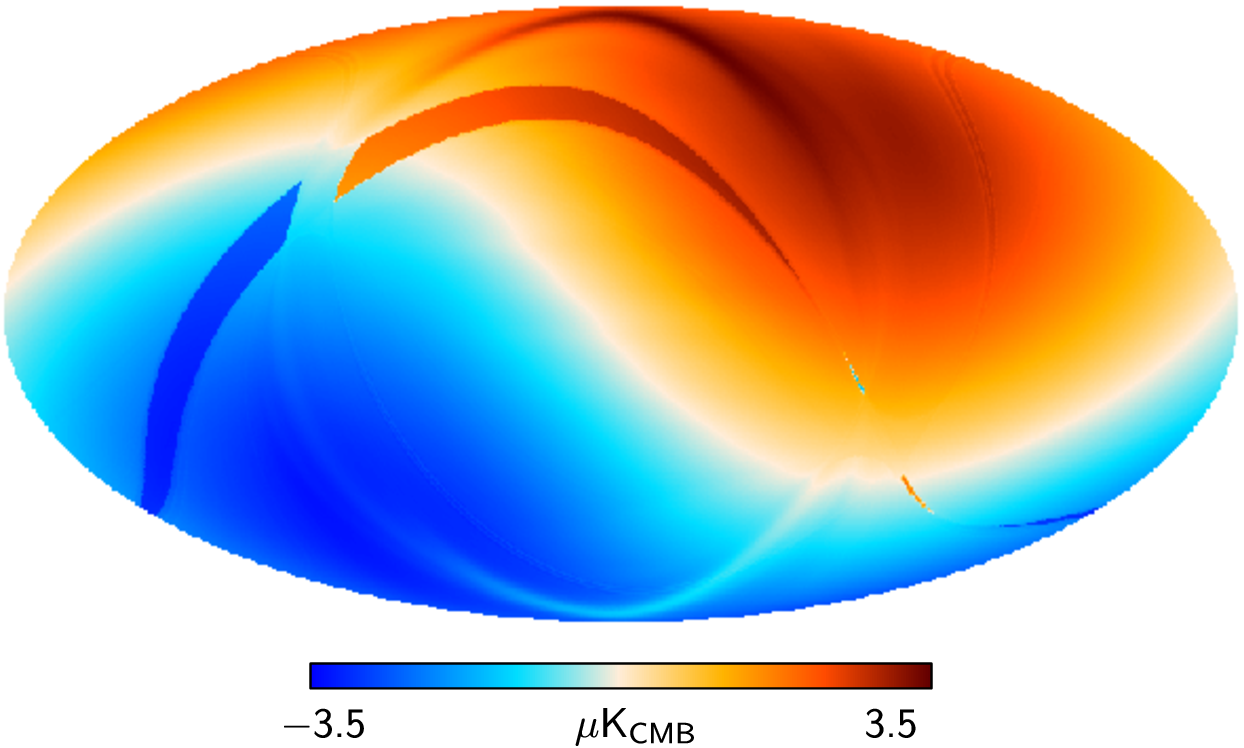}  
    \caption{Straylight contamination maps at 30\,GHz (top), 44\,GHz (middle) and 70\,GHz (bottom).}
    \label{fig_global_straylight_maps}
  \end{figure}

  Finally, to quantify the straylight effect on maps and power spectra, we have generated a global map per frequency, including the dipole and Galactic straylight signals for both surveys, as shown in Fig.~\ref{fig_global_straylight_maps}. 
  
  \subsubsection{ADC non-linearity}
  \label{sec_assessment_adc}

    \paragraph{Method.}
\label{sec_assessment_adc_method}

  The levels of the residuals due to ADC correction process are estimated by applying the correction algorithm to simulated data containing a known ADC effect and making difference maps with those produced from data with no ADC effect. {The starting point of this analysis are time-ordered data for all individual detectors based on (ring-based) sky and reference load simulations.} The noise component is simulated using average 1/$f$ noise parameters over the nominal mission for each detector. Galactic and CMB signals are based on the observed maps, converted into time-ordered data using real pointings and successively uncalibrated using the inverse gain table. The same is done with a map of the \textit{WMAP} dipole, while the orbital dipole is calculated from the pointing information and JPL ephemeris for the satellite velocity. Finally spline fits to the observed sky and reference voltage levels per pointing period are used together with estimated receiver temperatures, $T_{\rm CMB}=2.725$\,K and $T_{\rm ref}=4.5$\,K, as a model for the gain evolution. 

  The simulated ADC effect is induced by applying the inverse of the spline correction used in the real data. The same algorithm as used with the real data is then applied to the simulated data iteratively five times to ensure convergence. Intensity maps are constructed by simple binning into an $N_{\rm side}=1024$ map at each iteration, both for the simulation with and without the ADC effect. Some of these maps show a residual dipole caused by small changes in the overall slope of the temperature-voltage response curve due to the ADC correction. Since the calibration pipeline determines this response, and it does not give rise to a residual dipole, we correspondingly remove it here via a correlation fit with the input dipole map. The ADC effect maps are finally taken as the difference between the fifth iteration map and the no-ADC map. Maps for each frequency band are produced by averaging all maps for that frequency, taking into account the detector weighting. 

  Since some 70~GHz channels can not be corrected due to ``popcorn'' noise, a separate method was used to estimate the likely level of ADC error {for these channels}, using the white noise level on the difference data. This is immune to the ``pop-corn'' noise, but cannot be used to correct the ADC effect, since it is not known whether the effect is due to the sky or reference voltages. In these cases we only estimate the ADC effect and do not apply any correction.

\paragraph{Results.}
\label{sec_assessment_adc_results}
  
\begin{figure}[h!]
    \includegraphics[width=8.8cm]{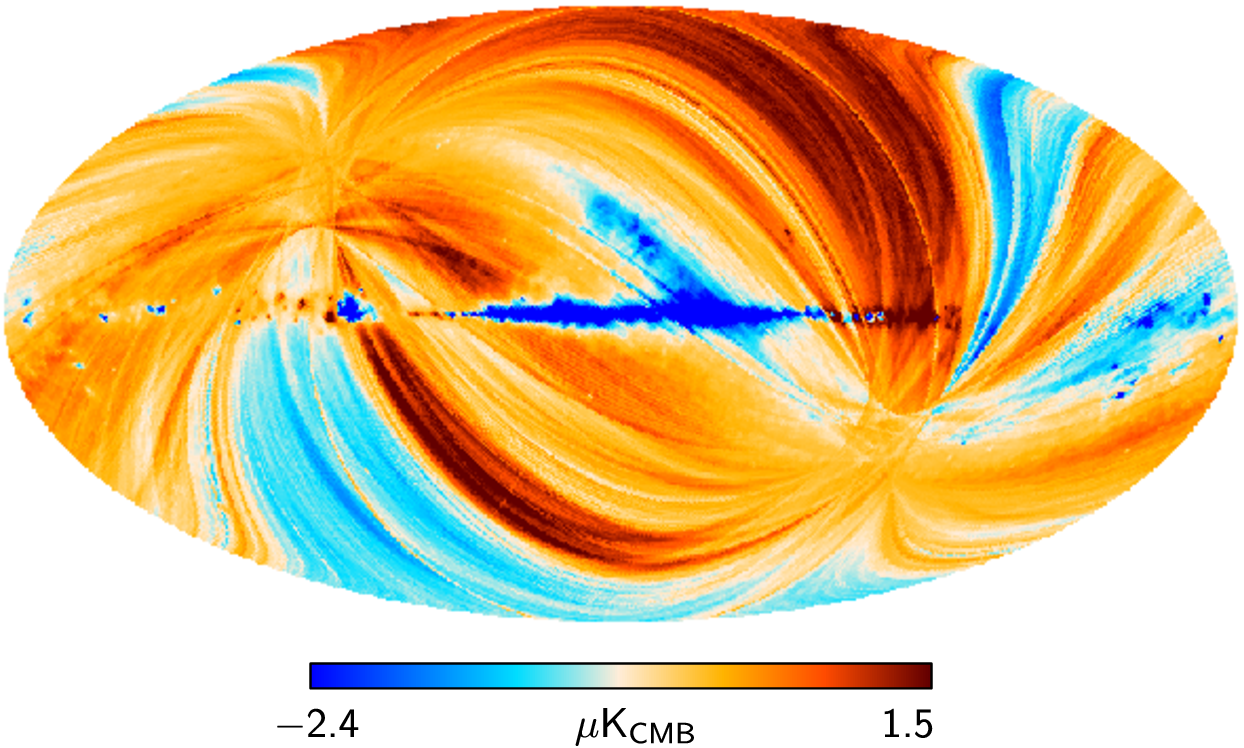} \\ 
    \\
    \includegraphics[width=8.8cm]{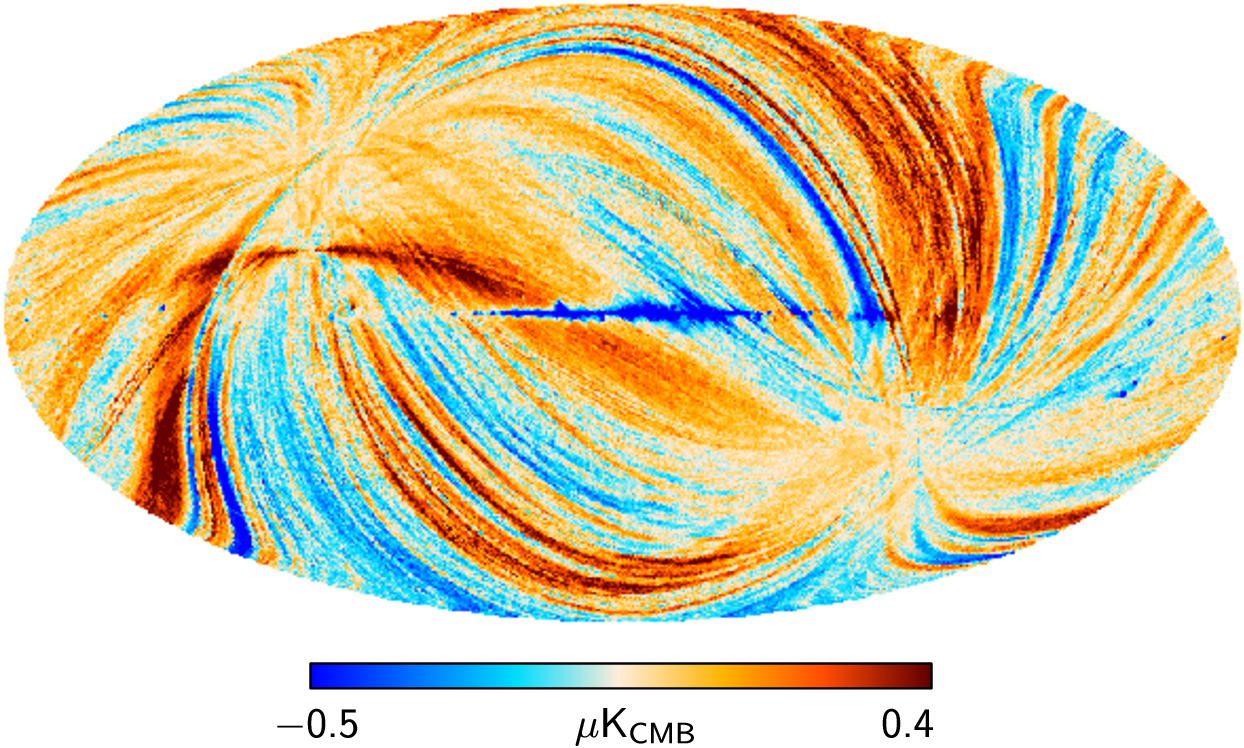} \\
    \\
    \includegraphics[width=8.8cm]{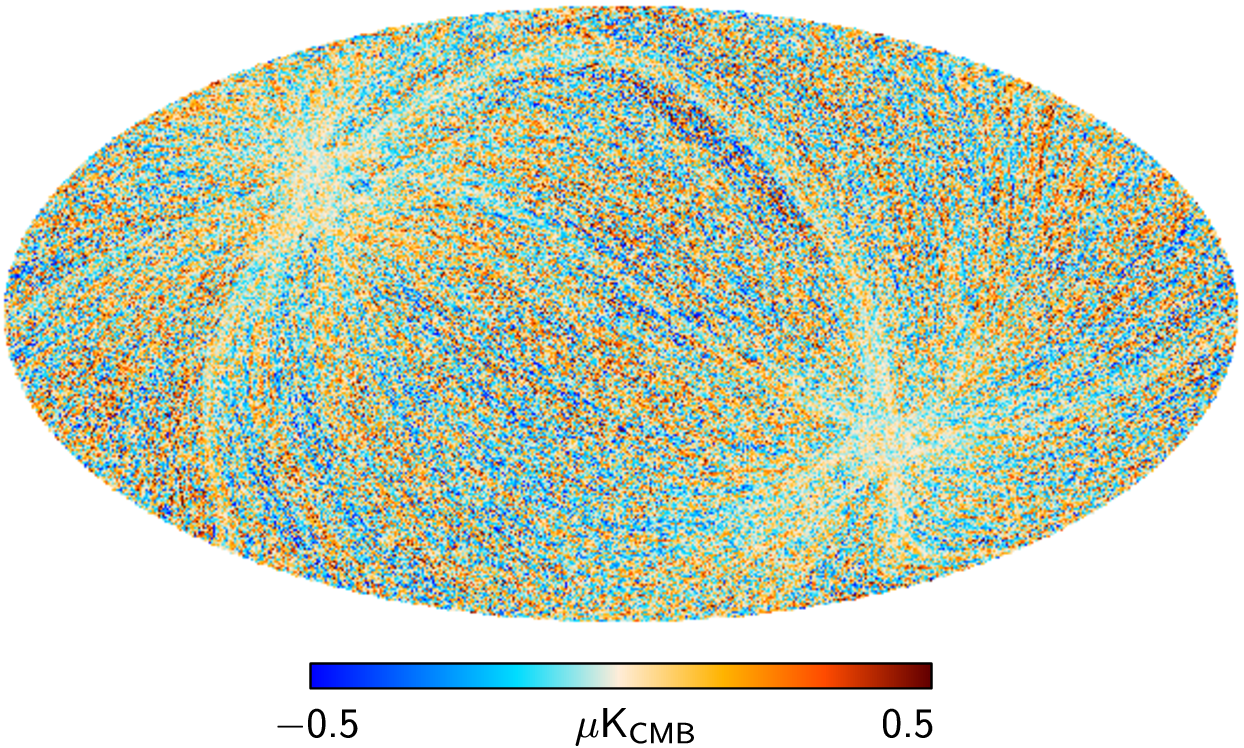}  
    \caption{Maps of the ADC non-linearity effect at 30\,GHz (top), 44\,GHz (middle), and 70\,GHz (bottom).}
    \label{fig_adc_maps}
  \end{figure}

  Maps of the ADC effect at the three LFI frequency channels are shown in Fig.~\ref{fig_adc_maps}. The main effect of the ADC residuals is a small ($<0.1\%$ of the dipole signal) ring-based gain error which appears in the maps as stripes in the scan direction. The residuals are generally larger where the sky signal is stronger, i.e., following the CMB dipole and Galactic plane. The contribution from the Galactic plane becomes weaker at higher frequencies as expected. Broad stripes in the 30\,GHz map are due to residual deviations from linearity on voltages ranges larger than the ADC peaks. These also occur at the other frequencies, but as the number of channels increases this effect averages out, leaving more uniform noise-limited, low-level residuals at 70\,GHz.  While the 44\,GHz channels have the strongest ADC effect due to lower detector voltages, they are also the best characterized, leading to a well-determined correction placing it between 30 and 70\,GHz in terms of the amplitude of residuals.
  
  \subsubsection{Imperfect photometric calibration}
  \label{sec_assessment_calibration}

    \paragraph{Method.}
\label{sec_assessment_calibration_method}

We have developed an analytical model of the impact of the uncertainty in the dipole calibration algorithm due to the radiometer white noise and the loss of integration time due to Galactic masking. We have run this model to estimate how this effect propagates through the calibration and mapmaking pipeline. Such simulations scan a sky map (the \emph{input map}) of pure astrophysical signal (without dipole) to produce a time-ordered data stream, which is then uncalibrated using gains inferred from the total-power output of the radiometers. These time-ordered data are then used as input in a simplified version of the LFI pipeline to produce a new calibrated map (the \emph{output map}). The difference between the input and output maps should be mainly due to dipole leakage, since the gains used in the decalibration phase differ from those calculated by the pipeline. Refer to Sect.~5 in \citep{planck2013-p02b} for more information.

\paragraph{Results.}
\label{sec_assessment_calibration_results}

\begin{figure}[h!]
  \includegraphics[width=8.8cm]{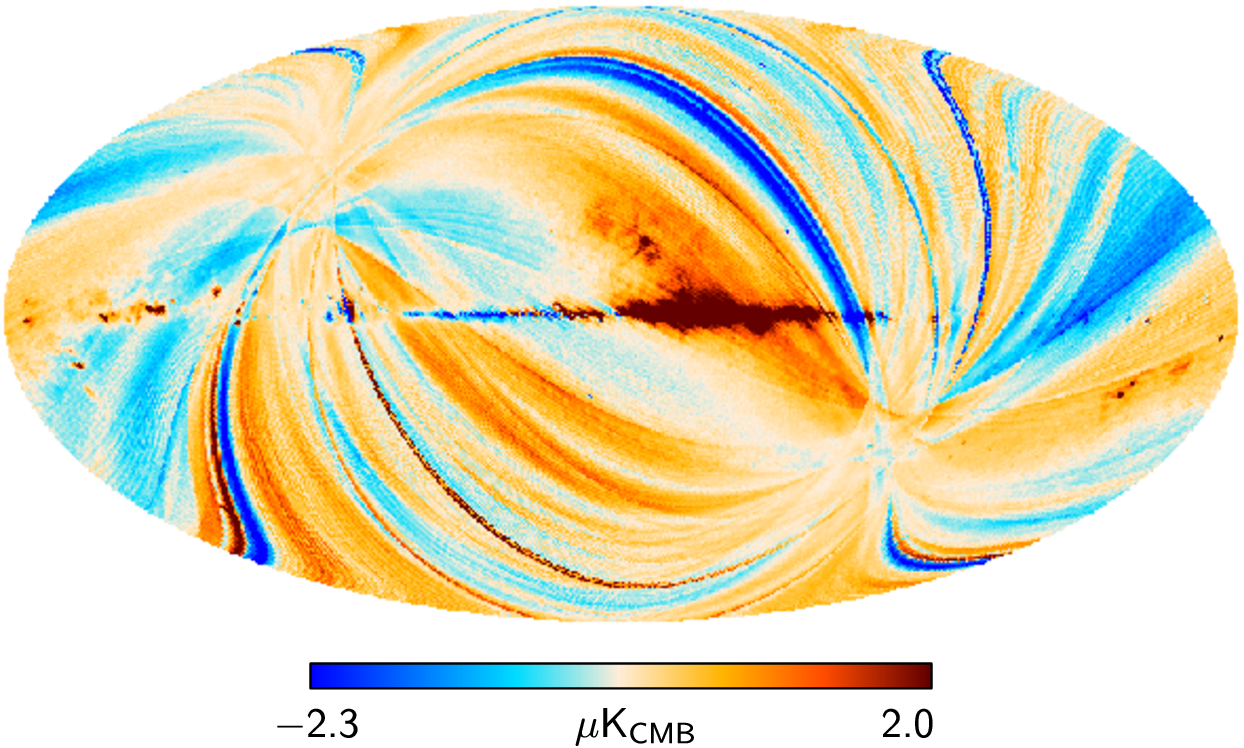} \\
  \\
  \includegraphics[width=8.8cm]{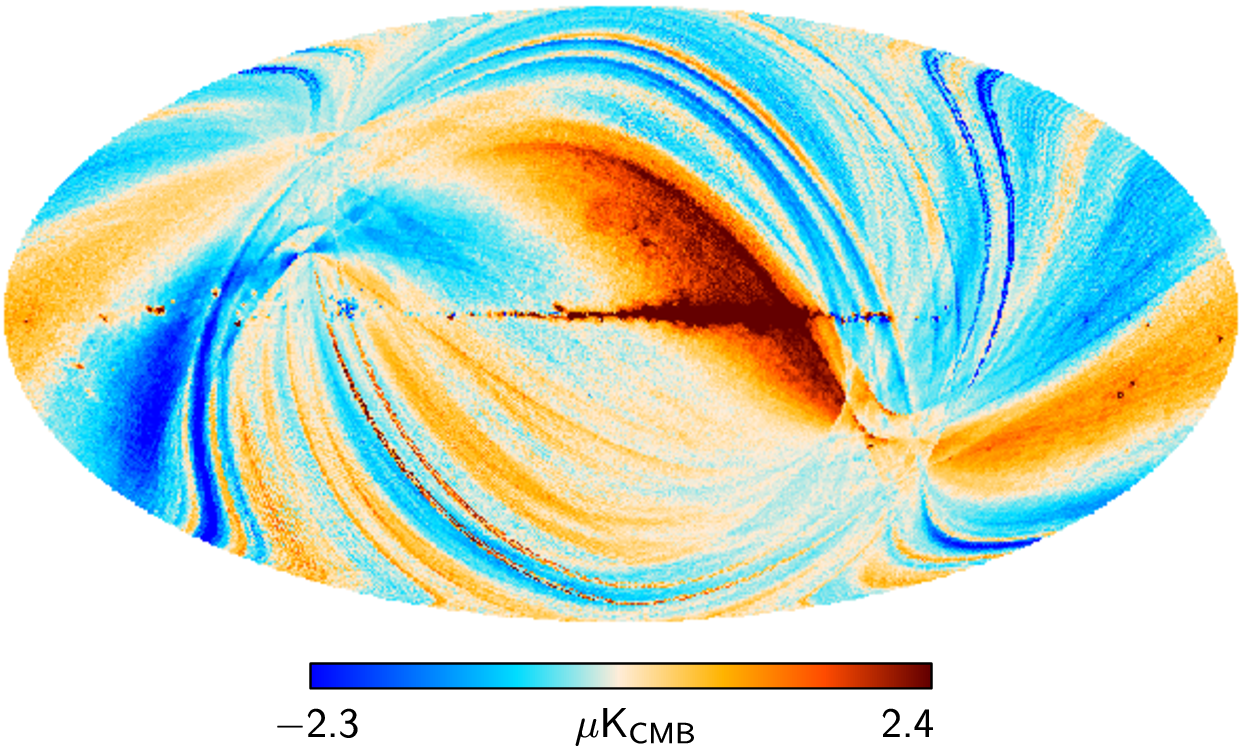} \\
  \\
  \includegraphics[width=8.8cm]{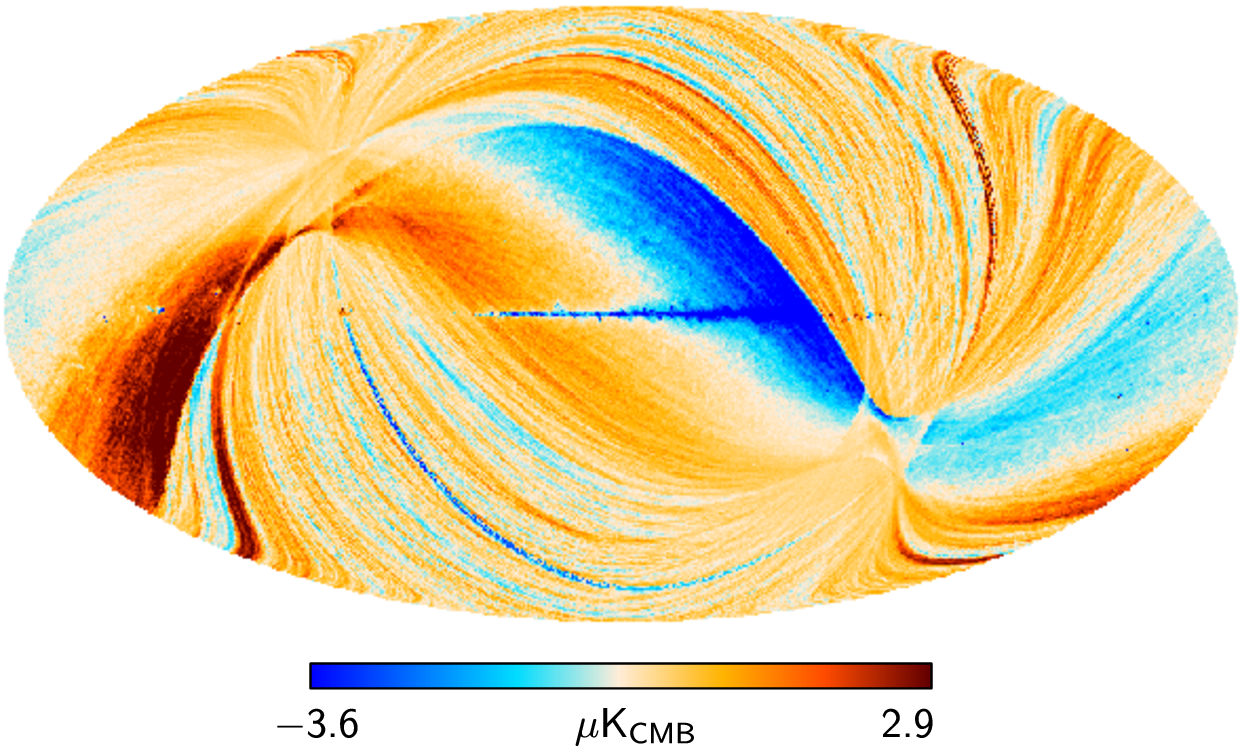}
  \caption{Maps of the effect of calibration uncertainties at 30\,GHz (top), 44\,GHz (middle), and 70\,GHz (bottom). 
  }
  \label{fig_gain_maps}
\end{figure}

Fig.~\ref{fig_gain_maps} shows the difference between the input and output maps. The shape of the features in these maps closely follows the scanning circles drawn by the pointing direction of the telescope towards the sky. (This is expected, since the calibration is performed on the time-ordered data.) The estimated impact of such systematic effects on the \Planck-LFI maps is of the order of a few $\mu$K per pixel.

\subsection{Pointing uncertainties}
\label{sec_assessment_pointing}

  \paragraph{Method.} To estimate the uncertainty introduced by the Gaussian fit in main beam measurements we perform tests using the radio-frequency model of the flight telescope \citep{planck2013-p02d} and compare the centre calculated by the fit with the beam maximum, which is uniquely determined in optical simulations. Typical differences between the centres are $1\arcs$ for all the 70\,GHz beams, $4\arcs$ for \texttt{LFI24}, $18\arcs$ for \texttt{LFI25} and \texttt{LFI26} (44\,GHz horns), and $6\arcs$ for the 30\,GHz beams. These estimates are all smaller than the statistical uncertainty in the determination of the beam centre, which ranges from $4\arcs$ at 70\,GHz to $10\arcs$ at 30\,GHz.

The focal plane geometry was reconstructed using four Jupiter transits labelled as J1, J2, J3 and J4 \citep{planck2013-p02,planck2013-p02d}. When we compare the focal plane geometry obtained from the combination of J1 and J2 with the one obtained from the combination of J3 and J4 we find a difference of about $15\arcs$ in pointing, mainly along the in-scan direction. On the other hand, the comparison of the focal plane geometries determined from single Jupiter transits (J1 against J2 and J3 against J4) shows differences within the expected uncertainty. The $15\arcs$ discrepancy, likely to be correlated with changes in the thermal control set-point of the data processing unit in the instrument digital electronics, is compensated using two different instrument databases in the data analysis pipeline, one for the period ranging from day 91 to day 539 after launch, and the other for the period between day 540 and day 563. Details of the focal plane reconstruction and related uncertainties can be found in \citet{planck2013-p02}.

We assess the impact of this effect using dedicated simulations constructed according to the following procedure:
\begin{enumerate}
  \item \label{pointing_simulation_1}Generate time ordered data by observing a CMB-only sky with flight detector pointing derived by applying the two focal plane database solution.
  \item \label{pointing_simulation_2}Reconstruct the CMB map from the time-ordered data generated in step~\ref{pointing_simulation_1}, applying each of the two focal plane database solution in map reconstruction.
  \item Repeat step \ref{pointing_simulation_2} using the single focal plane database solution.
  \item Compute the difference of the power spectra obtained from the two generated maps.
\end{enumerate}

\begin{figure}[!h]
  \includegraphics[width=8.8cm]{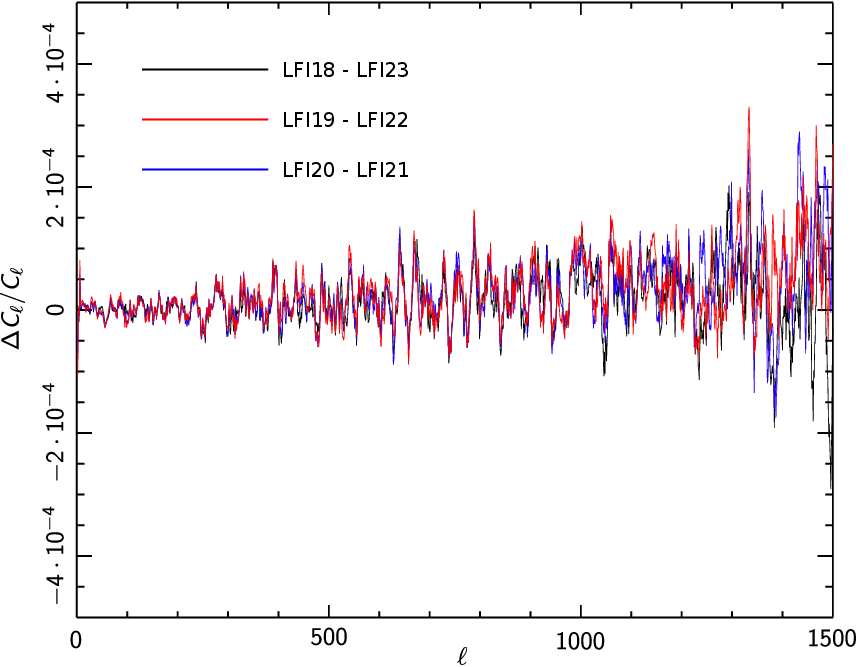}
  \caption{Relative difference between the spectra of the maps simulated with one and two instrument databases. The three curves represent power spectra relative to three different feed horn pairs in the 70\,GHz frequency channel. A running average smoothing kernel has been applied to reduce the scatter and enhance any larger-scale trends. The relative uncertainties are of the order $\Delta C_\ell / C_\ell \lesssim 10^{-4}$.}
  \label{fig_pointing_impact}
\end{figure}

\paragraph{Results.} Figure~\ref{fig_pointing_impact} shows that the relative difference of power spectra is of the order of $10^{-4}$, which is negligible. 

\subsection{Propagation of systematic uncertainties through component separation}
\label{sec_syseffect_compsep}

  A further step in our assessment has been to evaluate the impact of the various systematic effects on the CMB map independently from the frequency. In order to do this we have {computed a weighed sum of} the three maps for each effect using weights obtained derived with a pixel-based ILC (internal linear combination) component separation method \citep{leach2008,planck2013-p06}. {The ILC method implements direct variance minimisation exploiting the fact that the CMB component (in thermodynamic temperature units) is constant across frequencies, while foregrounds are characterized by non-thermal spectra.} The CMB temperature can then be {estimated} at each pixel, $p$, {in terms} of a simple weighted sum of the frequency maps,
\begin{equation}
  T_{\rm CMB}(p) = \sum_{i=1}^N w_i\, T_{\nu_i}(p)\quad\textrm{where}\quad \sum_{i=1}^N w_i = 1.
  \label{eq_tcmb}
\end{equation}
{The ILC coefficients are estimated including Planck frequencies between 30 and 353\~GHz. However, only the three LFI channels are included in the total systematic error map, as we are only interested in residual LFI systematic effects in the CMB products. To propagate systematic effects through component separation, we therefore replace the frequency maps in Eq.~\ref{eq_tcmb} with the corresponding systematic effect maps,}
\begin{equation}
  T_{\rm syst}(p) = \sum_{i=1}^3 w_i\, T_{\mathrm{syst},\,\nu_i}(p).
  \label{eq_tsyst}
\end{equation}
Note that for simplicity, the ILC weights are uniformly distributed in pixel and harmonic domains, whereas in the \Planck\ component separation pipeline, the variance minimisation is conducted in the needlet space, i.e., on sub-sets of the harmonic and pixel domains where foregrounds are relevant at various levels, resulting in a set of coefficients for each needlet domain \citep{planck2013-p06}.

\subsection{Gaussianity statistical tests }
\label{sec_gaussianity}

\begin{figure*}
  \begin{center}
  \includegraphics[width=5.86cm,angle=0] {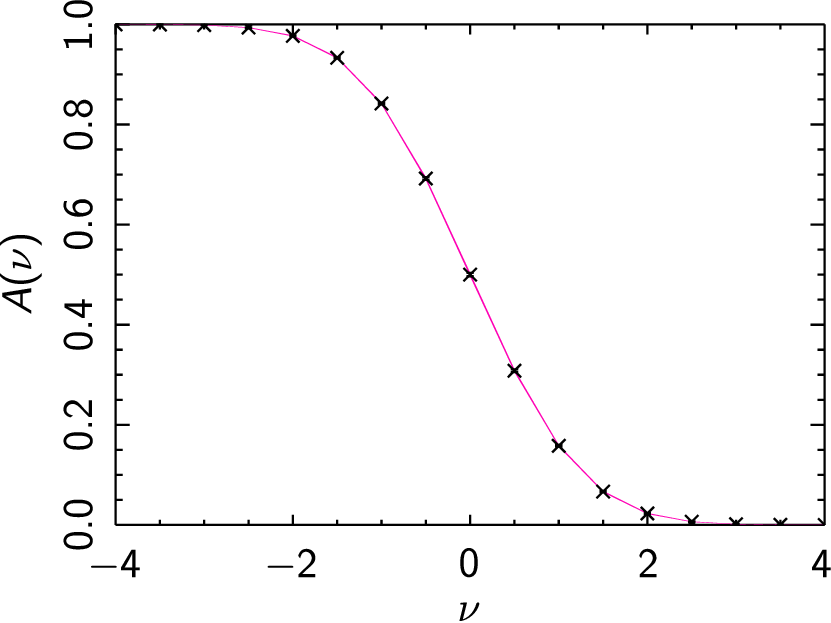}
  \includegraphics[width=5.86cm,angle=0] {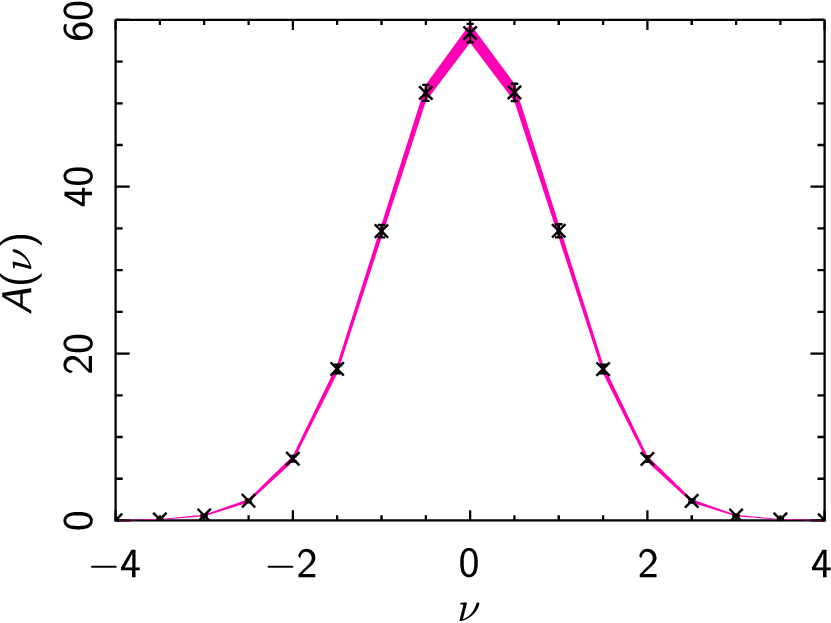}
  \includegraphics[width=5.86cm,angle=0] {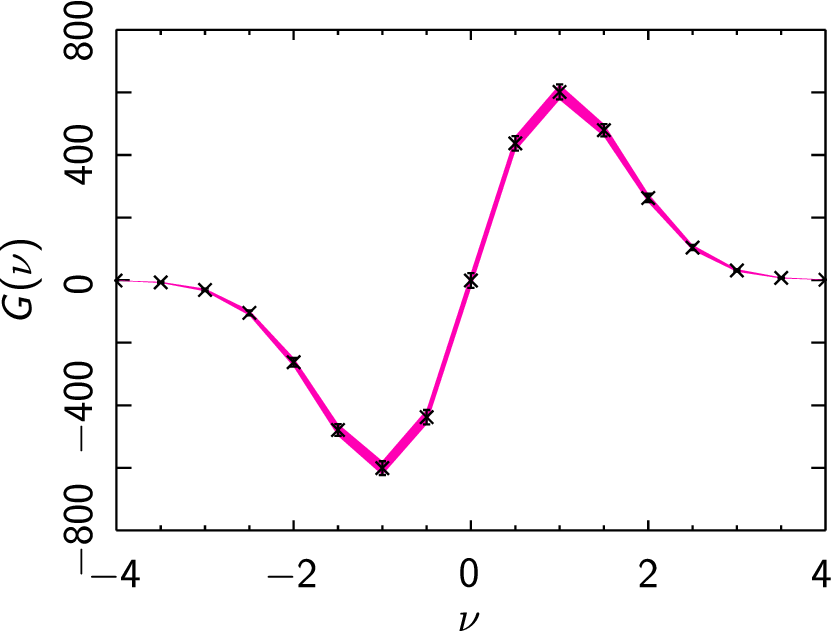}
  \includegraphics[width=5.86cm,angle=0] {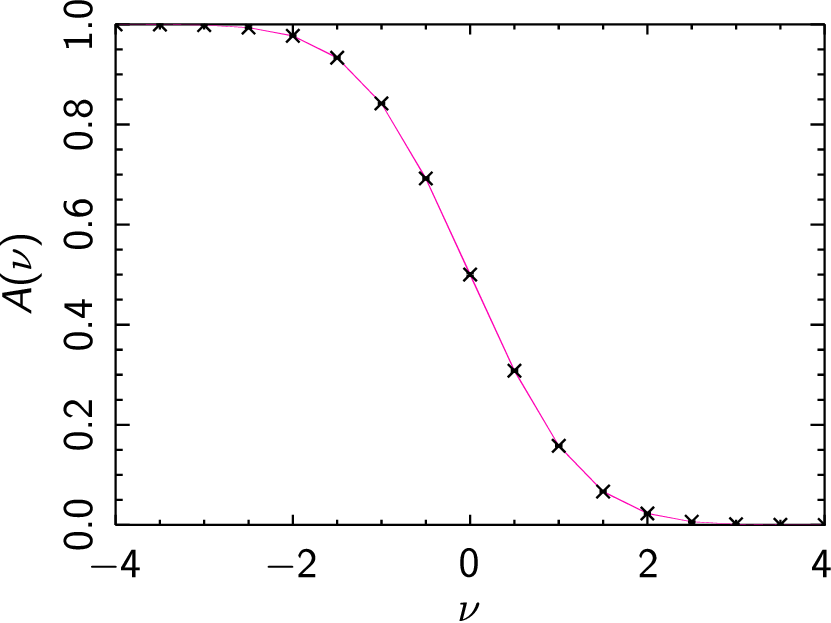}
  \includegraphics[width=5.86cm,angle=0] {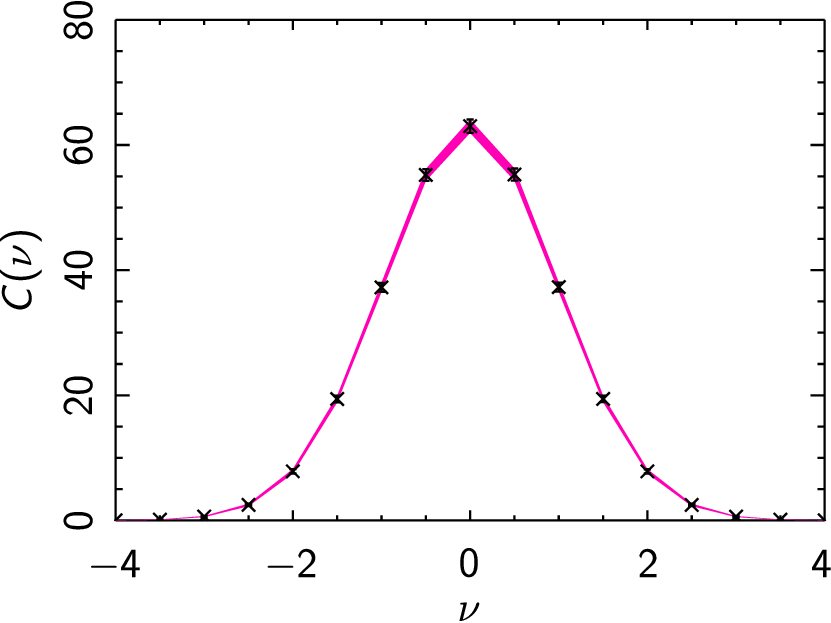}
  \includegraphics[width=5.86cm,angle=0] {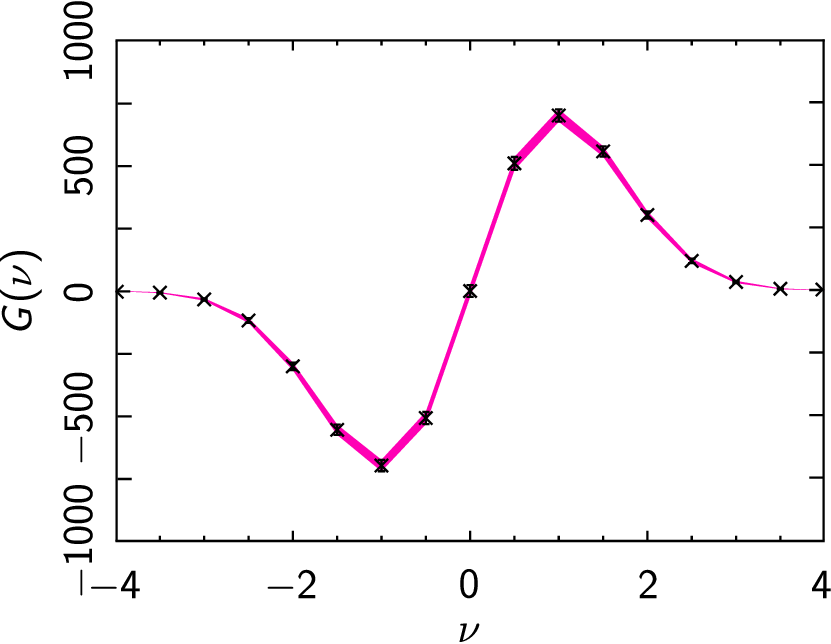}
  \includegraphics[width=5.86cm,angle=0] {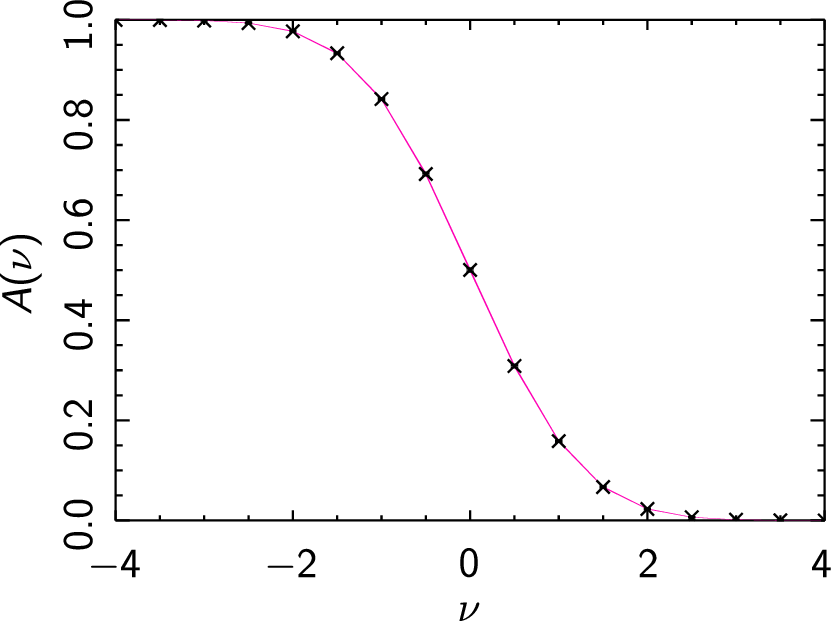}
  \includegraphics[width=5.86cm,angle=0] {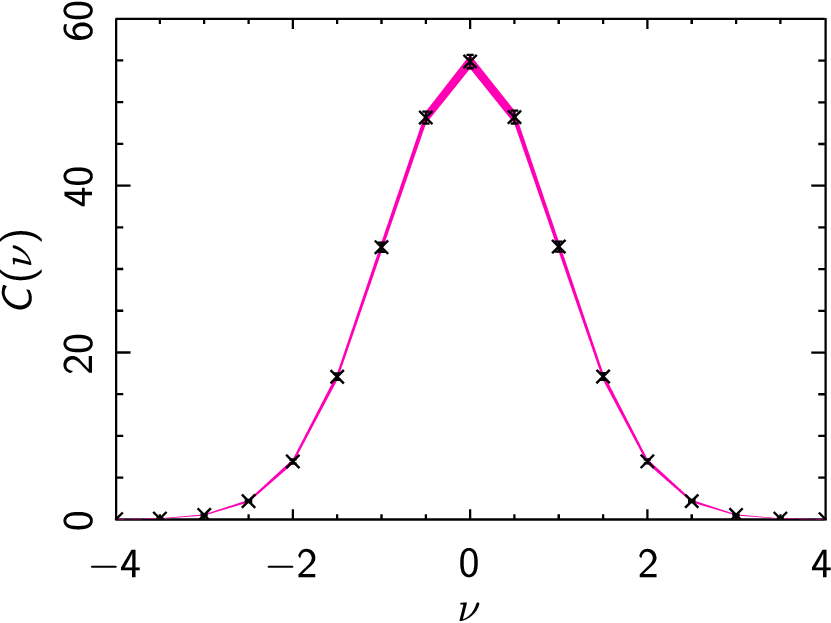}
  \includegraphics[width=5.86cm,angle=0] {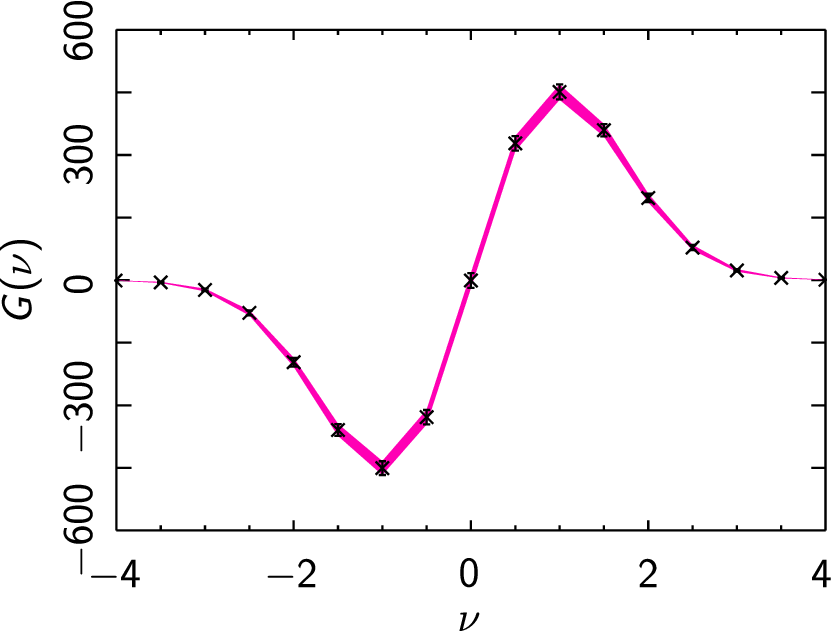}
  \caption{The three Minkowski functionals computed for Gaussian CMB and noise simulations (black symbols) compared with the Minkowski functionals computed for the same simulations with systematic effects added (solid magenta line). From left to right: the area, contour length or perimeter, and the genus. From top to bottom: the three LFI frequencies, 30, 44, and 70\,GHz. These are explicitly for: $N_{\rm side}~=~$512.  }
   \label{minkowski_functionals_global_512}
  \end{center}
\end{figure*}
{Finally we assess non-Gaussianity induced by known systematic effects in the LFI maps. We present results derived with different non-Gaussianity tests carried out at each frequency using the map obtained by summing the various systematic effects considered in this paper.}

For detailed information on the non-directional or targeted non-Gaussianity tests on the Planck data, we refer the interested reader to \citet{planck2013-p09,planck2013-p09a,planck2013-p14}. Here we consider the subset consisting of Minkowski functionals \citep{schmalzing1998}, statistical quantities derived from the 1-point PDF (variance, skewness, kurtosis and the Kolmogorov-Smirnov or KS distance) and the skewness and kurtosis of the spherical Mexican hat wavelet \citep[SMHW,][]{martinez2002}. The properties of these estimators are described in \citet{planck2013-p09} and references therein. 

We compare the values of our estimators derived from a set of ideal Gaussian CMB and noise realizations with those obtained from the same CMB and noise simulations to which the systematic effect maps are added. The CMB and noise maps were simulated following the \Planck-LFI data processing pipeline \citep{planck2013-p02}. Where the three estimators did not provide significant deviations between the maps with and without the systematic effects, we have carried out an additional test by rescaling the systematic effect maps with a constant factor in order to provide an estimate of the amplitude required to detect significant deviations with respect to the CMB signal (i.e., larger than $3 \sigma$ or 99\% confidence level).

Figure \ref{minkowski_functionals_global_512} shows the three Minkowski functionals for the three LFI frequency bands. In each panel we compare the $\pm 1\,\sigma$ (68\%) confidence band centred on the mean corresponding to the Gaussian CMB plus noise simulations, with the same simulations with systematic effects added for \texttt{HEALPix} resolutions $N_{\rm side}~=512,\,256$, and 128. 
Our analysis shows that the estimators based on the Minkowski functionals are not affected by the presence of systematic effects in the maps. 

Table \ref{table_1pdf_sys_global} contains the difference, $\Delta$, of the mean of the two distributions (maps with and without systematic effects), normalized by its dispersion {and multiplied by 100}, corresponding to the standard deviation, skewness and kurtosis for the three LFI frequency bands, {as detailed by Eq.~\ref{relative_mean_gauss_test}}:
\begin{equation}
  \Delta \equiv 100 \frac{\langle X_{\rm sys}\rangle - \langle X_{\rm clean}\rangle}{\sigma\left(X_{\rm sys}\right)}.
  \label{relative_mean_gauss_test}
\end{equation}
{Here $X_{\rm sys}$ represents each of the considered statistics corresponding to the maps with systematics effects, pixelized at a \texttt{HEALPix} resolution of $N_{\rm side}~=~$1024, and $X_{\rm clean}$ represents each of the considered statistics corresponding to the maps without systematics effects}. There are no significant deviations, as the distributions corresponding to the two types of map are virtually superimposed. 
%

%
\begin{table}[tmb]         
\begingroup
\newdimen\tblskip \tblskip=5pt
\caption{Impact of systematic effects on mean 1-point PDF estimators}
\label{table_1pdf_sys_global}
\nointerlineskip
\vskip -3mm
\footnotesize
\setbox\tablebox=\vbox{
   \newdimen\digitwidth 
   \setbox0=\hbox{\rm 0} 
   \digitwidth=\wd0 
   \catcode`*=\active 
   \def*{\kern\digitwidth}
   \newdimen\signwidth 
   \setbox0=\hbox{+} 
   \signwidth=\wd0 
   \catcode`!=\active 
   \def!{\kern\signwidth}
{\tabskip=0pt
\halign{ 
\hbox to 1.in{#\leaderfil}\tabskip=0em& 
\hfil#\hfil\tabskip=.6em& 
\hfil#\hfil& 
\hfil#\hfil
 \cr       
\noalign{\doubleline}
\omit Frequency [GHz] \hfil&
30& 
44&
70\cr    
\noalign{\vskip -3pt}
\noalign{\vskip 5pt\hrule\vskip 3pt}
\noalign{\vskip 2pt}
Standard deviation & $-${6.59}& $-${0.78}& $-${1.22} \cr
Skewness & $-${2.13}& $-${0.94}& $-${1.00}  \cr
Kurtosis & $-${2.46}&  $-${0.19}& $-${0.59} \cr
\noalign{\vskip 5pt\hrule\vskip 3pt}
}
}}
\endPlancktable    
\tablenote a { Values represent the normalized difference ({multiplied by 100}) of the mean for the skewness and the kurtosis for each scale of the SMHW, considering maps with and without systematic effects.}\par
\endgroup
\end{table}              

\newdimen\tblskip \tblskip=5pt

Table \ref{table_smhw_sys_global} shows the difference of the mean of the two distributions (maps with and without systematic effects), normalized by its dispersion {and multiplied by 100 (see Eq.~\ref{relative_mean_gauss_test})}, and corresponding to the skewness and the kurtosis of the SMHW for the LFI frequency bands. The list of angular scales selected for this analysis is the one used in \citet{planck2013-p09a} for the $f_{\rm nl}$ estimation, and comprises 16 angular scales between $1.3\arcm$ and $956.3\arcm$ with logarithmic spacing. Again, no significant deviations are seen. {As an additional check, the sum of the systematic effects at 70 GHz have been directly translated into a $f_{\rm nl}$ estimate for the local shape, resulting in a value of $\Delta f_{\rm nl} =-0.06$ and a relative deviation of $\Delta f_{\rm nl}/\sigma(f_{\rm nl})=-0.003$; the impact of known systematics effects at 70\~GHz on primordial non-Gaussianity is negligible. }

%

\begin{table}[tmb]       
\begingroup
\newdimen\tblskip \tblskip=5pt
\caption{Impact of systematic effects on skewness and kurtosis at various angular scales.}
\label{table_smhw_sys_global}
\nointerlineskip
\vskip -3mm
\footnotesize
\setbox\tablebox=\vbox{
   \newdimen\digitwidth 
   \setbox0=\hbox{\rm 0} 
   \digitwidth=\wd0 
   \catcode`*=\active 
   \def*{\kern\digitwidth}
   \newdimen\signwidth 
   \setbox0=\hbox{+} 
   \signwidth=\wd0 
   \catcode`!=\active 
   \def!{\kern\signwidth}
{\tabskip=0pt
\halign{ 
\hfil#\hfil\tabskip=.6em& 
\hfil#\hfil& 
\hfil#\hfil& 
\hfil#\hfil& 
\hfil#\hfil& 
\hfil#\hfil& 
\hfil#\hfil
 \cr         
\noalign{\doubleline}
Aungular scale&
\multispan3\hfil Skewness\hfil& 
\multispan3\hfil Kurtosis\hfil\cr 
$[\arcm]$&
\multispan3\hrulefill&
\multispan3\hrulefill\cr
&30\,GHz&
44\,GHz&
70\,GHz&
30\,GHz&
44\,GHz&
70\,GHz\cr   
\noalign{\vskip 5pt\hrule\vskip 3pt}
\noalign{\vskip 2pt}
**1.3&           $-$0.13&           $-$0.20&           $-$0.06&           $-$0.39&           $-$0.40 &           $-$0.25 \cr 
\noalign{\vskip 4pt}
**2.1&           $-$0.09&            !0.09&            !0.58&            !0.08&           $-$0.42 &           $-$0.10 \cr 
\noalign{\vskip 4pt}
**3.4&           $-$0.29&            !0.13&            !0.74&            !0.61&           $-$0.04 &           $-$0.06 \cr 
\noalign{\vskip 4pt}
**5.4&           $-$2.25&           $-$0.08&            !0.99&            !0.60&            !0.04 &           $-$0.88 \cr 
\noalign{\vskip 4pt}
**8.7&           $-$7.52&            !0.05&            !4.39&            !1.30&            !0.96 &           $-$6.62 \cr 
\noalign{\vskip 4pt}
*13.9&           $-$6.52&            !0.35&            !1.86&           $-$0.20&            !0.75 &           $-$2.38 \cr 
\noalign{\vskip 4pt}
*22.3&           $-$1.23&           $-$0.05&            !0.07&           $-$0.79&           $-$0.17 &           $-$0.32 \cr 
\noalign{\vskip 4pt}
*35.6&           $-$0.20&            !0.11&           $-$0.12&           $-$0.44&           $-$0.04 &           $-$0.06 \cr 
\noalign{\vskip 4pt}
*57.0&            !0.19&            !0.23&           $-$0.21&           $-$0.25&           $-$0.09 &           $-$0.16 \cr 
\noalign{\vskip 4pt}
*91.2&            !0.19&            !0.26&            !0.02&           $-$0.02&            !0.25 &           $-$0.08 \cr 
\noalign{\vskip 4pt}
146.0&           $-$0.07&            !0.09&            !0.14&           $-$0.13&            !0.07 &            !0.07 \cr 
\noalign{\vskip 4pt}
233.5&           $-$0.06&           $-$0.09&            !0.33&           $-$1.34&            !0.05 &           $-$0.33 \cr 
\noalign{\vskip 4pt}
373.6&            !0.45&            !0.28&            !0.20&           $-$1.56&            !0.08 &           $-$0.21 \cr 
\noalign{\vskip 4pt}
597.7&            !0.80&            !0.36&            !0.16&           $-$0.01&            !0.27 &           $-$0.20 \cr 
\noalign{\vskip 4pt}
956.3&            !0.08&            !0.21&            !0.25&           $-$0.05&           $-$0.28 &            !0.12 \cr 
\noalign{\vskip 5pt\hrule\vskip 3pt}
}
}}
\endPlancktable       
\tablenote a Values represent the normalized difference ({multiplied by 100}) of the mean for the skewness and the kurtosis for each scale of the SMHW, considering maps with and without systematic effects.\par
\endgroup
\end{table}

To conclude, we characterize the levels of detectability of the non-Gaussian contamination of these systematic effect maps. Adopting the 1-point PDF and Minkowski functionals statistics, we employ simulations with different levels of systematic effects,
\begin{equation}
  \Delta T({\bf n}) = \Delta T_{\rm CMB}({\bf n}) + \Delta T_{\rm noise}({\bf n}) + f \Delta T_{\rm syst}({\bf n}),
\end{equation}
to estimate the factor $f$ at which level the systematic effect is detectable. This level is taken to be the value of $f$ for which any of the estimators is outside the $3 \sigma$ confidence level of the values corresponding to maps without systematic effects.
The results indicate that the minimum values of $f$ are $f \sim 8$ at 30\,GHz, $f \sim 12$ at 44\,GHz, and $f \sim 7$ at 70\,GHz. To conclude, systematic effects do not generate significant levels of non-gaussianity for the temperature maps at the three LFI frequencies.

\section{Conclusions}

  In this paper we analyse and quantify the uncertainties on \Planck-LFI CMB temperature anisotropy measurements arising from systematic effects along two complementary approaches. On the one hand, we adopt a \textit{top-down} approach, in which spurious excess signals are highlighted by a series of dedicated null-tests in which maps containing the same sky signal are differenced to obtain maps containing noise and systematic effect residuals. On the other hand, we follow a \textit{bottom-up} approach in which each {known} effect is simulated in terms of timelines and maps.

Our analysis shows that systematic effect uncertainties are at least two orders of magnitudes below the CMB temperature anisotropy power spectrum. The two dominant effects are straylight pick-up from far sidelobes and imperfect photometric calibration. In this current data release the sidelobe signal is not removed from the data, although the CMB dipole pickup by far sidelobes is accounted for during the calibration process using a monochromatic model.

Statistical analyses performed on maps containing the sum of all the simulated systematic effects added to a simulated CMB map showed no detectable non Gaussianity levels unless their level was artificially increased by a factor ranging from 7 to 12. This confirms that instrumental effects do not significantly impact Gaussianity studies.

{Survey difference maps show a signal excess in the multipole range $\ell<20$ that is only partially accounted for in the simulated maps}. This excess could be caused by yet un-modelled straylight pick-up affecting the measurements both directly and in the photometric calibration process.

Currently, analysis focuses on understanding, and further reducing, the level of systematic uncertainties in view of the 2014 data release which will include polarisation data and results. Areas of activity include a more thorough in-band modelling of the sidelobe response at all frequencies, aimed at removing the spurious signal from timelines and a better correction in the calibration step.

\begin{acknowledgements}

The development of Planck has been supported by: ESA; CNES and CNRS/INSU-IN2P3-INP (France); ASI, CNR, and INAF (Italy); NASA and DoE (USA); STFC and UKSA (UK); CSIC, MICINN, JA and RES (Spain); Tekes, AoF and CSC (Finland); DLR and MPG (Germany); CSA (Canada); DTU Space (Denmark); SER/SSO (Switzerland); RCN (Norway); SFI (Ireland); FCT/MCTES (Portugal); and PRACE (EU). We acknowledge the computer resources and technical assistance provided by the Spanish Supercomputing Network nodes at Universidad de Cantabria and Universidad Polit\'ecnica de Madrid as well as the support provided by the Advanced Computing and e-Science team at IFCA. A description of the Planck Collaboration and a list of its members, including the technical or scientific activities in which they have been involved, can be found at \url{http://www.sciops.esa.int/index.php?project=planck&page=Planck_Collaboration}.
  
\end{acknowledgements}

\bibliographystyle{aa}
\bibliography{Planck_bib,custom}

\appendix
\section{Theory of the ADC non-linearity effect}
\label{app_adc_formalism}

ADC non-linearity arises when the measured detector voltage differs from the true voltage in some repeatable manner, depending on the exact values of the voltage thresholds of the chip. By mapping the apparent voltage, $V^{\prime}$, to the true voltage, $V$, the ADC effect can be corrected and this mapping is precisely the ADC response curve, $R(V^{\prime})$, as measured through the LFI acquisition system. In a perfect radiometer this voltage is the product of the system temperature, $T_{\rm sys}$, and radiometer gain, $G(t)$,
\begin{equation}
V = V^\prime R(V^\prime) = G(t)\,T_{\rm sys}.
\end{equation}

Probing the response function requires tracking small known input voltage variations, $\Delta V$, in terms of a measured $\Delta V^{\prime}$ at various working voltages, $V^\prime$. This can be illustrated by differentiating the above equation with respect to $V^{\prime}$,
\begin{equation}
\Delta V = \left( V^\prime \frac{\mathrm{d}R(V^\prime)}{\mathrm{d}V^\prime} + R(V^\prime) \right) \Delta V^\prime = G(t)\,\Delta T.
\label{eq_adv_deltav}
\end{equation}
Equation~\eqref{eq_adv_deltav} shows the relation between the differential input and output signals, and illustrates how a localized gradient change can dominate via the $\mathrm{d}R/\mathrm{d}V^\prime$ term. It also shows that small intrinsic thermal noise fluctuations, $\Delta T$, can be used as a test input temperature signal, assuming it is due to bandwidth limited noise power, $\Delta T = T_{\rm sys}/\sqrt{\Delta\nu\,\tau}$, where $\Delta\nu$ and $\tau$ are channel bandwidth and sample integration time, respectively. By combining the two previous equations, the differential response can be expressed as
\begin{equation}
\frac{\mathrm{d} R(V^\prime)}{\mathrm{d}V^\prime} =  \left( \frac{1}{\sqrt{\Delta\nu\,\tau} \Delta V^\prime}  - \frac{1}{V^\prime}  \right) R(V^\prime). 
\end{equation}

In the case of no ADC effects and voltage variations induced purely through gain fluctuations, we have $V^\prime=\sqrt{\Delta\nu\,\tau}\,\Delta V^\prime$, and the differential response $\mathrm{d}R(V^\prime)/\mathrm{d}V^\prime$ remains zero for all $V^\prime$, as expected. Non-linearities are signaled where the thermal white noise does not follow detector voltages, revealing variations in the response curve. Since the radiometer gains drift very slowly, many estimates of white detector noise by Fourier analysis from the one minute scan rings are available, and by binning and averaging signal-to-noises of $\approx$ 100 are achievable. The above equation can be integrated numerically, making use of these binned values, as $R(V^\prime)\approx1$ is a good approximation. A discrete set of corrected voltages $V_k$ for each binned measured voltages $V_i^\prime$ can be found via a trapezoidal summation,
\begin{equation}
  V_k =  V_0^\prime + \frac{\delta V^\prime}{2} \sum_{i=1}^k a \left( \frac{1}{\Delta V_{i-1}^\prime} + \frac{1}{\Delta V_i^\prime} \right) - \left( \frac{1}{V_{i-1}^\prime} + \frac{1}{V_i^\prime} \right).
\end{equation}
Here $V_0^\prime$ is lowest voltage bin, $\delta V^\prime$ is the voltage bin width, and $a = 1/\sqrt{\Delta\nu\,\tau}$ is fitted such that the top voltage bin, $V_{\rm max}$, is equal to $V_{\rm max}^\prime$ to maintain the same overall linear response. The tables of corrected voltages to measured voltages for each detector are stored in the  DPC database as the ADC correction, and are implemented as spline fits when correcting the time-ordered data.

\section{ADC error before and after correction}
\label{app_adc_error_befor_after_correction}
 
  {An estimator of the magnitude of the ADC effect is the relative variation in the white noise ratio of ``sky'' samples to the ``reference load'' samples. In fact, this removes the effect of the noise variations by comparing the ADC linearity between two well separated voltage levels. These estimates are given in Table~\ref{tab_adc_ratios_before_after_correction} for all the LFI detectors, before and after the correction. In boldface we show the channels that were actually corrected for this effect.}
  
  {Some channels that were not corrected for the ADC effect because the total power data were affected by so-called ``pop-corn'' noise, i.e. random jumps in the total power voltage that were irrelevant for the map-making\footnote{These jumps were present both in the sky and reference load data so that they were effectively removed by differencing.} but not for the ADC removal algorithm that is based on the total power data.}

\begin{table*}                    
\begingroup
\newdimen\tblskip \tblskip=5pt
\caption{Ratio of sky to ref white noise before and after correction}
\label{tab_adc_ratios_before_after_correction}
\nointerlineskip
\vskip -3mm
\footnotesize
\setbox\tablebox=\vbox{
\newdimen\digitwidth 
\setbox0=\hbox{\rm 0} 
\digitwidth=\wd0 
\catcode`*=\active 
\def*{\kern\digitwidth}
\newdimen\signwidth 
\setbox0=\hbox{+} 
\signwidth=\wd0 
\catcode`!=\active 
\def!{\kern\signwidth}
\halign{\hbox to 1.3in{#\leaderfil}\tabskip=2em&
\hfil#\hfil&
\hfil#\hfil&
\hfil#\hfil&
\hfil#\hfil&
\hfil#\hfil&
\hfil#\hfil&
\hfil#\hfil&
\hfil#\hfil&
\hfil#\hfil&
\hfil#\hfil&
\hfil#\hfil&
\hfil#\hfil\tabskip=0pt\cr 
\noalign{\doubleline}
\omit&\multispan2\hfil ADC E{\sc rror} B{\sc efore}\hfil & \multispan2\hfil ADC E{\sc rror} A{\sc fter}\hfil\cr
\omit&\hfil D{\sc etector} 0\hfil & \hfil D{\sc etector} 1\hfil & 
\hfil D{\sc etector} 0\hfil & \hfil D{\sc etector} 1\hfil \cr
\noalign{\vskip -4pt}
\omit&\hrulefill&\hrulefill&\hrulefill&\hrulefill\cr
\omit& Ratio & Ratio & Ratio & Ratio \cr
\omit\hfil \mbox{} \hfil&[\%]&[\%]&[\%]&[\%]\cr
\noalign{\vskip 3pt\hrule\vskip 5pt}
\omit{\bf 70\,GHz}\hfil\cr
\noalign{\vskip 4pt}
\hglue 2em LFI18M  & 0.27 & 0.20 & 0.56 & 0.42 \cr
\hglue 2em LFI18S  & 0.40 & 0.53 & 0.36 & 0.40 \cr
\hglue 2em LFI19M  & 0.38 & 0.44 & \bf 0.09 & \bf 0.18 \cr
\hglue 2em LFI19S  & 0.88 & 1.33 & 0.12 & 0.53 \cr
\hglue 2em LFI20M  & 0.30 & 0.28 & 0.20 & 0.39 \cr
\hglue 2em LFI20S  & 0.38 & 0.24 & 0.34 & 0.26 \cr
\hglue 2em LFI21M  & 0.69 & 0.77 & \bf 0.16 & \bf 0.29 \cr
\hglue 2em LFI21S  & 1.45 & 0.88 & \bf 0.52 & \bf 0.54 \cr
\hglue 2em LFI22M  & 0.60 & 1.51 & \bf 0.13 & \bf 0.16 \cr
\hglue 2em LFI22S  & 1.45 & 1.06 & 1.74 & 2.16 \cr
\hglue 2em LFI23M  & 0.86 & 0.65 & \bf 0.70 & \bf 0.42 \cr
\hglue 2em LFI23S  & 0.58 & 0.76 & \bf 0.24 & \bf 0.26 \cr
  \omit{\bf 44\,GHz}\hfil\cr
  \noalign{\vskip 4pt}
\hglue 2em LFI24M  & 2.18 & 0.62 & \bf 0.06 & \bf 0.10 \cr
\hglue 2em LFI24S  & 2.43 & 2.67 & \bf 0.49 & \bf 0.09 \cr
\hglue 2em LFI25M  & 1.04 & 6.95 & \bf 0.13 & \bf 0.11 \cr
\hglue 2em LFI25S  & 2.75 & 5.24 & \bf 0.10 & \bf 0.09 \cr
\hglue 2em LFI26M  & 0.57 & 3.27 & \bf 0.23 & \bf 0.10 \cr
\hglue 2em LFI26S  & 1.61 & 3.05 & \bf 0.12 & \bf 0.08 \cr
    \omit{\bf 30\,GHz}\hfil\cr
  \noalign{\vskip 4pt}
\hglue 2em LFI27M  & 1.39 & 0.45 & \bf 0.17 & \bf 0.15 \cr
\hglue 2em LFI27S  & 0.70 & 0.94 & \bf 0.18 & \bf 0.15 \cr
\hglue 2em LFI28M  & 0.64 & 1.29 & 0.12 & 0.16\cr
\hglue 2em LFI28S  & 0.55 & 0.95 & \bf 0.13 & \bf 0.18 \cr
  \noalign{\vskip 5pt\hrule\vskip 3pt}}}
  \endPlancktable
  \endgroup
  \end{table*}

\raggedright

\end{document}